\documentclass[prd,aps,superscriptaddress,preprintnumbers,eqsecnum,nofootinbib,notitlepage]{revtex4-2}
\usepackage[utf8]{inputenc}
\usepackage{amsfonts,amsmath,amssymb,bm,subfigure}
\usepackage{multirow,hyperref}
\usepackage{xcolor}

\usepackage[T1]{fontenc} 
\usepackage{amsmath, amssymb, bm}
\usepackage{braket}
\usepackage[normalem]{ulem}
\usepackage{latexsym}
\usepackage{wrapfig}
\usepackage{tabularx}
\usepackage{comment}
\usepackage{tikz}
\usetikzlibrary{calc,decorations.pathmorphing,decorations.markings}
\usepackage{tikz-feynhand}
\usepackage{cancel}
\usepackage{orcidlink}

\usepackage[most]{tcolorbox}
\usepackage{colortbl}

\hypersetup{
  colorlinks = true,
  linkcolor = red,
  urlcolor  = magenta,
  citecolor = blue,
  anchorcolor = blue
}

\newcommand{\rd}{{\rm d}}
\newcommand{\be}{\begin{equation}}
\newcommand{\ee}{\end{equation}}
\newcommand{\ba}{\begin{eqnarray}}
\newcommand{\ea}{\end{eqnarray}}

\newcommand{\nn}{\nonumber \\}
\newcommand{\Mpl}{M_{\rm Pl}}
\newcommand{\D}{{\rm d}}

\newcommand{\bX}{X_{\rm BG}}
\newcommand{\btX}{\tilde{X}_{\rm BG}}

\newcommand{\pim}{\pi_{\text{m}}}
\newcommand{\alpham}[1]{\alpha_{m_#1}}
\newcommand{\Omegac}{\Omega_{\text{c}}}
\newcommand{\deltac}{\delta_{\text{c}}}
\newcommand{\deltacp}{\dot{\delta}_{\text{c}}}
\newcommand{\deltacpp}{\ddot{\delta}_{\text{c}}}

\newcommand{\Rs}{{}^{(3)}\!R}

\begin{document}
\baselineskip=12pt

\preprint{YITP-25-56, WUCG-25-04}

\title{Effective field theory of coupled 
dark energy and dark matter}

\author{Katsuki Aoki\orcidlink{0000-0002-9616-096X}}
\email{katsuki.aoki@yukawa.kyoto-u.ac.jp}
\affiliation{Center for Gravitational Physics and Quantum Information, Yukawa Institute for Theoretical Physics, Kyoto University, 606-8502, Kyoto, Japan}

\author{Jose Beltr\'an Jim\'enez\orcidlink{0000-0002-3053-9213
}}
\email{jose.beltran@usal.es}
\affiliation{Departamento~de~F{\'i}sica~Fundamental~and~IUFFyM,~Universidad~de~Salamanca,~E-37008~Salamanca,~Spain}

\author{Masroor C. Pookkillath\orcidlink{0000-0002-7199-8037}}
\email{masroorcp@sogang.ac.kr}
\affiliation{Centre for Theoretical Physics and Natural Philosophy, Mahidol University, Nakhonsawan Campus,  Phayuha Khiri, Nakhonsawan 60130, Thailand}
\affiliation{Center for Quantum Spacetime, Sogang University, 35 Baekbeom-ro, Mapo-gu, Seoul 04107, Korea}

\author{Shinji~Tsujikawa\orcidlink{0000-0002-9378-2229}}
\email{tsujikawa@waseda.jp}
\affiliation{Department of Physics, Waseda University, 3-4-1 Okubo, Shinjuku, Tokyo 169-8555, Japan}




\date{\today}

\begin{abstract}
We formulate an effective field theory (EFT) of coupled dark energy (DE) and dark matter (DM) interacting through energy and momentum transfers. In the DE sector, we exploit the 
EFT of vector-tensor theories with the 
presence of a preferred time direction 
on the cosmological background.
This prescription allows one to accommodate 
shift-symmetric and non-shift-symmetric 
scalar-tensor theories by taking a particular 
weak coupling limit, with and without  
consistency conditions respectively.
We deal with the DM sector as 
a non-relativistic perfect fluid, 
which can be described by a system 
of three scalar fields. By choosing 
a unitary gauge in which the perturbations 
in the DE and DM sectors are eaten by the 
metric, we incorporate the leading-order 
operators that characterize the energy and 
momentum transfers besides those present 
in the conventional EFT of vector-tensor 
and scalar-tensor theories and the 
non-relativistic perfect fluid. 
We express the second-order action of 
scalar perturbations in real space 
in terms of time- and scale-dependent dimensionless EFT parameters and derive the linear perturbation 
equations of motion by taking into account additional matter (baryons, radiation). In the small-scale 
limit, we obtain conditions for the 
absence of both ghosts and Laplacian 
instabilities and discuss how they are 
affected by the DE-DM interactions. 
We also compute the effective DM gravitational 
coupling $G_{\rm eff}$ by using a quasi-static 
approximation for perturbations deep inside 
the DE sound horizon and show that the 
existence of momentum and energy transfers 
allow a possibility to realize $G_{\rm eff}$ 
smaller than in the uncoupled case 
at low redshift.
\end{abstract}


\maketitle
\flushbottom

\section{Introduction}

Despite the tremendous observational development 
over the past few decades, the origins of 
neither dark energy (DE) nor dark matter (DM)    
have been identified yet. DE is the source
for late-time cosmic acceleration, while DM 
is responsible for the gravitational clustering 
of large-scale structures. 
The properties of DE and DM can be probed 
not only by the distance measurements 
like supernovae type-Ia~\cite{SupernovaSearchTeam:1998fmf,SupernovaCosmologyProject:1998vns,Brout:2022vxf,Rubin:2023ovl,DES:2024jxu}
and baryon acoustic 
oscillations~\cite{SDSS:2005xqv,2dFGRS:2005yhx,Blake_2011,eBOSS:2020yzd,DESI:2024mwx,DESI:2025zgx}  
but also by the observations of inhomogeneities in the Universe such as Cosmic Microwave Background (CMB) temperature 
anisotropies~\cite{WMAP:2003elm,Planck:2013pxb,Planck:2018vyg}, galaxy clustering, 
and weak lensing~\cite{SDSS:2003eyi,Kuijken:2015vca,Hildebrandt:2016iqg,DES:2017myr,KiDS:2020suj,DES:2021wwk,Dalal:2023olq,Li:2023tui,Wright:2025xka}.

The standard cosmological paradigm for the dark sector of the Universe is known as 
the $\Lambda$CDM model~\cite{Peebles:1982ff,Peebles:1984ge}. 
In this scenario, DE and DM are described by the cosmological constant $\Lambda$ and 
Cold Dark Matter (CDM), 
respectively, without their direct interactions. This model is highly successful in explaining our Universe with minimal six parameters.
However, it has started showing some tensions in estimating its parameters from different observations~\cite{Perivolaropoulos:2021jda,Abdalla:2022yfr,DiValentino:2025sru}. In particular, the estimation 
of today’s Hubble expansion 
rate $H_0$ between the CMB data (Planck 2018~\cite{Planck:2018vyg} and the Atacama Cosmology Telescope DR6~\cite{ACT:2025fju})
and low-redshift measurements~\cite{Riess:2021jrx,Wong:2019kwg,Riess:2020fzl,Verde:2019ivm,DiValentino:2021izs,Perivolaropoulos:2021jda,Freedman:2021ahq} exhibits different levels of tension depending on the probes, being the biggest discrepancy of more than $5\sigma$
with SH0ES value~\cite{Riess:2021jrx}.

On top of that, the amplitude of matter density perturbations, measured by the parameter $\sigma_8$, shows some tension within $\Lambda$CDM, between the CMB observations and low-redshift probes like 
shear-lensing~\cite{Heymans:2012gg,Hildebrandt:2016iqg,Abbott:2017wau}
and redshift-space distortions of 
galaxies~\cite{Samushia:2013yga,Macaulay:2013swa} (see e.g. \cite{Perivolaropoulos:2021jda,Freedman:2021ahq} for a comprehensive review), a discrepancy that can be up to $3\sigma$. The statistical significance of this discrepancy is not firmly established and one could argue that it is a mild tension that could be explained by a statistical fluke or some systematics, but it could also be an incipient signal of some unknown physical mechanism.\footnote{It is worth mentioning that the original tension found by the KiDS collaboration with Planck2018 data has disappeared in the latest KiDS-Legacy analysis~\cite{Wright:2025xka}.} 

Finally, recent observations from the Dark Energy Spectroscopic Instrument (DESI) collaboration suggest that the nature of  DE may deviate from the standard cosmological constant~\cite{DESI:2024mwx} and shows a preference for dynamical dark energy 
over $\Lambda$CDM at a significance of 
$2.8-4.2\sigma$~\cite{DESI:2025zgx}. These tensions as well as the DESI observation motivate investigating alternatives to $\Lambda$CDM with a potential dynamical origin of DE featuring propagating 
degrees of freedom (DOFs).\footnote{There are also models that can modify the background dynamics with respect to $\Lambda$CDM without introducing any additional degrees of freedom, see e.g. Refs.~\cite{DeFelice:2015hla, Aoki:2018brq, Aoki:2020oqc,Aoki:2020lig,DeFelice:2020cpt, DeFelice:2022mcd}.}
In these scenarios, a pertinent question to be addressed is whether these DOFs exhibit direct interactions with the other component of the dark sector, namely DM, through energy and momentum exchanges.

A representative class of dynamical DE models 
is based on a scalar field 
$\phi$, whose potential or kinetic energies can drive the cosmic acceleration~\cite{Fujii:1982ms,Ratra:1987rm,Wetterich:1987fm,Chiba:1997ej,Ferreira:1997au,Caldwell:1997ii,Armendariz-Picon:1999hyi,Chiba:1999ka,Armendariz-Picon:2000nqq,Arkani-Hamed:2003pdi,Piazza:2004df}. 
For instance, Horndeski theories~\cite{Horndeski:1974wa} 
are the most general scalar-tensor theories 
with second-order field equations 
of motion~\cite{Deffayet:2011gz,Kobayashi:2011nu,Charmousis:2011bf}. In such theories, there is one scalar propagating DOF besides the two  
polarizations of tensor modes 
arising from the gravity sector. 
These Horndeski theories can be extended to construct healthy theories beyond Horndeski without increasing 
the propagating DOFs ~\cite{Zumalacarregui:2013pma,Gleyzes:2014dya,Langlois:2015cwa,Crisostomi:2016czh,
BenAchour:2016fzp}. 
Instead of a scalar field, we can also 
consider a massive vector field $A_{\mu}$ 
as the source for DE. In generalized Proca (GP) theories~\cite{Tasinato:2014eka,Heisenberg:2014rta, Allys:2015sht, BeltranJimenez:2016rff, Allys:2016jaq},\footnote{The GP interactions can be geometrically constructed from generalized 
Lovelock terms in Weyl and vector-distorted geometries~\cite{BeltranJimenez:2014iie,BeltranJimenez:2015pnp}. These geometrical realizations were in turn the original motivation for the construction of general GP interactions.} 
for example, the accelerated cosmic expansion 
can be realized by derivative interactions of 
a massive vector field with broken $U(1)$ 
gauge invariance~\cite{BeltranJimenez:2016wxw,DeFelice:2016yws}. 
In GP theories, there are 
five propagating DOFs arising from 
one longitudinal scalar mode, two transverse 
vector modes, and two tensor polarizations. 
Healthy extensions of GP theories are also possible without invoking the Ostrogradski-type 
instability~\cite{Heisenberg:2016eld,Kimura:2016rzw,deRham:2020yet}.

The observational signatures of dynamical 
scalar or vector DE models have been extensively studied in the literature. 
At the background level, 
quintessence~\cite{Zlatev:1998tr,Steinhardt:1999nw,Caldwell:2005tm}
and k-essence~\cite{Chiba:1999ka,Armendariz-Picon:2000nqq,Armendariz-Picon:2000ulo,Piazza:2004df} give rise to the time-varying 
DE equation of state in the range 
$w_{\rm DE} \ge -1$. 
In several subclasses of Horndeski theories, 
GP theories, and their extensions, it is 
possible to realize $w_{\rm DE} \le -1$ 
without having ghosts~\cite{Gannouji:2006jm,Hu:2007nk,Starobinsky:2007hu,Tsujikawa:2007xu,Tsujikawa:2008uc,DeFelice:2010pv,Peirone:2019aua,deFelice:2017paw,Nakamura:2017dnf, DeFelice:2020sdq,deRham:2021efp}. 
Thus, these models can be observationally distinguished from the cosmological constant. 
At the level of perturbations, the effective gravitational coupling of sub-horizon CDM  
perturbations in Horndeski theories is 
enhanced by the scalar-CDM interaction 
mediated by 
gravity~\cite{DeFelice:2011hq,Tsujikawa:2015mga,Amendola:2017orw,Kase:2018aps}. 
This property also holds 
for a subclass of GP theories in which the 
speed of tensor perturbations $c_{\rm T}$ 
is luminal~\cite{DeFelice:2016uil,Amendola:2017orw,Nakamura:2018oyy}. Since the observations of gravitational waves along with the electromagnetic 
counterpart~\cite{LIGOScientific:2017zic} 
impose that $c_{\rm T}$ is very close to the speed of light (see also~\cite{BeltranJimenez:2015sgd} for a constraint based on binary pulsar observations), the gravitational coupling of 
CDM in GP theories 
is enhanced compared to the $\Lambda$CDM model. 
However, the resolution of 
the $\sigma_8$ tension requires that 
the cosmic growth rate at low redshift 
is weaker than in the $\Lambda$CDM model.\footnote{The fact that the resolution to the $\sigma_8$ tension should come from low-redshift effects is motivated by the ACT measurements~\cite{ACT:2023kun} that are compatible with Planck2018 data and whose sensitivity peaks at $z\simeq2$.} 
Thus, in Horndeski and GP theories without 
direct interactions between DE and DM, 
it is difficult to realize a suppression of the DM clustering that could alleviate the $\sigma_8$ tension.

In interacting scenarios where DE and DM have 
direct couplings, the growth of DM perturbations 
is subject to modifications 
(see e.g., Ref.~\cite{Wang:2024vmw} for a status report). For example, 
the CDM density $\rho_{\rm c}=m_{\rm c} 
n$ can have an interaction with the 
DE scalar field $\phi$ of the form 
$f_1(\phi) n$, where $n$ 
and $m_{\rm c}$ are the number density 
and the mass of CDM, respectively, and 
$f_1$ is a function of 
$\phi$~\cite{Pourtsidou:2013nha,Boehmer:2015kta,Kase:2019veo,Amendola:2020ldb,Kase:2020hst}. 
This mediates the energy transfer between 
DE and CDM. The CDM four-velocity 
$u^{\mu}$ can have a coupling 
with the scalar-field derivative 
$\nabla_{\mu} \phi$ through the scalar 
product $Z=u^{\mu} \nabla_{\mu} \phi$, 
where $\nabla_{\mu}$ is a covariant 
derivative operator~\cite{Pourtsidou:2013nha,Boehmer:2015sha,Skordis:2015yra,Koivisto:2015qua,Pourtsidou:2016ico,Dutta:2017kch,Linton:2017ged,Kase:2019veo,Kase:2019mox,Chamings:2019kcl,Amendola:2020ldb,Kase:2020hst,Linton:2021cgd,Liu:2023mwx}.
Indeed, the Lagrangian $f_2(Z)$ can suppress 
the growth rate of CDM perturbations at 
low redshift through the momentum exchange 
between CDM and DE. A similar suppression for the growth of structures is possible in 
perfect fluid models of coupled DE and DM~\cite{Asghari:2019qld,Figueruelo:2021elm,BeltranJimenez:2020iyx,BeltranJimenez:2020qdu,BeltranJimenez:2021wbq,BeltranJimenez:2024lml}. 
For the DE vector field 
$A_{\mu}$, we may consider other forms of 
momentum transfers characterized by the 
products $A_{\mu}u^{\mu}$~\cite{DeFelice:2020icf} 
and $-A^{\mu}F_{\mu \nu}u^{\nu}$~\cite{Pookkillath:2024ycd}.
They can also lead to weak cosmic growth 
rate, whose property is desirable 
for alleviating the $\sigma_8$ tension. 

The aforementioned theories, including both 
uncoupled and coupled DE-DM scenarios, 
give rise to widely spread cosmological 
predictions. The effective field theory (EFT) approach can provide a unified description for dealing with the dynamics of background and perturbations systematically.
The EFT on a Friedmann-Lema\^{i}tre-Robertson-Walker (FLRW) background was first developed in the context of inflation with 
a scalar field~\cite{Weinberg:2008hq,Creminelli:2006xe,Cheung:2007st}. 
By choosing a unitary gauge in which the preferred time is identified with the time-dependent scalar field, the EFT action can be systematically constructed under the invariance of spatial diffeomorphism. Later, the EFT was applied to the scalar-field DE in the presence 
of additional matter like DM~\cite{Creminelli:2008wc,Gubitosi:2012hu,Bloomfield:2012ff,Gleyzes:2013ooa,Piazza:2013coa,Gleyzes:2014rba,Hu:2013twa,Raveri:2014cka,Kase:2014cwa,Frusciante:2015maa}.
Since the dimensionless EFT parameters called the 
$\alpha$-basis 
parameters~\cite{Bellini:2014fua}
have explicit relations with the coupling functions in concrete scalar-tensor theories (Horndeski and its extensions), 
the observational bounds on the former translate to constraints on each DE model.

The EFT of vector-tensor theories has been developed by assuming the existence of 
a preferred direction characterized 
by a timelike vector field 
$v_{\mu}=\nabla_{\mu} \tilde{t}+g_M A_{\mu}$, 
where $\tilde{t}$ is a St\"uckelberg field associated with the $U(1)$ gauge 
transformation and $g_M$ is the gauge 
coupling constant~\cite{Aoki:2021wew} (see 
also Refs.~\cite{Lagos:2016wyv,Lagos:2017hdr}).
The preferred vector $v_{\mu}$ 
is invariant under the local transformations 
$\tilde{t} \to \tilde{t}-g_M \theta$ and 
$A_{\mu} \to A_{\mu}+\nabla_{\mu} \theta$, 
where $\theta$ is an arbitrary scalar 
that depends on the spacetime coordinates. 
By choosing the unitary gauge 
where $\tilde{t}$ coincides with the time 
coordinate $t$, the preferred vector 
reduces to $v_{\mu}=\delta^{0}_{\mu}
+g_M A_{\mu}$. Since $v_{\mu}$ is not 
orthogonal to spacelike hypersurfaces 
in general, this leads to a different 
symmetry-breaking pattern in comparison 
to scalar-tensor theories. 
The EFT of vector-tensor theories not 
only accommodates GP theories and their 
extensions as specific cases but also 
it also recovers the EFT of shift-symmetric 
scalar-tensor 
theories~\cite{Finelli:2018upr} by taking a weak coupling 
limit with certain consistency conditions. 
The conventional EFT of inflation/DE 
in non-shift-symmetric scalar-tensor theories can also be recovered within this framework without imposing the latter consistency conditions.

So far, apart from the multi-component fluid scenarios in Ref.~\cite{Ballesteros:2013nwa} and the formulation based on a disformally coupled 
metric performed in Refs.~\cite{Gleyzes:2015pma,Chibana:2019jrf}, 
the EFT of DE has been formulated
by assuming the absence of direct couplings
between DE and DM. In this paper, we will construct the EFT of coupled DE and DM by introducing leading-order operators 
associated with the energy and momentum 
transfers. For consistency with the 
observation of gravitational waves, 
we focus on the EFT operators leading to the luminal propagation speed of tensor perturbations. 
We will be mainly interested in scalar and tensor perturbations, so we will leave the description of vector perturbations out of our framework. Although vector modes can have interesting effects in some scenarios (see e.g.~Ref.~\cite{Cembranos:2019jlp}), they are largely irrelevant for most models. 
We deal with DM as a pressureless perfect fluid, 
which can be described by 
a specific case of the gravitating continuum. 
Indeed, the EFT for hydrodynamics 
and solids have been developed in 
Refs.~\cite{Endlich:2010hf,Dubovsky:2011sj,Endlich:2012pz,Ballesteros:2012kv,Ballesteros:2014sxa,Kovtun:2014hpa}. 
In particular, Ref.~\cite{Aoki:2022ipw} constructed the EFT of a non-dissipative gravitating continuum, providing a unified framework of fluids, solids, 
and Aether fields. 
For the DE sector, we adopt the unified EFT description of vector-tensor and scalar-tensor 
theories advocated in Ref.~\cite{Aoki:2021wew}.
The DE-DM couplings introduced in our EFT framework can accommodate a large class of concrete coupled DE-DM theories studied in the literature. This formulation includes not only 
GP, shift-symmetric, non-shift-symmetric Horndeski theories but also perfect fluid 
DE models~\cite{BeltranJimenez:2020qdu,BeltranJimenez:2021wbq} by using a purely k-essence 
Lagrangian for scalar perturbations~\cite{Giannakis:2005kr,Arroja:2010wy}.

We construct the EFT action of 
coupled DE and DM by choosing a unitary 
gauge in which the perturbations in the DE 
and DM sectors are eaten by the metric. 
We then derive the 
full scalar perturbation equations of motion 
in Fourier space containing time- and scale-dependent 
dimensionless 
EFT parameters. 
We also obtain conditions for 
the absence of ghosts and Laplacian instabilities of scalar perturbations in the small-scale limit. 
Finally, we compute the effective 
gravitational coupling of DM by using a
quasi-static approximation for the 
modes deep inside the DE sound horizon. Then, we show that the direct couplings between 
DE and DM can lead to a gravitational interaction weaker than in the $\Lambda$CDM model during the epoch of DE domination and also during matter domination. Our EFT approach to coupled DE and DM 
will be useful to provide a general 
framework for addressing the 
$\sigma_8$-tension through DE-DM interactions as well as a systematic study of the presence of non-gravitational interactions in the dark sector that will eventually permit a more optimal exploitation of stage IV experiments data.

This paper is organized as follows. In section~\ref{sec:Buliding_block}, 
after reviewing the EFT of DE and DM fluid 
described by three scalar fields, 
we construct the EFT action of 
coupled DE and DM 
in a general scheme accommodating both vector-tensor 
and scalar-tensor theories for the DE sector. 
We also discuss the consistency conditions 
for the EFT coefficients, which need to be 
imposed for some classes of theories.
Then, we derive the background equations of motion in section~\ref{backsec}, whose explicit forms are 
given in vector-tensor theories and 
scalar-tensor theories (both shift-symmetric and non-shift-symmetric cases). Subsequently, in section~\ref{sec:EFTaction_dimensionless}, 
we express the second-order action of perturbed fields 
in the Arnowitt-Deser-Misner (ADM) language and introduce dimensionless 
$\alpha$-basis parameters in our EFT framework of 
coupled DE and DM.
Then, in section~\ref{sec:perturbation_EOMs}, 
we study the tensor mode propagation and express the second-order EFT action by using scalar 
metric perturbations in the presence of standard matter. 
Subsequently, in section \ref{stasec}, 
we derive conditions for the presence of neither ghosts nor Laplacian 
instabilities in scalar perturbations. 
In the same section, we also obtain the effective gravitational coupling of CDM density perturbations 
$G_{\rm eff}$ for 
the modes deep inside the DE sound horizon.
Then, in section~\ref{intsec}, we investigate
how each EFT coupling function associated 
with energy and momentum transfers affects 
$G_{\rm eff}$. In particular, we show that the momentum transfer EFT coefficients can realize the cosmic growth 
rate weaker than in the $\Lambda$CDM. 
We also briefly discuss the validity of the quasi-static approximation and show that the suppression of structures can also occur beyond 
the quasi-static regime.

In Appendix~\ref{sec:Dictionary}, we provide dictionaries that establish the correspondence between the EFT parameters and concrete theoretical frameworks, specifically Horndeski theories and GP theories. 
In Appendix~\ref{vecsec}, we derive the linear stability conditions of intrinsic vector perturbations 
in the decoupling limit.
Furthermore, in Appendix~\ref{sec:3scalarMatter}, 
we discuss the second-order action of standard matter 
perturbations by using the three-scalar EFT description. 

The relevant results of this paper are summarized 
in Table~\ref{tab:summary_of_results}, where we point to the relevant equations.
Table~\ref{tab:int_alpha_summary} summarizes the interaction coefficients ($\alpha$-basis 
parameters), their roles, and the possibility of 
realizing the CDM gravitational coupling 
weaker than in the $\Lambda$CDM model.
\begin{table}[h]
    \centering
    \begin{tabular}{|c|c|}
        \hline
        \rowcolor[gray]{0.3}
        \textcolor{white}{Relevant results} & \textcolor{white}{Equation numbers} \\
        \hline
        \hline
        Building blocks for coupled DE and DM & 
        Eq.~(\ref{blocks})\\
        \hline
        \rowcolor[gray]{0.9}
        EFT action for coupled DE and DM & Eq.~(\ref{Sfull}) - Eq.~\eqref{EFTD} \\
        \hline
        Consistency conditions & 
        Eq.~(\ref{consistency1Mr}) - 
        Eq.~(\ref{consistency4MD}) \\
        \hline
        \rowcolor[gray]{0.9}
        Total second-order action in $\alpha$-basis parameters & 
        Eq.~(\ref{eq:EQ_EFT_alpha_action}) \\
        \hline
        Tensor mode stability & Eq.~(\ref{eq:NGNLtensor_mode}) \\
        \hline
        \rowcolor[gray]{0.9}
        Second-order scalar action with matter in metric variables & Eq.~(\ref{Ssfinal}) \\
        \hline
        No-ghost conditions for DM and DE & Eq.~(\ref{snoghost1}) - Eq.~(\ref{snoghost2}) \\
        \hline
        \rowcolor[gray]{0.9}
        Speed of propagation for DE & Eq.~(\ref{cst}) \\
        \hline
        Vector mode stability & Eq.~(\ref{eq:Vector_stability}) \\
        \hline
        \rowcolor[gray]{0.9}
        Effective gravitational coupling of CDM & Eq.~(\ref{Geff}) \\
        \hline
        Dictionary to Horndeski theories & Eq.~(\ref{fre}) - Eq.~(\ref{eq:gMmap}) \\
        \hline
        \rowcolor[gray]{0.9}
        Dictionary to GP theories & Eq.~(\ref{eq:DfP}) - Eq.~(\ref{bm1}) \\
        \hline
    \end{tabular}
    \caption{This table summarizes the relevant results and their corresponding equations.}
    \label{tab:summary_of_results}
\end{table}

\begin{table}[h]
    \centering
       \begin{tabular}{|c|c|c|}
       \hline
       \rowcolor[gray]{0.3}
        \textcolor{white}{$\alpha$-basis parameters} & \textcolor{white}{Roles} & \textcolor{white}{Possibility of weak gravity for CDM} \\ 
        \hline
        \hline
        $\alpha_{m_{c}}$ & Energy transfer & No \\ 
        \hline
        \rowcolor[gray]{0.9}
        $\alpha_{m_{1}}$ & Energy transfer & Yes \\ 
        \hline
        $\alpha_{m_{2}}$ & Momentum transfer & Yes \\
        \hline
        \rowcolor[gray]{0.9}
        $\alpha_{\bar{m}_{1}}$ & Momentum transfer & Yes \\
        \hline
    \end{tabular}
    \caption{Summary of the interaction coefficients ($\alpha$-basis parameters), their roles, and 
    the possibility of realizing the CDM gravitational 
    coupling weaker than in the $\Lambda$CDM model.}
    \label{tab:int_alpha_summary}
\end{table}

\section{Building blocks for EFT}\label{sec:Buliding_block}

The EFT action of DE coupled to DM together with minimally coupled matter can be generally 
expressed in the form 
\begin{align}
{\cal S} &=\int \D^4 x \sqrt{-g} \big( 
\mathcal{L}_{\rm DE} + \mathcal{L}_{\rm DM}+ \mathcal{L}_{\rm int} 
+ \mathcal{L}_{\rm m} \big)\,,
\label{Stotal}
\end{align}
where $g$ is the determinant of the metric 
tensor $g_{\mu \nu}$, $\mathcal{L}_{\rm DE}$ 
and $\mathcal{L}_{\rm DM}$ are 
the Lagrangians for DE (including gravity) 
and DM sectors respectively, 
$\mathcal{L}_{\rm int}$ describes 
non-gravitational interactions between DE and DM, and 
the Lagrangian $\mathcal{L}_{\rm m}$ 
corresponds to contributions from other matter fields $\psi_m$ 
such as baryons and radiation, which are assumed to be minimally coupled to gravity, 
$\mathcal{L}_{\rm m} = 
\mathcal{L}_{\rm m}(g, \psi_{\rm m})$. 

Before going into the details of the EFT 
formulation, it would be worth 
mentioning the differences between the EFT presented in this paper and the EFT of interacting DE developed in Refs.~\cite{Gleyzes:2015pma,Chibana:2019jrf}. 
The latter assumes that DE is described by a scalar field $\phi$ and that the DM field $\psi_{\rm DM}$ is minimally coupled to a disformally transformed metric,
\begin{align}
\hat{g}_{\mu \nu}:=C(\phi)g_{\mu\nu}
+ D(\phi) \nabla_{\mu}\phi \nabla_{\nu}\phi\,,
\end{align}
where $C$ and $D$ are functions of $\phi$. 
The EFT of interacting DE is then described by the following action 
\begin{align}
\hat{{\cal S}} =\int \D^4 x \sqrt{-g} \left[\hat{\mathcal{L}}_{\rm DE}(g,\phi) + \hat{\mathcal{L}}_{\rm DM}(\hat{g}, \psi_{\rm DM}) + \mathcal{L}_{\rm m}(g, \psi_{\rm m}) \right]\,.
\end{align}
However, there would be no a priori reason for the interaction between DE and DM being described by the disformally transformed metric. 
In addition, DE does not necessarily need to be a scalar field. 
In this paper, we consider more general cases 
in which the source for DE can also be a vector field. 
We incorporate all possible 
leading-order interactions 
between DE and DM mediating the energy 
and momentum transfers. 
We call our scenario the EFT 
of coupled DE and DM.

In the following, we briefly review 
our EFT constructions of the DE 
and DM sectors in Secs.~\ref{DEsec} and \ref{DMsec}, respectively, and 
then introduce their interactions 
in Sec.~\ref{DEDMint}.

\subsection{DE sector}
\label{DEsec}

We begin with reviewing the 
EFT of DE in which the DE sector 
is described by a scalar field $\phi$. 
We choose the unitary gauge in which the scalar field perturbation 
$\delta \phi$ vanishes. 
Then, the scalar field plays 
a role of the time coordinate 
$t$, such that 
\be
\phi=t\,.
\ee
The residual symmetry of the EFT is 
the invariance under spatial 
diffeomorphisms (diffs):
\begin{align}
\bm{x} \to \bm{x}'(t,\bm{x})\,.
\end{align} 
In this case, the EFT of DE 
coupled to gravity has been 
formulated in Ref.~\cite{Gubitosi:2012hu} 
by using the $3+1$ ADM decomposition of spacetime. 
The line element in the ADM formalism 
is given by 
\be
\rd s^2=g_{\mu \nu} \rd x^{\mu}\rd x^{\nu}=
-N^2 \rd t^2+h_{ij} \left( \rd x^i+N^i \rd t 
\right) \left( \rd x^j+N^j \rd t 
\right)\,, \label{eq:3+1metric}
\ee
where $N$ is the lapse, $N^i$ is the shift, and $h_{ij}$ is the three-dimensional metric 
on constant-$t$ hypersurfaces $\Sigma_t$. The inverse metric is given by
\begin{eqnarray}
 g^{\mu\nu}=   \begin{pmatrix}
-1/N^2 & N^j/N^2 \\
N^i/N^2 & h^{ij}-N^iN^j/N^2 & 
\end{pmatrix}.
\end{eqnarray}
For later purposes, it is convenient to use the covariant notation of the ADM formalism with the spatial metric 
given by $h_{\mu\nu}=g_{\mu\nu}+n_{\mu}n_{\nu}$, where 
\be
n_{\mu}=-\frac{\delta^0_{\mu}}{\sqrt{-g^{00}}}
=-N\delta^0_{\mu}
\label{nmudef}
\ee
is the unit normal vector orthogonal to $\Sigma_t$.
On the three-dimensional hypersurface characterized by the metric $h_{ij}$, 
we define the extrinsic curvature 
\be
K_{ij}=h_{i}^{\alpha} 
h_{j}^{\beta} \nabla_{\alpha} n_{\beta}
=\frac{1}{2N} \Big( \partial_{t} h_{ij}
-D_{i} N_{j} -D_{j} N_{i} 
\Big)\,,
\label{Kij}
\ee
and the Ricci tensor (intrinsic curvature) ${}^{(3)}R_{ij}$ on the 
three-dimensional hypersurface, 
where $D_i$ is the covariant derivative 
operator with respect 
to the metric $h_{ij}$.

For simplicity, we consider the EFT corresponding to a subclass of Horndeski theories~\cite{Horndeski:1974wa} 
with the luminal speed of 
gravitational waves. 
The explicit form of such an action is given by 
\be
{\cal S}=\int \rd^4 x \sqrt{-g} \Big[ 
G_2(\phi,X)+G_3(\phi, X) \square \phi 
+G_4(\phi)R \Big]\,,
\label{Horndeski}
\ee
where $G_2$ and $G_3$ are functions of 
$\phi$ and 
$X=-\nabla_{\mu}\phi 
\nabla^{\mu}\phi/2$, 
$G_4$ is a function of $\phi$, 
and $R$ is the four-dimensional Ricci scalar. 
For this class of scalar-tensor 
theories, in terms of the EFT perspective, 
we only need to take 
into account the following 
quantities as the EFT 
building blocks:
\begin{align}
t,\quad n_{\mu}, \quad g^{00}, \quad K_{\mu\nu}, 
\quad \Rs_{\mu\nu}\,,
\label{block_DE}
\end{align}
on top of the metric $h_{\mu\nu}$. 
Since $K_{\mu\nu}$ and $\Rs_{\mu\nu}$ satisfy the relations  $K_{\mu\nu}n^{\mu}=\Rs_{\mu\nu}n^{\mu}=0$ and $n^{\mu}$ is normalized to be $n_{\mu}n^{\mu}=-1$, one cannot find 
any non-trivial scalar quantities from $n_{\mu}$. We note that the four-dimensional Ricci scalar can be 
expressed as 
\be
R={}^{(3)}R+K_{\mu \nu}K^{\mu \nu}-K^2
+2\nabla_{\nu} \left( n^{\nu} 
\nabla_{\mu}n^{\mu}-n^{\mu} \nabla_{\mu} 
n^{\nu} \right)\,,
\ee
where ${}^{(3)}R:={}^{(3)}{R^{\mu}}_{\mu}$ and 
$K:={K^{\mu}}_{\mu}$.
The gravitational (Einstein-Hilbert) action, 
which is multiplied by a time-dependent function $f(t)$ and a constant term $M_*^2/2$, is given by 
\be
\frac{M_*^2}{2}\int {\rm d}^4 x \sqrt{-g} f(t)R
=\int {\rm d}^4 x\, N\sqrt{h} 
\left[ \frac{M_*^2}{2} f(t) \left( {}^{(3)}R+K_{\mu \nu}K^{\mu \nu}-K^2 \right) -M_*^2 \frac{\rd f(t)}{\rd t} \frac{K}{N} \right]\,,
\label{Raction}
\ee
up to boundary terms, where $h$ is the determinant 
of the three-dimensional tensor $h_{ij}$.
In theories with $\rd f(t)/\rd t=0$, which is the case for General Relativity, 
the last term in Eq.~(\ref{Raction}) vanishes.
When $\rd f(t)/\rd t \neq 0$, there is 
a contribution arising from the term 
$-M_*^2 ({\rm d}f(t)/{\rm d}t)K/N$.

The action (\ref{Raction}) contains 
the dependence of $t$, $N=1/\sqrt{-g^{00}}$, $K_{\mu \nu}$, and ${}^{(3)}R_{\mu \nu}$. 
The general Lagrangian constructed from the quantities in Eq.~\eqref{block_DE} is given by
\begin{align}
\mathcal{L}_{\rm DE}=\mathcal{L}_{\rm DE}(t, g^{00}, 
K_{\mu\nu}, \Rs_{\mu\nu})\,, \label{eq:Genaction}
\end{align}
where the indices are contracted with $h^{\mu\nu}$ (or equivalently with $g^{\mu\nu}$). Since $t$ is one of the EFT building blocks, one can subtract the background parts of $K_{\mu\nu}$ and $g^{00}$, while keeping the invariance under the spatial diffs:
\begin{align}
\delta K_{\mu\nu}:=K_{\mu\nu}-H(t)h_{\mu\nu}\,, \qquad
\delta g^{00}:= 
g^{00}-g^{00}_{\rm BG}(t)\,,
\label{eq:deltaK_g00}
\end{align}
where $H(t):=(\rd a/\rd t)/(N a)$ and $g^{00}_{\rm BG}(t)$ are the background 
Hubble expansion rate and the background value of $g^{00}$, respectively. 

By expanding the general Lagrangian~(\ref{eq:Genaction}) up to quadratic order in perturbations and also only keeping terms that lead to 
the luminal speed of gravitational 
waves, the EFT Lagrangian for DE is given by
\begin{align}
\mathcal{L}_{\rm DE} 
=\frac{M_*^2}{2}f(t) 
\left[ \Rs+K_{\mu\nu}K^{\mu\nu}-K^2 \right] 
-\hat{\Lambda}(t) - \hat{c}(t) g^{00} - d(t) K+\frac{1}{2}\hat{M}_{2}^4(t) \left(\frac{\delta g^{00}}{-g^{00}_{\rm BG}} \right)^2 
-\frac{1}{2}\bar{M}_1^3(t) \left(\frac{\delta g^{00}}{-g^{00}_{\rm BG}} \right) \delta K 
+\cdots,
\label{EFTofDE}
\end{align}
where $\hat{\Lambda}$, $\hat{c}$, $d$, $\hat{M}_2^4$, $\bar{M}_1^3$ 
are functions of $t$, and 
$\delta K=\delta K_{\mu\nu}g^{\mu\nu}= \delta K_{\mu\nu}h^{\mu\nu}$. 
The ellipsis in Eq.~(\ref{EFTofDE}) 
stands for terms corresponding to 
the higher-order perturbations. 
We note that the last term 
$-M_*^2 ({\rm d}f(t)/{\rm d}t)K/N$ in 
Eq.~(\ref{Raction}) has been absorbed 
into the definitions of $\hat{\Lambda}$, 
$\hat{c}$, $d$, and $\hat{M}_2^4(t)$. 
In the conventional EFT of 
scalar-field inflation and DE~\cite{Creminelli:2006xe,Cheung:2007st,Creminelli:2008wc,Gubitosi:2012hu}, we usually set  $d(t)=0$ by absorbing the Lagrangian 
$-d(t)K$ into 
other terms after integration by parts. We will retain $-d(t)K$, as it is more useful for the EFT framework that accommodates vector-tensor theories~\cite{Aoki:2021wew}.

The EFT can be extended to theories 
with different symmetry-breaking patterns~\cite{Aoki:2021wew, Aoki:2024ktc}. 
Let us first consider the EFT 
invariant under the transformations
\begin{align}
t \to t'=t+\chi_0 \,,\qquad
\bm{x} \to \bm{x}'(t,\bm{x})\,,
\label{shift_sym}
\end{align} 
with $\chi_0$ is a constant parameter. 
The global time-translation is nothing but the global shift of the scalar field when it is reintroduced by the St\"uckelberg 
trick, $t \to \phi$. 
More generally, the EFT of shift-symmetric theories is obtained by imposing a residual invariance under time-diffs with 
$\xi^0=-c_{\phi}/\dot{\bar{\phi}}$, because this preserves the unitary gauge $\phi(t)=\bar{\phi}(t)$ in combination with a shift $\phi\to\phi+c_{\phi}$. When $\phi=t$, this reduces to the 
global time-translation mentioned above. 
Therefore, the EFT with the symmetry \eqref{shift_sym} corresponds to 
shift-symmetric scalar-tensor theories~\cite{Finelli:2018upr}. 
In the subclass of Horndeski 
theories (\ref{Horndeski}), 
the symmetry (\ref{shift_sym}) translates to the couplings 
$G_2(\phi, X)=G_2(X)$, $G_3(\phi, X)=G_3(X)$, and $G_4(\phi)={\rm constant}$.
In this case, the general form of the invariant Lagrangian 
is given by 
\begin{align}
\mathcal{L}_{\rm DE}=\mathcal{L}_{\rm DE}( g^{00}, K_{\mu\nu}, \Rs_{\mu\nu})\,.
\label{shifts}
\end{align}
Without the $t$ dependence in the Lagrangian (\ref{shifts}), 
the background parts of $g^{00}$ and $K_{\mu\nu}$ cannot be 
subtracted by keeping the covariance. 
Nonetheless, one can perform the Taylor expansion 
of the Lagrangian to arrive at the same form as 
Eq.~\eqref{EFTofDE}, provided that the EFT coefficients satisfy the following consistency conditions~\cite{Aoki:2021wew} (see also~\cite{Finelli:2018upr}):
\begin{align}
\dot{\hat{\Lambda}}+3H\dot{d} + \dot{\hat{c}} g^{00}_{\rm BG} &= 0\,, \label{consistency1MST} \\
2\hat{M}_2^4 \frac{\D}{\bar{N} \D t} \ln (- g^{00}_{\rm BG}) + 3\bar{M}_1^3 \dot{H} + 2 \dot{\hat{c}} g^{00}_{\rm BG}
&= 0\,, \label{consistency2MST}\\
\dot{d} +\frac{1}{2}\bar{M}_1^3 \frac{\D}{\bar{N} \D t} \ln (- g^{00}_{\rm BG})
&= 0\,, \label{consistency3MST} \\
\dot{f}&= 0\,, 
\label{consistency4MST}
\end{align}
where $\bar{N}$ is the background value of $N$,  
and a dot denotes the derivative with respect to 
the cosmic time $\tilde{t}=\int \bar{N}\,\rd t$, 
e.g.,~$\dot{\Lambda}=\D\Lambda/(\bar{N} \D t)$.
The condition (\ref{consistency4MST}) means that the last term in Eq.~(\ref{Raction}) vanishes and hence the Lagrangian (\ref{EFTofDE}) can be used without the modifications to the EFT coefficients induced by the term $-M_*^2 ({\rm d}f(t)/{\rm d}t)K/N$.

The EFT of vector-tensor theories is obtained by 
promoting the shift symmetry to a local one 
with the help of a gauge field $A_{\mu}$, as
\begin{align}
t &\to t'=t-g_M \theta(t,\bm{x}) \,,\qquad A_{\mu}\to A'_{\mu}=A_{\mu}
+\nabla_{\mu}\theta(t, \bm{x})
\,, \label{shift_sym2}\\
\bm{x} &\to \bm{x}'(t,\bm{x})
\,,
\end{align} 
with $g_M$ being the gauge 
coupling constant. 
The EFT building blocks are given by the ``tilded'' quantities as examined in Ref.~\cite{Aoki:2021wew}. 
If one is interested in the irrotational solution (scalar and tensor perturbations only) as a consistent truncation, however, one can make an ansatz 
\be
A_{\mu}=\Big[ A_0(t, \bm{x}), \bm{0} 
\Big]\,,
\label{Amua}
\ee
by using the freedom of the combined $U(1)$ and time diffs \eqref{shift_sym2}. 
Note that we are assuming the existence of 
a preferred vector field 
$v_{\mu}=\nabla_{\mu} \tilde{t}+g_M A_{\mu}$, 
where we choose the unitary gauge in which 
the St\"uckelberg field $\tilde{t}$ is 
equivalent to $t$.
For this gauge choice with the vector field 
configuration (\ref{Amua}), we have 
$v_{\mu}=\tilde{\delta}^{0}_{\mu}=
\delta^{0}_{\mu}+g_M A_{\mu}=(1+g_M A_0, {\bm 0})$ 
and hence $v_{\mu}$ is orthogonal to constant 
$t$ hypersurfaces. The norm of the preferred 
vector is 
\be
\tilde{g}^{00}=\tilde{\delta}^{0}_{\alpha}
\tilde{\delta}^{0}_{\beta}g^{\alpha \beta}
=\left( 1+g_M A_0 \right)^2 g^{00}\,.
\ee
In this case, the unit timelike vector 
$\tilde{n}_{\mu}=-\tilde{\delta}^{0}_{\mu}
/\sqrt{-\tilde{g}^{00}}$ is equivalent to 
$n_{\mu}$ defined in Eq.~(\ref{nmudef}).

The vector-tensor EFT building blocks together with 
the ansatz~(\ref{Amua}) consist of 
the following quantities:
\begin{align}
n_{\mu}, \quad \tilde{g}^{00}, \quad 
F_{\mu}\,,\quad
K_{\mu\nu}, \quad \Rs_{\mu\nu}, 
\end{align} 
where we have defined
\begin{align}
F_{\mu}:=n^{\alpha}F_{\mu\alpha}\,,\qquad 
F_{\mu\alpha}:=2\nabla_{[\mu}A_{\alpha]}\,.
\end{align}
The magnetic part of the field strength 
$h^{\alpha}{}_{\mu}h^{\beta}{}_{\nu}F_{\alpha\beta}$ vanishes identically under the ansatz (\ref{Amua}). This trivializes the possible building block $\tilde{F}_\mu=n^\alpha \tilde{F}_{\alpha\mu}$, where $\tilde{F}_{\alpha\mu}$ is the spatially covariant dual of the field strength.
The EFT action, which subtracts the 
last term in Eq.~(\ref{Raction}), 
is then given by
\begin{align}
\mathcal{L}_{\rm DE}&=
\frac{M_*^2}{2}f(t) \left[ 
\Rs+K_{\mu\nu}K^{\mu\nu}-K^2 \right] 
- \hat{\Lambda}(t) - \hat{c}(t) \tilde{g}^{00}- d(t) K \nn
&~~~+ 
\frac{1}{2}\hat{M}_2^4(t) \left(\frac{\delta \tilde{g}^{00}}{-\tilde{g}^{00}_{\rm BG}} \right)^2 
-\frac{1}{2}\bar{M}_1^3(t) \left(\frac{\delta \tilde{g}^{00}}{-\tilde{g}^{00}_{\rm BG}} \right) \delta K +\frac{1}{2}\gamma_1(t) F_{\mu}F^{\mu}\,,
\label{UEFTofDE}
\end{align}
where we have taken into account the term 
$\gamma_1(t) F_{\mu}F^{\mu}/2$ associated 
with the electromagnetic field strength. 
To avoid the ghost in the vector sector, 
we require that~\cite{Aoki:2021wew,Aoki:2024ktc} (see also Appendix~\ref{vecsec})
\be
\gamma_1(t)>0\,.
\ee
In this case, there are the following 
consistency conditions
\begin{align}
    \dot{\hat{\Lambda}}+3H\dot{d} + \dot{\hat{c}} \tilde{g}^{00}_{\rm BG} &= 0
    \,, \label{consistency1M} \\
    2\hat{M}_2^4 \frac{\D}{\bar{N} \D t} \ln (- \tilde{g}^{00}_{\rm BG}) + 3\bar{M}_1^3 \dot{H} + 2 \dot{\hat{c}} \tilde{g}^{00}_{\rm BG}
    &= 0\,, \label{consistency2M}\\
    \dot{d} +\frac{1}{2}\bar{M}_1^3 \frac{\D}{\bar{N} \D t} \ln (- \tilde{g}^{00}_{\rm BG})
    &= 0\,, \label{consistency3M} \\
    \dot{f}&= 0\,. \label{consistency4M}
\end{align}
Again, the relation (\ref{consistency4M}) 
means that the last term in Eq.~(\ref{Raction}) 
does not contribute to the EFT action.
The Lagrangian \eqref{UEFTofDE} can describe 
the cosmological evolution of the background and 
perturbations in GP theories~\cite{Heisenberg:2014rta,Tasinato:2014eka,Allys:2015sht,BeltranJimenez:2016rff,Allys:2016jaq} with the luminal speed of gravitational waves.
The explicit action of such a subclass 
of GP theories is given by 
\be
{\cal S}=\int \rd^4 x \sqrt{-g} \left[ 
F+G_2(\tilde{X})+G_3(\tilde{X}) \nabla_{\mu} A^{\mu}
+\frac{\Mpl^2}{2}R \right]\,,
\label{eq:GP}
\ee
where $F:=-F_{\mu \nu}F^{\mu \nu}/4$, $G_2$ and $G_3$ are functions of 
$\tilde{X}=-A_{\mu}A^{\mu}/2$, and $\Mpl$ is the reduced Planck mass. Taking the limit 
$A_{\mu} \to \nabla_{\mu} \phi$, 
we recover the shift-symmetric Horndeski
theories with $F=0$, 
$\tilde{X} \to X=-\nabla_{\mu} 
\phi \nabla^{\mu} \phi/2$, 
and $\nabla_{\mu} A^{\mu} \to 
\nabla_{\mu} \nabla^{\mu} \phi$.
The constancy of the coefficient 
of $R$ is consistent with the condition (\ref{consistency4M}). 
Let us also clarify that we only need to consider 
a term linear in $F$ owed to our ansatz in Eq.~\eqref{Amua} that trivialises $F_{\mu\nu}$ at the background level, so $F_{\mu\nu}$ will only appear in the perturbation sector.

The usage of the EFT Lagrangian \eqref{UEFTofDE} 
is not restricted to vector-tensor theories, 
but it serves as a unified framework for 
both scalar-tensor and vector-tensor theories~\cite{Aoki:2021wew, Aoki:2024ktc}. 
Taking the limit $g_M \to 0$, one recovers 
shift-symmetric scalar-tensor theories with a decoupled vector field. 
The conventional EFT of DE, which is 
given by the action \eqref{EFTofDE},
is also reproduced by omitting the consistency conditions further. 
For example, the absence of the consistency 
condition $\dot{f}=0$ allows the 
time-dependent non-minimal coupling 
$G_4(\phi) R$ in 
non-shift-symmetric scalar-tensor theories. 
In this way, one can accommodate vector-tensor, 
shift-symmetric scalar-tensor, and 
non-shift-symmetric scalar-tensor 
theories in a unified manner. 
The gauge coupling and consistency 
conditions characterize the boundaries 
of those theories.

\subsection{DM sector}
\label{DMsec}

We assume that the dynamics in the 
DM sector is described by a dust 
fluid (CDM) on cosmological scales. 
While there are several descriptions of fluid dynamics, it will be convenient for our purposes to adopt the one based on three scalar fields 
$\phi^i(t,\bm{x})~(i=1,2,3)$.\footnote{A dual formulation in terms of 2-forms exists where the cosmological symmetries are also realized 
in a dual manner \cite{Aoki:2022ylc}. This dual formulation also exists for solids.}
These scalar fields can be understood as comoving coordinates of the fluid. 
Since the spatial diffs are preserved in the EFT of DE, one can use this freedom to move to the comoving gauge $\phi^i=x^i$, where the perturbations of $\phi^i$ are eaten by the metric.

Among the general action for the scalar fields $\phi^i$, 
the fluid phase is defined by an invariance under internal volume-preserving 
diffs~\cite{Dubovsky:2005xd,Endlich:2012pz}
\begin{align}
\phi^i \to \phi'{}^i \quad {\rm s.t.} 
\quad \det \frac{\partial \phi'{}^i}{\partial \phi^j} =1\,.
\end{align}
The invariant building blocks  
are given by 
\begin{align}
n &:= \sqrt{\det g^{\mu\nu}\partial_{\mu}\phi^i \partial_{\nu} \phi^j} \,,
\label{eq:number_density}\\
u^{\mu}&:=-\frac{1}{6n}\varepsilon_{ijk}\epsilon^{\mu\nu\rho\sigma}\partial_{\nu}\phi^i \partial_{\rho} \phi^j \partial_{\sigma}\phi^k\,,
\label{eq:four-velocity}
\end{align}
and their derivatives. Here, $n$ and $u^{\mu}$ 
are the number density and four-velocity of the fluid, respectively, 
$\varepsilon_{ijk}$ is the 
anti-symmetric symbol with 
$\varepsilon_{123}=1$, and $\epsilon^{\mu\nu\rho\sigma}$ 
is the spacetime Levi-Civita tensor
with $\epsilon^{0123}=-1/\sqrt{-g}$. 
In this formulation, the conservation 
of the current $\mathcal{J}^{\mu}
=n u^{\mu}$ is an off-shell identity, $\nabla_{\mu}\mathcal{J}^{\mu}=0$. 
When we use the comoving gauge $\phi^i=x^i$, 
we find
\begin{align}
n=\sqrt{\det g^{ij}}\,,\qquad 
u^{\mu}=\frac{\delta^{\mu}_0}{\sqrt{-g_{00}}}\,,
\end{align}
where the residual symmetry is the spatial 
volume-preserving diffs:
\begin{align}
\bm{x} \to \bm{x}'(\bm{x}) \quad {\rm s.t.} \quad \det \frac{\partial \bm{x}'}{\partial \bm{x}} 
=1\,.
\end{align}
The general action of the fluid is given by a function of $n$ at leading order in a derivative expansion. We are interested in the DM dynamics which should be approximated by the dust fluid with the Lagrangian linear in the number density $n$. As a result, the action of the DM sector 
is given by ${\cal S}_{\rm DM}=\int \D^4 x \sqrt{-g}\,{\cal L}_{\rm DM}$, 
with the Lagrangian 
\begin{align}
{\cal L}_{\rm DM}=
-\hat{m}_{\rm c} n \,.
\label{SDM}
\end{align}
Here, $\hat{m}_{\rm c} n$ 
is the energy density of the dust fluid, with $\hat{m}_{\rm c}$ being a constant.

Before closing this subsection, it is worthwhile mentioning the possibility of 
a solid DM.\footnote{The possibility of having solid DM was first explored in~Ref.~\cite{Bucher:1998mh}.} Using three scalars, the solid phase means that the symmetry is downgraded to the internal $ISO(3)$ 
symmetry~\cite{Dubovsky:2005xd, Endlich:2012pz}. 
The solidity generates a non-vanishing sound speed of the transverse mode, $c_{V} \neq 0$. On the other hand, in the decoupling limit of 
gravity, the squared sound speed of 
the longitudinal mode 
$c_{S}^2$ is given by the following 
form~\cite{Aoki:2022ipw}
\begin{align}
c_{S}^2=\frac{\partial \bar{p}}
{\partial \bar{\rho}} 
+\frac{4}{3}c_{V}^2\,,
\end{align}
where $\bar{\rho}$ and $\bar{p}$ are the 
background energy density and pressure 
of the solid. In the following, we will use 
a bar to represent background quantities. 
Even for the case in which the equation of state $w=\bar{p}/\bar{\rho}$ is close to 0, 
the solid $(c_V^2 \neq 0)$ leads to the deviation of $c_S^2$ from 0 and hence it does not generally 
work as CDM. While it would be interesting to discuss how the solidity affects the small-scale dynamics of DM, the solidity should be well negligible on cosmological scales which we are interested in. 
Hence, we only consider the fluid DM throughout the paper. 

Besides DM, we incorporate other matter fields (baryons, radiation) described by the perfect fluids. While they can be 
dealt as the three-scalar field description 
explained above (as we do in Appendix~\ref{sec:3scalarMatter}), it is also possible 
to accommodate them as a 
Schutz-Sorkin action~\cite{Schutz:1977df,Brown:1992kc,DeFelice:2009bx,Pookkillath:2019nkn, Pookkillath:2024ycd} or 
a purely k-essence~\cite{Giannakis:2005kr,Arroja:2010wy} 
on the time-dependent cosmological background.
In the following, we will adopt the 
Schutz-Sorkin action without the 
rotational mode, which is given by 
\be
{\cal S}_{\rm m}=-\int {\rm d}^4 x
\left[ \sqrt{-g}\,\rho_{\rm m} (n_{\rm m})
+J^{\mu} \nabla_{\mu}\ell \right]\,,
\label{Smatter}
\ee
where the matter density $\rho_{\rm m}$ 
is a function of its number density 
$n_{\rm m}$, and $\ell$ is a Lagrange 
multiplier. The vector field $J^{\mu}$ 
in the action (\ref{Smatter}) is related to 
$n_{\rm m}$, as $n_{\rm m}=
\sqrt{g_{\mu \nu}J^{\mu}J^{\nu}/g}$. 
The fluid four-velocity is defined by 
$(u_{\rm m})^{\mu}=J^{\mu}/(n_{\rm m} \sqrt{-g})$, satisfying the normalization 
$(u_{\rm m})^{\mu}(u_{\rm m})_{\mu}=-1$. 
Varying (\ref{Smatter}) with respect to $\ell$, 
we obtain $\nabla_{\mu}J^{\mu}=0$. 
This equation can be expressed as 
\be
(u_{\rm m})^{\mu} \nabla_{\mu} \rho_{\rm m}
+\left( \rho_{\rm m}+p_{\rm m} \right) 
\nabla_{\mu}(u_{\rm m})^{\mu} =0\,,
\label{coneq}
\ee
where $p_{\rm m}$ is the matter pressure 
defined by 
\be
p_{\rm m}:=n_{\rm m} 
\rho_{{\rm m},n_{\rm m}}
-\rho_{\rm m}\,,
\ee
where $\rho_{{\rm m},n_{\rm m}}:=
\rd \rho_{\rm m}/\rd n_{\rm m}$.
The energy-momentum tensor 
$T_{\mu \nu}^{(m)}$ associated with 
the variation of (\ref{Smatter}) with respect to $g^{\mu \nu}$ 
is given by~\cite{Amendola:2020ldb}
\be
T_{\mu \nu}^{(m)}=(\rho_{\rm m}+p_{\rm m})
(u_m)_{\mu} (u_n)_{\nu}+p_m g_{\mu \nu}\,, \label{eq:SS_PF_action}
\ee
which corresponds to that of the perfect 
fluid.

\subsection{EFT action for coupled DE and DM}
\label{DEDMint}

Given that we have introduced basic tools for the EFT of dark sectors, it is now straightforward to describe the interactions between DE and DM. 
We start our formulation from the vector DE because we can easily accommodate the scalar DE by taking the appropriate limit and omitting the 
consistency conditions. 
For simplicity, we focus on the 
leading-order terms in derivative expansions. 

The EFT building blocks associated with 
the interactions between DE and DM are
\begin{tcolorbox}[colback=black!5!white,colframe=black!75!white,title=Building blocks for coupled DE and DM]
\begin{align}
\underbrace{n_{\mu}, \quad \tilde{g}^{00}, 
\quad F_{\mu}}_{\rm DE}, \qquad \underbrace{n, \quad u^{\mu}}_{\rm DM}\,.
\label{blocks}
\end{align}
\end{tcolorbox}
\noindent
There are several DE-DM interacting 
Lagrangians constructed from the ingredients (\ref{blocks}). 
One of them is the product between $\tilde{g}^{00}$
and $n$, which characterizes the energy exchange between DE and DM. 
To avoid the non-vanishing CDM sound speed, we require that the coupling 
is linear in $n$~\cite{Kase:2020hst}, i.e., 
$\mathcal{L}_n (\tilde{g}^{00}) n$, 
where $\mathcal{L}_n$ is a function 
of $\tilde{g}^{00}$.
The others are the scalar products 
$n_{\mu} u^{\mu}$ and $F_{\mu} u^{\mu}$, 
which mediate the momentum exchanges 
between DE and DM. 
To quantify the latter interactions, 
it is convenient to decompose $u^{\mu}$ into 
the temporal and spatial components 
with respect to $n_{\mu}$, as
\begin{align}
{\cal U}:=n_{\nu}u^{\nu}
\,, \qquad
q^{\mu}:=u^{\mu}+n^{\mu}{\cal U}\,,
\end{align}
where $q^{\mu}$ satisfies the orthogonal 
relation $n_{\mu}q^{\mu}=0$.
These two quantities are related to 
each other, as
\begin{align}
q^{\mu}q_{\mu}=-1+{\cal U}^2\,,
\label{eq:q2}
\end{align}
so that we only need to take the spatial 
components of $q^{\mu}$ as the building block. 
Note that $q^{\mu}$ and $F_{\mu}$ start at 
linear order in perturbations. 
In the comoving gauge, we find
\begin{align}
q^{\mu}=\left(0, \frac{N^i}{\sqrt{N^2-N_jN^j}} \right)\,,
\label{qmuex}
\end{align}
meaning that $q^{\mu}$ is related to the shift vector in the ADM language.
Instead of using $u^{\mu}$, 
we will use $q^{\mu}$ to quantify 
the DE-DM momentum transfer and consider 
the interactions $q^{\mu}q_{\mu}$ 
and $q^{\mu} F_{\mu}$. 
These scalar products can be further 
multiplied by 
$\mathcal{L}_{q^2}(\tilde{g}^{00})$ and 
$\mathcal{L}_{q\cdot F}(\tilde{g}^{00})$, respectively, 
where $\mathcal{L}_{q^2}$ and $\mathcal{L}_{q\cdot F}$ 
are functions of $\tilde{g}^{00}$. 
Then, the general interacting Lagrangian is given by 
\begin{align}
\mathcal{L}_{\rm int}&=\mathcal{L}_n (\tilde{g}^{00}) n + \mathcal{L}_{q^2}(\tilde{g}^{00}) q^{\mu}q_{\mu} + \mathcal{L}_{q\cdot F}(\tilde{g}^{00}) F_{\mu}q^{\mu}  + \cdots 
\nn
&=- \Delta \Lambda(t) - \Delta c(t) \tilde{g}^{00} 
- \Delta m_{\rm c} (t) n
\nn
&~~~+\frac{1}{2} \Delta M_2^4(t)\left(\frac{\delta \tilde{g}^{00}}{-\tilde{g}^{00}_{\rm BG}} \right)^2  - m_1^4(t) \frac{\delta n}{\bar{n}}   \left(\frac{\delta \tilde{g}^{00}}{-\tilde{g}^{00}_{\rm BG}} \right) - m_2^4(t) q^{\mu} q_{\mu} - \bar{m}_1^2(t)q^{\mu}F_{\mu} 
+ \cdots\,,
\label{Lint}
\end{align}
where dots represent terms 
for higher-order perturbations, and
\begin{align}
\Delta \Lambda(t) &= \bar{\mathcal{L}}_{n \tilde{g}^{00}}\,
\bar{n}\,\tilde{g}^{00}_{\rm BG}
\,, \qquad
\Delta c(t) = -\bar{\mathcal{L}}_{n \tilde{g}^{00}}
\,\bar{n}\,, \qquad
\Delta m_{\rm c}(t)=-\bar{\mathcal{L}}_n
\,,\nonumber \\
\Delta M_2^4 (t) &= \bar{\mathcal{L}}_{n \tilde{g}^{00} \tilde{g}^{00}}\,\bar{n}\,
(\tilde{g}_{\rm BG}^{00})^2\,, \qquad
m_1^4(t) = \bar{\mathcal{L}}_{n\tilde{g}^{00}}\,
\tilde{g}_{\rm BG}^{00}\,\bar{n}
\,, \qquad
m_2^4(t) =-\bar{\mathcal{L}}_{q^2}
\,,  \qquad
\bar{m}_1^2(t) = - \bar{\mathcal{L}}_{q\cdot F}\,,
\end{align}
and 
\be
\delta n=n-\bar{n}(t)\,,\qquad
\mathcal{L}_{n \tilde{g}^{00}}= 
\frac{\D \mathcal{L}_n}{\D \tilde{g}^{00}}, 
\qquad 
\mathcal{L}_{n \tilde{g}^{00}\tilde{g}^{00}}= 
\frac{\D \mathcal{L}_{n\tilde{g}^{00}}}{
\D \tilde{g}^{00}}\,.
\ee
The functions with the bar correspond to their background parts, which are 
implicit functions of time. 
The following two relations need to 
be satisfied
\begin{align}
\frac{\D}{\D t} \bar{\mathcal{L}}_n(t) &=\bar{\mathcal{L}}_{n \tilde{g}^{00}}(t) \frac{\D}{\D t} \tilde{g}^{00}_{\rm BG}(t)
\,, 
\label{consistency_int1} \\
\frac{\D}{\D t} \bar{\mathcal{L}}_{n\tilde{g}^{00}}(t) &=\bar{\mathcal{L}}_{n \tilde{g}^{00} \tilde{g}^{00} }(t) \frac{\D}{\D t} \tilde{g}^{00}_{\rm BG}(t)
\,.
\label{consistency_int2}
\end{align}

The total action ${\cal S}$, 
which is given by Eq.~(\ref{Stotal}), 
is composed of the Lagrangians 
(\ref{EFTofDE}), (\ref{SDM}), 
(\ref{Lint}), and the matter action (\ref{Smatter}). Then, we can express it 
in the following form: 
\begin{tcolorbox}[colback=black!5!white,colframe=black!75!white,title=EFT action for coupled DE and DM]
\be
{\cal S}=\int {\rm d}^4 x \sqrt{-g}\,
\left( \mathcal{L}_{\rm D}^{\rm NL} 
+ \mathcal{L}_{\rm D}^{(2)}\right)
+{\cal S}_{\rm m}\,,
\label{Sfull}
\ee
where 
\begin{align}
\mathcal{L}_{\rm D}^{\rm NL} &=
\frac{M_*^2}{2}f(t) \left( 
\Rs+K_{\mu\nu}K^{\mu\nu}-K^2 
\right)
-\Lambda(t) - \tilde{c}(t) \tilde{g}^{00}-d(t)K - m_{\rm c}(t) n\,,
\label{eq:non-linearpart}
\\
\mathcal{L}_{\rm D}^{(2)}& = 
\frac{1}{2}M_2^4{} (t) \left(\frac{\delta \tilde{g}^{00}}
{-\tilde{g}^{00}_{\rm BG}} \right)^2 
-\frac{1}{2}\bar{M}_1^3(t) 
\left(\frac{\delta \tilde{g}^{00}}
{-\tilde{g}^{00}_{\rm BG}} \right) 
\delta K +\frac{1}{2}\gamma_1(t) F_{\mu}F^{\mu} 
 \nn
&~~~-m_1^4(t) \frac{\delta n}{\bar{n}} \left(\frac{\delta \tilde{g}^{00}}{-\tilde{g}^{00}_{\rm BG}} \right) - m_2^4(t) q^{\mu} q_{\mu} 
-\bar{m}_1^2(t) q^{\mu} F_{\mu}\,, 
\label{EFTLD2} \\
{\cal S}_{\rm m} &=-\int {\rm d}^4 x
\left[ \sqrt{-g}\,\rho_{\rm m} (n_{\rm m})
+J^{\mu} \nabla_{\mu}\ell \right]\,.
\label{EFTD}
\end{align}
\end{tcolorbox}
\noindent
Here, the coefficients are normalized 
EFT coefficients, which are given by 
\be
\Lambda (t) = \hat{\Lambda}(t) 
+ \Delta \Lambda(t)\,, 
\qquad
\tilde{c} (t)= \hat{c}(t) + \Delta c(t) \,, 
\qquad
m_{\rm c} (t) = \hat{m}_{\rm c} 
+ \Delta m_{\rm c}(t)\,,
\qquad
M_2^4 (t) = \hat{M}_2^4(t) 
+ \Delta M_2^4(t) \,.
\ee
The effective CDM mass $m_{\rm c} (t)$ acquires the time dependence through the term $\Delta m_{\rm c}(t)$, 
mediating the energy transfer between DE and DM.
The operator associated with the EFT function 
$m_1^4(t)$ also leads to the energy exchange, 
whereas the momentum transfer is weighed 
by the two functions 
$m_2^4(t)$ and $\bar{m}_1^2(t)$.

The consistency conditions in terms of 
the normalized coefficients are given by
\begin{tcolorbox}[colback=black!5!white,colframe=black!75!white,title=Consistency conditions]
\begin{align}
\dot{\Lambda}+3H\dot{d} + \dot{\tilde{c}} \tilde{g}^{00}_{\rm BG} - m_1^4 \frac{\D}{\bar{N} \D t} \ln (- \tilde{g}^{00}_{\rm BG}) &= 0
\,, \label{consistency1Mr} \\
2M_2^4 \frac{\D}{\bar{N} \D t} \ln (- \tilde{g}^{00}_{\rm BG}) + 3\bar{M}_1^3 \dot{H} + 2 \dot{\tilde{c}} \tilde{g}^{00}_{\rm BG} -6H m_1^4
&= 0\,, \label{consistency2Mr}\\
\dot{m}_{\rm c} \bar{n} + m_1^4  \frac{\D}{\bar{N} \D t}  \ln (- \tilde{g}^{00}_{\rm BG})  &=0\,, 
\label{consistency3Mr}\\
\dot{d} +\frac{1}{2}\bar{M}_1^3 \frac{\D}{\bar{N} \D t} \ln (- \tilde{g}^{00}_{\rm BG})
&= 0\,, \label{consistency3MD} \\
\dot{f}&= 0\,, 
\label{consistency4MD}
\end{align}
\end{tcolorbox}
\noindent
where we used the property 
$\bar{n}\propto a^{-3}$. 
Note that the coefficient $\bar{\mathcal{L}}_{n\tilde{g}^{00}
\tilde{g}^{00}}$ 
does not explicitly appear in the 
action \eqref{Sfull} up to quadratic order. 
Hence, the consistency condition \eqref{consistency_int2}, 
which is regarded as a constraint on $\bar{\mathcal{L}}_{n\tilde{g}^{00}\tilde{g}^{00}}$, can be safely neglected at quadratic order in perturbations.

As we already explained, the action \eqref{Sfull} can accommodate both scalar and vector DE scenarios. 
The shift-symmetric scalar-tensor theories coupled to DM are obtained by taking the limit $g_M \to 0$ (corresponding to $\tilde{g}^{00} \to g^{00}$) and $\bar{m}_1^2 \to 0$, 
while keeping the consistency conditions 
(\ref{consistency1Mr})-(\ref{consistency4MD}). 
The generic non-shift-symmetric 
scalar-tensor theories are obtained 
by omitting the consistency conditions. 
Note that the recovery of scalar-tensor theories requires the limit $\bar{m}_1^2\to 0$ in addition to the decoupling limit $g_M \to 0$, because the coupling $q^{\mu}F_{\mu}$ 
can survive even for $g_M \to 0$. 
The condition (\ref{consistency_int1}) 
translates to Eq.~(\ref{consistency3Mr}), 
which relates $\dot{m}_{\rm c}$ with $m_1^4$. 
This means that, for $m_1^4=0$, we have that 
$m_{\rm c}={\rm constant}$. 
In non-shift-symmetric scalar-tensor theories, 
the consistency condition (\ref{consistency3Mr}) 
does not need to hold and hence, the time variation of $m_{\rm c}$ is allowed even for $m_1^4=0$.

\section{Background Equations of motion}
\label{backsec}

Let us consider a spatially-flat FLRW 
background given by the line element 
\begin{equation}
\D s^{2} = - \bar{N}^2(t) \D t^{2} 
+ a^2(t) \delta_{ij} \D x^{i} \D x^{j} \,,
\label{FLRW}
\end{equation}
together with the vector field configuration 
\be
\bar{A}_{\mu}=\left[ \bar{A}_0(t), {\bm 0} 
\right]\,,
\ee
where a bar represents the time-dependent 
background quantities.
Since the Lagrangian 
${\cal L}_{\rm D}^{(2)}$ is of second order 
in perturbations, it does not contribute 
to the background equations of motion. 
We exploit the following properties 
\be
K=3H\,,\qquad 
K_{\mu \nu} K^{\mu \nu}=
3H^2\,,\qquad
{}^{(3)}R=0\,,\qquad 
\tilde{g}^{00}=-\bar{N}^{-2} \left( 1+g_M A_0 \right)^2\,,
\ee
with the Hubble expansion rate 
$H=\dot{a}/a=(\rd a/\rd t)/(\bar{N}a)$.
The action relevant to the background 
dynamics, which arises from the Lagrangian 
${\cal L}_{\rm D}^{\rm NL}$ with the matter 
action ${\cal S}_{\rm m}$, is given by 
\ba
{\cal S} &=&
\int {\rm d}^4 x \left[ 
-\frac{3M_*^2 f}{\bar{N}} 
a \left( \frac{\rd a}{\rd t} \right)^2 
-\bar{N} a^3 \Lambda-\bar{N} m_{\rm c} n_0
+\frac{\tilde{c} a^3}{\bar{N}} \left( 1+g_M A_0 \right)^2
-3d\,a^2 \left(\frac{\rd a}{\rd t}\right) 
\right] 
+{\cal S}_{\rm m}\,,
\label{Sback}
\ea
where we used the property $\bar{n}=n_0/a^3$, 
with $n_0$ being today's value of 
$\bar{n}$ (the scale factor is normalized 
to be $a=1$ today).
Since $(u_{\rm m})^{\mu}=(1/\bar{N},0,0,0)$ 
and $\nabla_{\mu}(u_{\rm m})^{\mu}=3H$ 
on the background (\ref{FLRW}), 
the matter continuity equation 
(\ref{coneq}) reduces to
\be
\dot{\bar{\rho}}_{\rm m}
+3H ( \bar{\rho}_{\rm m}
+\bar{p}_{\rm m})=0\,,
\label{coneq2}
\ee
where $\bar{\rho}_{\rm m}$ and 
$\bar{p}_{\rm m}$ are the background 
energy density and pressure, respectively.
Varying the action (\ref{Sback}) with 
respect to $\bar{N}$ and $a$, respectively, 
we obtain
\ba
& &
3M_*^2 f H^2=\Lambda+\frac{\tilde{c}}{\bar{N}^2}
\left( 1+g_M A_0 \right)^2
+m_{\rm c} \bar{n}+\bar{\rho}_{\rm m}\,,\label{back1}\\
& &
M_*^2 \left( 2f \dot{H}+2\dot{f}H+3f H^2 
\right)=\Lambda-\frac{\tilde{c}}{\bar{N}^2}
\left( 1+g_M A_0 \right)^2-\dot{d}
-\bar{p}_{\rm m}\,.
\label{back2}
\ea
We note that the energy transfer between DE 
and DM gives rise to a time-dependent 
mass $m_{\rm c}$.
Defining the background CDM density as 
\be
\rho_{\rm c}:=m_{\rm c}\bar{n}\,,
\ee
it follows that 
\be
\dot{\rho}_{\rm c}+3H\rho_{\rm c}
=\dot{m}_{\rm c} \bar{n}\,,
\ee
where we used $\dot{\bar{n}}+3H\bar{n}=0$.
At the background level, the interaction 
between DE and DM appears as the terms 
$\Delta \Lambda$, $\Delta c$, and 
$\Delta m_{\rm c}$ in $\Lambda$, 
$\tilde{c}$, and $m_{\rm c}$, 
respectively. 
As one should expect, the momentum transfer interactions do not affect 
the background equations.
In the following, we consider 
three different theories in turn.

\subsection{Vector-tensor theories}

In vector-tensor theories ($g_M \neq 0$), 
we obtain the equation of motion for 
$A_0$ by varying Eq.~(\ref{Sback}) 
with respect to $A_0$. 
This leads to 
$2\tilde{c} a^3 g_M (1+g_M A_0)/\bar{N}=0$, 
i.e., $\tilde{c}\,(1+g_M A_0) = 0$. 
If $1+g_M A_0=0$, it contradicts the EFT setup based on a preferred vector field. 
Then, we require that $1+g_M A_0 \neq 0$, 
and hence 
\be
\tilde{c}=0\,,
\label{cprimecon}
\ee
so that the $\tilde{c}$-dependent terms 
in Eqs.~(\ref{back1}) and (\ref{back2}) 
vanish. Moreover, the consistency 
condition (\ref{consistency4MD}) 
holds for vector-tensor theories under 
consideration now, so that 
$M_*^2 f=\Mpl^2={\rm constant}$.
Then, the background Eqs.~(\ref{back1}) and (\ref{back2}) reduce, respectively, to
\ba
& &
3\Mpl^2 H^2=\Lambda
+m_{\rm c} \bar{n}+\bar{\rho}_{\rm m}\,,\label{back1a}\\
& &
\Mpl^2 \left( 2 \dot{H}+3 H^2 
\right)=\Lambda-\dot{d}-\bar{p}_{\rm m}\,,
\label{back2a}
\ea
where $\dot{d}$ is related to $\bar{M}_1^3$ 
according to Eq.~(\ref{consistency3MD}). 
At the background level, the effect of 
interactions between vector DE and DM  
appears as the two terms
$\Delta \Lambda$ and $\Delta m_{\rm c}$ 
in $\Lambda$ and $m_{\rm c}$, respectively. 

Let us consider the subclass of GP 
theories (\ref{eq:GP}) 
with the interacting Lagrangian (\ref{GPint}) 
given in Appendix~\ref{sec:GP}, i.e., the action 
\ba
{\cal S}
&=&\int \rd^4 x \sqrt{-g} \left[ 
F+G_2(\tilde{X})+G_3(\tilde{X}) 
\nabla_{\mu} A^{\mu}
+\frac{\Mpl^2}{2}R 
-f_1 (\tilde{X}, \tilde{Z}, \tilde{E})n
+f_2 (\tilde{X}, \tilde{Z}, \tilde{E}) 
\right]+{\cal S}_{\rm m}\,,
\label{eq:GP2}
\ea
where $\tilde{X}=-A_{\mu}A^{\mu}/2$, 
$\tilde{Z}=A_{\mu}u^{\mu}$, and 
$\tilde{E}=-A^{\mu}F_{\mu\nu}u^{\nu}$.
In Appendix \ref{sec:GP}, we show the correspondence between the background quantities 
$\Lambda$, $\tilde{c}$, $d$ and the coupling 
functions $G_2$, $G_3$, $f_1$, $f_2$.
Note that the function $g_3$ appearing in the 
dictionary of the correspondence is related to 
$G_3$ according to $G_3=g_3+2\tilde{X} 
g_{3,\tilde{X}}$. From the correspondence (\ref{tildec}), Eq.~(\ref{cprimecon}) 
translates to 
\be
G_{2,\tilde{X}} 
-3H\sqrt{2\btX} G_{3,\tilde{X}}  
-\bar{n}
\left[ f_{1,\tilde{X}}+(2\btX)^{-1/2} 
f_{1,\tilde{Z}} \right] 
+f_{2,\tilde{X}} 
+ (2\btX)^{-1/2}f_{2,\tilde{Z}}=0\,,
\label{backGP1}
\ee
where we used the relation 
$G_{3,\tilde{X}}=3g_{3,\tilde{X}}
+2 \tilde{X}g_{3,\tilde{X}\tilde{X}}$, 
and $\btX$ is the background 
value of $\tilde{X}$ given by $\btX=\bar{A}_0^2/(2\bar{N}^2)$. 
With the dictionaries (\ref{Lambda}) and (\ref{dcorre}), the EFT functions appearing in Eqs.~(\ref{back1a}) and (\ref{back2a}) can be expressed as
\ba
\Lambda &=&
-G_2-f_2\,,\label{backGP2} \\
\dot{d} &=& \sqrt{2\btX}\,
\dot{\tilde{X}}_{\rm BG}
G_{3,\tilde{X}}\,,
\label{backGP3}
\ea
where Eq.~(\ref{backGP1}) has been used to simplify $\Lambda$.
Eq.~(\ref{backGP2}) describes how the interaction 
$f_2$ contributes to $\Lambda$. 
We also note the presence of the $G_3(\tilde{X})$ coupling makes $d$ non-vanishing. 

\subsection{Shift-symmetric 
scalar-tensor theories}

The shift-symmetric scalar-tensor theories 
with the luminal speed of gravitational waves 
can be obtained by taking the limit $g_M \to 0$ 
in Eqs.~(\ref{back1})-(\ref{back2})
and keeping the consistency conditions \eqref{consistency1Mr}-\eqref{consistency4MD}. 
From Eq.~(\ref{consistency4MD}), we have
$M_*^2 f=\Mpl^2={\rm constant}$. 
Using the consistency conditions 
(\ref{consistency1Mr}) and (\ref{consistency3Mr}) together with 
the background Eqs.~(\ref{back1}) and (\ref{back2}), we obtain 
\be
\frac{\rd }{\rd t} \left( 
\frac{a^3 \tilde{c}}{\bar{N}} \right)=0\,,
\label{tcevo}
\ee
whose integrated solution is given by 
\be
\tilde{c}=\tilde{c}_0 
\frac{\bar{N}}{a^3}\,,
\ee
where $\tilde{c}_0$ is an integration constant. 
The background equations of motion (\ref{back1}) and (\ref{back2}) reduce to 
\ba
& &
3\Mpl^2 H^2=\Lambda+\frac{\tilde{c}}{\bar{N}^2}
+m_{\rm c} \bar{n}+\bar{\rho}_{\rm m}\,,\label{back1b}\\
& &
\Mpl^2 \left( 2 \dot{H}+3 H^2 
\right)=\Lambda-\frac{\tilde{c}}{\bar{N}^2}-\dot{d}-\bar{p}_{\rm m}\,.
\label{back2b}
\ea
The difference from GP theories is that 
$\tilde{c}$ does not necessarily vanish. 
As $a$ increases with the cosmic 
expansion, $\tilde{c}$ 
eventually approaches 0 
in an ever-expanding Universe. 
The limit 
$\tilde{c} \to 0$ can be identified 
as a tracker solution first recognized in 
Ref.~\cite{DeFelice:2010pv} for 
uncoupled covariant Galileons.
The couplings between DE and DM appear 
as the correction terms $\Delta \Lambda$, 
$\Delta m_{\rm c}$, and $\Delta c$ 
in Eqs.~(\ref{back1b})-(\ref{back2b}).

As an example, we consider the subclass of shift-symmetric Horndeski theories with the luminal propagation of gravitational waves 
in the presence of the interacting Lagrangian (\ref{Hointer}), i.e., 
\be
{\cal S}=\int \rd^4 x \sqrt{-g} \left[ 
G_2(X)+G_3(X) \square \phi 
+\frac{\Mpl^2}{2} R 
-f_1 (\phi, X, Z)n+f_2(\phi, X, Z)
\right]+{\cal S}_{\rm m}\,,
\label{Horndeski2}
\ee
where $X=-\nabla_{\mu}\phi 
\nabla^{\mu}\phi/2$ and 
$Z=u^{\mu}\nabla_{\mu}\phi$.
In the context of more general 
non-shift-symmetric Horndeski theories, 
the relations between 
$\Lambda$, $\tilde{c}$, $d$ 
and $G_2$, $G_3$, $f_1$, 
$f_2$ are presented in 
Eqs.~(\ref{Lambda0})-(\ref{d0}) in 
Appendix~\ref{Hocorres}. 
The shift-symmetric Horndeski theories 
correspond to taking the limits 
$G_2(\phi,X) \to G_2 (X)$, 
$g_3(\phi,X) \to g_3(X)$, and 
$G_4(\phi) \to \Mpl^2/2$, where 
$G_3=g_3+2Xg_{3,X}$.
Then, the functions $\tilde{c}$, $\Lambda$, 
$\dot{d}$ are given by 
\ba
\tilde{c} &=& \frac{1}{2}G_{2,X}
-\frac{3}{2}H \sqrt{2X_{\rm BG}}\,G_{3,X}
-\frac{1}{2}\bar{n} \left[ f_{1,X}
+(2X_{\rm BG})^{-1/2} f_{1,Z} \right] 
+\frac{1}{2} \left[ f_{2,X}
+(2X_{\rm BG})^{-1/2} f_{2,Z} \right]
=\tilde{c}_0\frac{\bar{N}}{a^3}
\,,\\
\Lambda &=& -G_2-f_2+2\tilde{c} X_{\rm BG}\,,\\
\dot{d} &=& \sqrt{2X_{\rm BG}} 
\dot{X}_{\rm BG}\,G_{3,X}\,,
\ea
where $X_{\rm BG}=\dot{\bar{\phi}}^2/(2\bar{N}^2)$. 
In the limit $\tilde{c} \to 0$, the structure 
of the background equations is similar to 
that in GP theories.

\subsection{Non-shift-symmetric 
scalar-tensor theories}

In non-shift-symmetric scalar-tensor theories, 
we do not need to impose the consistency condition $\dot{f}=0$. 
Taking the limit $g_M \to 0$ 
in Eqs.~(\ref{back1}) and (\ref{back2}), 
we obtain
\ba
& &
3M_*^2 f H^2=\Lambda
+\frac{\tilde{c}}{\bar{N}^2}
+m_{\rm c} \bar{n}+\bar{\rho}_{\rm m}\,,\label{nonback1}\\
& &
M_*^2 f \left( 2\dot{H}+3 H^2 
\right)=\Lambda-\frac{\tilde{c}}{\bar{N}^2}
-\dot{d}-2M_*^2\dot{f}H
-\bar{p}_{\rm m}\,.
\label{nonback2}
\ea
The absence of consistency conditions 
(\ref{consistency1Mr}) and (\ref{consistency3Mr}) means that the 
evolution of $\tilde{c}$ is not constrained 
to be as Eq.~(\ref{tcevo}).

Let us consider the subclass of 
non-shift-symmetric Horndeski theories with 
the interacting Lagrangian (\ref{Hointer}), i.e., 
\be
{\cal S}=\int \rd^4 x \sqrt{-g} \Big[ 
G_2(\phi,X)+G_3(\phi,X) \square \phi 
+G_4(\phi) R 
-f_1 (\phi, X, Z)n+f_2(\phi, X, Z)
\Big]+{\cal S}_{\rm m}\,.
\label{Horndeski3}
\ee
In this case, the EFT functions $f$, $\Lambda$, 
$\tilde{c}$, $d$ in Eqs.~(\ref{nonback1}) and (\ref{nonback2}) are related to $G_2$, $G_3$, $G_4$, $f_1$, $f_2$ according to  
Eqs.~(\ref{fre})-(\ref{d0}) in Appendix~\ref{Hocorres}. 
While $\Lambda$, $\tilde{c}$, and $d$ 
contain the $\phi$, $X$ derivatives 
of $g_3$, the combinations 
$\Lambda+\tilde{c}/\bar{N}^2$ and 
$\Lambda-\tilde{c}/\bar{N}^2-\dot{d}$ can 
be expressed in terms of the $\phi$, $X$ derivatives of $G_3=X+2XG_{3,X}$. 
Therefore, the right-hand sides of Eqs.~(\ref{nonback1}) and (\ref{nonback2}) 
do not possess the $g_3$-dependent quantities.
The resulting background equations of motion 
are consistent with those derived in 
the literature, e.g., Eqs.~(2.35) and 
(2.36) in Ref.~\cite{Kase:2020hst}.

\section{EFT action with the 
$\alpha$-basis parameters} \label{sec:EFTaction_dimensionless}

In this section, we express the EFT action (\ref{Sfull}) in terms of the 
$\alpha$-basis dimensionless parameters 
often used in the EFT of DE~\cite{Bellini:2014fua,Frusciante:2019xia,Aoki:2021wew,Aoki:2024ktc}. 
As we already mentioned in 
Sec.~\ref{DEsec}, we choose the gauge 
for the vector field 
in which the spatial components 
of $A_{\mu}$ vanish. Taking into account the perturbation $\delta A_{0}$ for the temporal vector component, 
we consider the following configuration 
\begin{equation}
A_{\mu} = \left[ \bar{A}_{0}(t) + \delta A_{0}, 
{\bm 0} \right]\,,
\end{equation}
where $\bar{A}_{0}(t)$ is the background 
value of $A_0$ and 
$\delta A_{0}$ is the perturbation.
For the lapse function $N$, we split it into the background and 
perturbed parts, as 
$N=\bar{N}+\delta N$.
The perturbation of the metric component 
$g^{00}=-1/N^2$ is given by 
\be
\delta g^{00}=\frac{2 \delta N}{\bar{N}^3} 
-3\frac{(\delta N)^2}{\bar{N}^4}
+{\cal O}\left(\frac{(\delta N)^3}{\bar{N}^5} \right)\,.
\ee
Now, we express the second-order Lagrangian 
${\cal L}_{\rm D}^{(2)}$ in Eq.~(\ref{EFTLD2})
by using the perturbations $\delta A_0$, $\delta N$, and $N^i$. In doing so, we exploit the following property
\begin{equation}
\frac{\delta\tilde{g}^{00}}
{-\tilde{g}^{00}_{\rm BG}} = 
\frac{\delta g^{00}}{-g^{00}_{\rm BG}} 
-\frac{2 g_M}{(1+ g_M A_{0})} \delta A_{0}\,,
\end{equation}
where $g^{00}_{\rm BG}=-1/\bar{N}^2$. 

\subsection{Second-order action from 
${\cal L}{}_{\rm D}^{(2)}$}

Let us first study the contribution to the action (\ref{Sfull}) arising from ${\cal L}_{\rm D}^{(2)}$.
Up to second order in perturbations, 
the scalar products $F_{\mu} F^{\mu}$, $q^{\mu}q_{\mu}$, and $q^{\mu}F_{\mu}$ 
are given, respectively, by 
\ba
F_{\mu} F^{\mu} &=& (n^0 \nabla_{i}A_0)
(n^0 \nabla^{i}A_0)=
\frac{\delta^{ij} \nabla_{i} \delta A_0
\nabla_{j} \delta A_0}{a^2 \bar{N}^2}\,,\\
q^{\mu}q_{\mu} &=& \frac{N^i N_i}
{N^2-N_j N^j} \simeq 
\frac{N^i N_i}{\bar{N}^2}\,,\\
q^{\mu}F_{\mu} &=& \frac{n^0 N^i 
\nabla_i \delta A_0}
{\sqrt{N^2-N_j N^j}}
\simeq \frac{N^i \nabla_{i}\delta A_{0}}{\bar{N}^2}\,.
\ea
Then, the Lagrangian (\ref{EFTLD2})
can be expressed in the form
\begin{equation}
\mathcal{L}^{(2)}_{\rm D} = \mathcal{L}^{(2)}_{\rm ST} + \mathcal{L}^{(2)}_{\delta \hat{A}_{0}} \,,
\end{equation}
where 
\ba
\mathcal{L}^{(2)}_{\rm ST} &=& 
2M_2^4 \left( \frac{\delta N}{\bar{N}} 
\right)^2-\bar{M}_1^3 \frac{\delta N}{\bar{N}}
\delta K-2 m_1^4 \frac{\delta N}{\bar{N}}\frac{\delta n}{\bar{n}} 
-m_2^4 \frac{N^i N_i}
{\bar{N}^2}\,,
\label{L2a} \\
\mathcal{L}^{(2)}_{\delta \hat{A}_{0}}
&=&
\frac{1}{2} \left( 
\nabla_i\delta \hat{A}_0 \nabla^i\delta \hat{A}_0+g_{\rm eff}^2 M_2^4 \delta \hat{A}_0^2 \right) 
+\left[ 
\frac{\bar{m}_1^2 \nabla_i N^i}
{\sqrt{\gamma}_1 \bar{N}}
+g_{\rm eff} \left( \frac{1}{2} \bar{M}_{1}^{3} \delta K - 2M_{2}^{4} 
\frac{\delta N}{\bar{N}} 
+m_{1}^{4} \frac{\delta n}{\bar{n}} \right)
\right]\delta \hat{A}_0\,,
\label{L2b}
\ea
up to boundary terms. 
Here, $g_{\rm eff}$ and $\delta \hat{A}_{0}$ are 
defined by 
\begin{equation}
g_{\rm eff}:=\frac{2 g_{M} \bar{N}}{\sqrt{\gamma_{1}} (1 + g_{M}\bar{A}_{0})} \,, \qquad 
\delta \hat{A}_{0}:= 
\frac{\sqrt{\gamma_{1}}}{\bar{N}} 
\delta A_{0} \,.
\label{hatA0}
\end{equation}
As we will see in Appendix~\ref{vecsec}, we require that $\gamma_1>0$ to avoid the ghost for intrinsic vector perturbations. We have used this property
for the definition of $\delta \hat{A}_{0}$ in Eq.~(\ref{hatA0}).

Under Fourier transformation, any perturbation $X$ in real space 
is decomposed as 
\be
X(t,{\bm x})=\int \frac{{\rm d}^3 k}
{(2\pi)^3} e^{i {\bm k} \cdot {\bm x}}
X_k (t)\,,
\ee
where ${\bm k}$ is a comoving wavenumber 
with $k=|{\bm k}|$. When we refer to quantities 
in Fourier space, we omit the subscript $k$ 
in the following. For example, the first 
two terms in Eq.~(\ref{L2b}) can be interpreted as 
$(1/2) ( k^2/a^2+g_{\rm eff}^2 M_2^4 )
\delta \hat{A}_0^2$ 
in Fourier space.

The Lagrangian (\ref{L2a}) does not contain any dependence of $\delta \hat{A}_0$. Furthermore, in vector-tensor theories, the term $-\tilde{c}(t) \tilde{g}^{00}$ vanishes due to 
the equation of motion for $\bar{A}_0$, so that 
${\cal L}_{\rm D}^{\rm NL}$ does not have 
any dependence on $\delta \hat{A}_0$ either. 
Then, the variation of the full action 
${\cal S}^{(2)}=\int {\rm d}^4 x \sqrt{-g}
({\cal L}_{\rm D}^{\rm NL}+\mathcal{L}^{(2)}_{\rm ST} + \mathcal{L}^{(2)}_{\delta \hat{A}_{0}})
+{\cal S}_{\rm m}$ 
with respect to $\delta \hat{A}_0$ amounts to 
varying ${\cal S}^{(2)}_{\delta \hat{A}_0}=
\int {\rm d}^4 x \sqrt{-g}\,
\mathcal{L}^{(2)}_{\delta \hat{A}_{0}}$. 
In real space, this leads to a non-dynamical equation for $\delta \hat{A}_0$ that can be formally solved as 
\be
\delta \hat{A}_0=-\frac{1}{-\nabla_i\nabla^i
+g_{\rm eff}^2 M_2^4} \left[ 
\frac{\bar{m}_1^2 \nabla_i N^i}
{\sqrt{\gamma}_1 \bar{N}}
+g_{\rm eff} \left( \frac{1}{2} \bar{M}_{1}^{3} 
\delta K - 2M_{2}^{4} 
\frac{\delta N}{\bar{N}} 
+m_{1}^{4} \frac{\delta n}{\bar{n}} \right)
\right]\,.
\label{delAcon}
\ee
In Fourier space, the non-local pre-factor 
$\nabla_i \nabla^i$ should be understood as 
$-k^2/a^2$.

The shift-symmetric scalar-tensor theories 
correspond to the limits
$g_{\rm eff} \to 0$ and 
$\bar{m}_{ 1}^2 \to 0$, in which case 
we have $\delta \hat{A}_{0}=0$.
In vector-tensor theories, 
$\delta \hat{A}_{0}$ is non-vanishing 
and affects the dynamics of perturbations 
through the constraint equation (\ref{delAcon}).
Substituting Eq.~(\ref{delAcon}) into 
${\cal L}_{\rm D}^{(2)}$ to eliminate  
$\delta \hat{A}_{0}$, the resulting 
second-order action is given by 
\begin{eqnarray}
\hspace{-0.3cm}
{\cal S}_{\rm D}^{(2)}
&=& \int \rd^4 x 
\sqrt{-g}\,{\cal L}_{\rm D}^{(2)} \nonumber \\
&=&
\int \rd^4 x\,
\bar{N}a^3 \biggl[
2 M_{2, {\rm eff}}^{4} 
\left( \frac{\delta N}{\bar{N}} \right)^{2} 
-\bar{M}_{1, {\rm eff}}^{3} 
\frac{\delta N}{\bar{N}} \delta K  
- 2 m_{1, {\rm eff}}^{4} \frac{\delta N}{\bar{N}}
\frac{\delta n}{\bar{n}} 
-m_2^4 \frac{N_{i}N^{i}}{\bar{N}^2} 
-\bar{m}_{1,{\rm eff}}^4 
\frac{(\nabla_i N^i)^2}{\bar{N}^2}
\nonumber \\
\hspace{-0.3cm}
& &\qquad \qquad \quad
-\frac{\mu_{1}^{4}}{2}  
\left( \frac{\delta n}{\bar{n}} \right)^{2} 
- \frac{\mu_{2}^{2}}{8} 
(\delta K)^2 
- \frac{\mu_{3}^{3}}{2} \delta K
\frac{\delta n}{\bar{n}}
-\frac{\mu_{4}^{4}}{2} 
\delta K \frac{\nabla_{i} N^{i}}{\bar{N}}
+2\mu_{5}^{5} \frac{\delta N}{\bar{N}}  \frac{\nabla_{i} N^{i}}{\bar{N}}
-\mu_{6}^{5} \frac{\delta n}{\bar{n}}  \frac{\nabla_{i}N^{i}}{\bar{N}} \biggr],
\label{eq:EFT_action_no_dA}
\end{eqnarray}
where
\begin{eqnarray}
& &
M_{2,{\rm eff}}^{4}:=
\left( 1 - \mathcal{G} \right) M_{2}^{4}\,, 
\qquad
\bar{M}_{1, {\rm eff}}^{3}:=
\left( 1 - \mathcal{G} \right) \bar{M}_{1}^{3}\,, 
\qquad
m_{1, {\rm eff}}^{4}:= 
\left( 1 - \mathcal{G} \right) m_{1}^{4}\,, 
\qquad 
\bar{m}_{1,{\rm eff}}^4:=\frac{{\cal G}\bar{m}_1^4}
{2 \gamma_1 g_{\rm eff}^2 M_2^4}\,, 
\nonumber \\
& &
\mu_{1}^{4}:=\mathcal{G} 
\frac{m_{1}^{8}}{M_{2}^{4}} \,,\qquad
\mu_{2}^{2}:=\mathcal{G} 
\frac{\bar{M}_{1}^{6}}{M_{2}^{4}} \,, 
\qquad
\mu_{3}^{3}:=\mathcal{G}\frac{\bar{M}_{1}^{3} m_{1}^{4}}{M_{2}^{4}}\,, \qquad
\mu_{4}^{4}:=  
\frac{\mathcal{G}\bar{M}_{1}^{3}\bar{m}_{1}^{2}}{\sqrt{\gamma_{1}}\,g_{\rm eff} M_{2}^{4}} \,, \qquad
\mu_{5}^{5}:= \frac{\mathcal{G}\bar{m}_{1}^{2}}
{\sqrt{\gamma_{1}}\,g_{\rm eff}} \,, 
\nonumber \\
& &
\mu_{6}^{5}:=\frac{\mathcal{G} m_{1}^{4}\bar{m}_{1}^{2}}{\sqrt{\gamma_{1}}\,g_{\rm eff} 
M_{2}^{4}}\,,
\label{coeff}
\end{eqnarray}
with
\begin{equation}
\mathcal{G}:= 
\frac{g_{\rm eff}^{2} M_{2}^{4}}{-\nabla_i\nabla^i 
+ g_{\rm eff}^{2} M_{2}^{4}} \,.
\end{equation}
After integrating out the field 
$\delta \hat{A}_{0}$, 
the coefficients in Eq.~(\ref{coeff}) 
depend not only on $t$ but they are also non-local operators in space, which means that they feature a characteristic scale dependence ($k$-dependence 
in Fourier space) for the corresponding Fourier modes.  
This reflects the effect of the vector DE 
arising from the non-vanishing value of $\mathcal{G}$ 
for $g_M \neq 0$.
In particular, the coefficients 
$\mu_{i}$ ($i=1,2,\cdots,6$)
emerge from the interaction of 
DM with the vector DE. 
In the long-wavelength limit 
($k^2/a^{2} \ll g_{\rm eff}^{2} M_{2}^{4}$ 
in Fourier space), 
we have $\mathcal{G} \rightarrow 1$ and hence 
$M_{2,{\rm eff}}^4$, 
$\bar{M}_{1,{\rm eff}}^3$, and 
$m_{1,{\rm eff}}^4$ vanish, while
the other coefficients in Eq.~(\ref{coeff})  
survive. In the small-scale limit 
($k^2/a^2 \gg g_{\rm eff}^{2}M_{2}^{4}$ 
in Fourier space), 
we have ${\cal G} \to 0$ and hence 
$M_{2,{\rm eff}}^{4} \to M_{2}^{4}$, 
$\bar{M}_{1,{\rm eff}}^{3} \to \bar{M}_{1}^{3}$, and 
$m_{1,{\rm eff}}^{4} \to m_{1}^{4}$, 
while the other coefficients in Eq.~(\ref{coeff}) are suppressed. Nonetheless, this does not mean that these operators are negligible on small scales. For example, $(\nabla_i N^i)^2\sim k^2 N^i N_i$ is a higher derivative term and apparently becomes more important than $N^i N_i$. However, the coefficient scales as $\bar{m}_{1,{\rm eff}}^4 \propto k^{-2}$, so $\bar{m}_{1,{\rm eff}}^4 (\nabla_i N^i)^2$ and $m_2^4 N_i N^i$ contribute to the dynamics to the same extent. These non-local operators feature the vector DE.

As in the case of the EFT of uncoupled 
DE in vector-tensor theories~\cite{Aoki:2021wew}, 
we have three unified descriptions of 
the EFT of coupled DE and DM: 
\begin{enumerate}
\item $g_{\rm eff} \neq 0$, it corresponds 
to the EFT of coupled vector DE, 
\item $g_{\rm eff} = 0$, together with the 
consistency conditions, we obtain the EFT 
of coupled scalar DE in shift-symmetric 
scalar-tensor theories, 
\item $g_{\rm eff} = 0$, without the consistency 
conditions, we have the EFT of coupled scalar DE 
in non-shift-symmetric scalar-tensor theories.
\end{enumerate}
\subsection{Second-order action from 
${\cal L}_{\rm D}^{\rm NL}$}

To compute the contribution to the action (\ref{Sfull}) arising from ${\cal L}_{\rm D}^{\rm NL}$, we introduce the following 
perturbed quantities
\begin{equation}
\delta K_{\mu \nu} = 
K_{\mu \nu} - Hh_{\mu \nu} \,, 
\qquad \delta K = K - 3 H\,.  
\end{equation}
We also exploit the following properties
\ba
K_{\mu \nu} K^{\mu \nu}
&=& 
3H^2+2H \delta K+ \delta K_{\mu \nu}
\delta K^{\mu \nu}\,,\\
\int \rd^4 x \sqrt{-g}\,{\cal F}(t) K
&=& -\int \rd^4 x \sqrt{-g} 
\frac{1}{N} \frac{\rd {\cal F}(t)}{\rd t}\,,
\ea
where the latter property holds up to a boundary 
term for an arbitrary function ${\cal F}(t)$. 
Then, the action 
${\cal S}=\int \rd^4 x \sqrt{-g}\, 
{\cal L}_{\rm D}^{\rm NL}$ can 
be expressed as 
\ba
{\cal S}_{\rm D}^{\rm NL}
&=& 
\int \rd^4 x
\sqrt{-g} \biggl[ 
\frac{M_*^2 f}{2} \left( {}^{(3)}R+\delta K_{\mu \nu} \delta K^{\mu \nu}
-\delta K^2 \right)
-\Lambda - \tilde{c} \tilde{g}^{00}
-m_{\rm c} \bar{n}-m_{\rm c}  \delta n 
+3M_*^2 f H^2+\frac{\bar{N}}{N} 
\left( 2M_*^2 fH +d \right)^{\cdot}
\biggr] \nonumber \\
&=& 
\int \rd^4 x\, N \sqrt{h} 
\Biggl[ \frac{M_*^2 f}{2} \left( {}^{(3)}R+\delta K_{\mu \nu} \delta K^{\mu \nu}
-\delta K^2 \right)
-\frac{\bar{N}}{N}m_{\rm c} \bar{n} 
-m_{\rm c} \delta n 
+\bar{\rho}_{\rm m}-\frac{\bar{N}}{N} 
(\bar{\rho}_{\rm m}+\bar{p}_{\rm m})
+\tilde{c} \left( \frac{1}{N} 
-\frac{1}{\bar{N}} \right)^2\Biggr],
\nonumber\\
\label{SD2}
\ea
where, in the second equality, we used the background Eqs.~(\ref{back1}) and (\ref{back2}). 
We have also taken the limit $g_M \to 0$ for the derivation of the last term in Eq.~(\ref{SD2}). 
This reflects the fact that $\tilde{c}$ is generally non-vanishing in scalar-tensor theories, 
but it vanishes in vector-tensor theories.
The second-order contribution to 
${\cal S}_{\rm D}^{\rm NL}$ is given by 
\ba
\hspace{-0.1cm}
{\cal S}_{\rm D}^{{\rm NL}(2)}
&=&
\int \rd^4 x  
\bar{N} a^3 
\biggl\{ \frac{M_*^2 f}{2} 
\biggl[ \delta K_{\mu \nu} \delta K^{\mu \nu}
-\delta K^2+\frac{\delta N}{\bar{N}} 
\delta^{(3)}R
+\delta_2 \biggl( \frac{\sqrt{h}}{a^3}
{}^{(3)}R \biggr)
\biggr] -m_{\rm c} \frac{\delta N}{\bar{N}} \delta n 
-m_{\rm c} \delta_2 \biggl( \frac{\sqrt{h}}{a^3}n \biggr)
+\tilde{c}\,\frac{(\delta N)^2}{\bar{N}^4}
\biggr\} \nonumber\\
\hspace{-0.1cm}
& &
+{\cal S}_{\rm m}^{{\rm NL}(2)},
\label{Scon2}
\ea
with
\be
{\cal S}_{\rm m}^{{\rm NL}(2)}
=\delta_2 \int \rd^4 x\,\bar{N} \sqrt{h} 
\left( \frac{\delta N}{\bar{N}} 
\bar{\rho}_{\rm m}-\bar{p}_{\rm m} \right)\,,
\label{SmNL2}
\ee
where $\delta_2$ represents the second-order part.

\subsection{Total second-order action}\label{subsec:2ndorder_EFTaction}

Now, we take the sum of ${\cal S}_{\rm D}^{(2)}$, 
${\cal S}_{\rm D}^{{\rm NL}(2)}$, 
and the second-order part ${\cal S}_{\rm m}^{(2)}$ of the matter action (\ref{EFTD}).
On using Eqs.~(\ref{eq:EFT_action_no_dA}) 
and (\ref{Scon2}), the total second-order action 
${\cal S}^{(2)}={\cal S}_{\rm D}^{(2)}+
{\cal S}_{\rm D}^{{\rm NL}(2)}
+{\cal S}_{\rm m}^{(2)}$ yields
\begin{tcolorbox}[colback=black!5!white,colframe=black!75!white,title=Total second-order action in $\alpha$-basis parameters]
\begin{eqnarray}
{\cal S}^{(2)} & = & \int \rd^4 x\,  \bar{N} a^{3} 
\frac{M^{2}}{2} \nn
&\times& \biggl[ \delta K_{\mu \nu} 
\delta K^{\mu \nu} - \delta K^{2} 
+\delta_{2}\left( \frac{\sqrt{h}}{a^3} 
\Rs \right) + \frac{\delta N}{\bar{N}}\delta \Rs - 6\Omega_{\rm c} H^{2} \delta_{2} \left( \frac{\sqrt{h}}{a^{3}} \frac{n}{\bar{n}} 
\right)+H^2 \tilde{\alpha}_{K} 
\left(\frac{\delta N}{\bar{N}} \right)^2 
+ 4 \tilde{\alpha}_{B} H \delta K \frac{\delta N}{\bar{N}}  \nn
& &\,
+(\tilde{\alpha}_{m_{1}}-6\Omega_{\rm c})
H^2\frac{\delta N}{\bar{N}}\frac{\delta n}{\bar{n}}
+ \alpha_{m_{2}} H^2 \frac{N^i N_i}{\bar{N}^2}
+\tilde{\alpha}_{\bar{m}_1}  
\frac{(\nabla_i N^i)^2}{\bar{N}^2}
+ \alpha_{\mu_{1}} H^{2} 
\left( \frac{\delta n}{\bar{n}} \right)^2  
+ \alpha_{\mu_{2}} \delta K^2 
+ \alpha_{\mu_{3}} H \delta K 
\frac{\delta n }{\bar{n}}
\nonumber \\
& &  
\,-\alpha_{\mu_{4}} \delta K 
\frac{\nabla_i N^i}{\bar{N}}
-\alpha_{\mu_{5}} H 
\frac{\delta N}
{\bar{N}}\frac{\nabla_i N^i}{\bar{N}}
-\alpha_{\mu_{6}} H \frac{\delta n}{\bar{n}}
\frac{\nabla_i N^i}{\bar{N}} \biggr] 
+ \tilde{\cal S}_{\rm m}^{(2)}\,,
\label{eq:EQ_EFT_alpha_action}
\end{eqnarray}
\end{tcolorbox}
\noindent
where
\be
\tilde{\cal S}_{\rm m}^{(2)}:=
{\cal S}_{\rm m}^{(2)}
+\delta_2 \int \rd^4 x\,\bar{N} \sqrt{h} 
\left( \frac{\delta N}{\bar{N}} 
\bar{\rho}_{\rm m}-\bar{p}_{\rm m} \right)\,,
\ee
and the $\alpha$-basis dimensionless parameters are defined by 
\ba
\hspace{-0.1cm}
& &
\tilde{\alpha}_K := 
\alpha_K (1-{\cal G})
+6 \Omega_{\tilde{c}}\,,
\qquad 
\alpha_K:=\frac{4M_2^4}{H^2 M^2}\,,\qquad
\tilde{\alpha}_B :=\alpha_B (1-{\cal G})\,,
\qquad \alpha_B:=-\frac{\bar{M}_1^3}{2H M^2}\,,
\nonumber \\ 
\hspace{-0.1cm}
& & 
\tilde{\alpha}_{m_1} :=
\alpha_{m_1} (1-{\cal G})\,,
\qquad
\alpha_{m_1}:=-\frac{4m_1^4}{H^2 M^2}\,,
\qquad
\alpha_{m_2}:=-\frac{2m_2^4}{H^2 M^2}\,,\qquad
\tilde{\alpha}_{\bar{m}_1} 
:= -\frac{2\bar{m}_{1,{\rm eff}}^4}{M^2}
=-\frac{{\cal G}}{\alpha_K \alpha_g^2} 
\alpha_{\bar{m}_1}^2 \,,\nonumber \\
\hspace{-0.1cm}
& &
\alpha_{\bar{m}_1}:=\frac{\bar{m}_1^2}
{\sqrt{\gamma_1} H M},
\qquad
\alpha_{\mu_1}:= -\frac{\mu_1^4}
{H^2 M^2}
=-{\cal G} \frac{\alpha_{m_1}^2}{4\alpha_K},\qquad 
\alpha_{\mu_2}:=-\frac{\mu_2^2}{4M^2}
=-4{\cal G} \frac{\alpha_B^2}{\alpha_K},
\qquad
\alpha_{\mu_3}:= -\frac{\mu_3^3}{H M^2}
=-2{\cal G} \frac{\alpha_B \alpha_{m_1}}{\alpha_K}, \nonumber \\
\hspace{-0.1cm}
& &
\alpha_{\mu_4}:=
\frac{\mu_4^4}{M^2}
=-4 {\cal G}  
\frac{\alpha_{\bar{m}_1} \alpha_B}
{\alpha_K \alpha_g}\,,\qquad
\alpha_{\mu_5}=
-\frac{4\mu_5^5}{H M^2}
=-2{\cal G}  
\frac{\alpha_{\bar{m}_1}}{\alpha_g} \,,
\qquad 
\alpha_{\mu_6}=\frac{2\mu_6^5}
{H M^2}=-{\cal G} 
\frac{\alpha_{\bar{m}_1} \alpha_{m_1}}
{\alpha_K \alpha_g}\,,
\label{aldef1}
\ea
with\footnote{In Ref.~\cite{Aoki:2021wew}, 
the notation $\alpha_V=M^2 g_{\rm eff}^2/4=\alpha_g^2$ was adopted. 
In this paper, we will use $\alpha_g$ 
instead of $\alpha_V$.}
\be
{\cal G}=\frac{\alpha_K \alpha_g^2}
{-H^{-2}\nabla_i \nabla^i+\alpha_K \alpha_g^2}\,,
\qquad
\alpha_g:=\frac{M}{2} g_{\rm eff}\,,
\label{aldef2}
\ee
and 
\be
M^2:= M_*^2 f\,,\qquad 
\Omega_{\rm c}:=
\frac{m_{\rm c} \bar{n}}{3H^2 M^2}\,,\qquad 
\Omega_{\tilde{c}}:=\frac{\tilde{c}}{3H^2 \bar{N}^2M^2}\,.
\ee
We note that $\Omega_{\rm c}$ is the effective 
density parameter for CDM, which is 
different from the other density parameter 
$\Omega_{\tilde{c}}$ arising from the EFT 
function $\tilde{c}$. We recall that $\tilde{c}=0$ 
in vector-tensor theories ($\alpha_g \neq 0)$ and $\tilde{c}=\tilde{c}_0 \bar{N}/a^3$ 
in shift-symmetric scalar-tensor theories.
In non-shift-symmetric scalar-tensor theories, 
there is no particular constraint on $\tilde{c}$. 
We caution that, in scalar-tensor theories, 
the term $1/\bar{N}^2$ in $\Omega_{\tilde{c}}$ 
should be interpreted as 
$1/\bar{N}^2 \to 2X \to \dot{\phi}^2$, 
so that $\Omega_{\tilde{c}} \to \tilde{c} \dot{\phi}^2/(3H^2 M^2)$.
The $\alpha$-basis EFT parameters appearing 
as the coefficients in the action (\ref{eq:EQ_EFT_alpha_action}) feature 
non-localities (except $\alpha_{m_2}$) 
through the non-local operator ${\cal G}$, 
which in turn leads to the corresponding 
$k$-dependence in Fourier space. 
In vector-tensor theories, we have 
$\alpha_g \neq 0$ and 
${\cal G} \neq 0$, so that  
$\tilde{\alpha}_{\bar{m}_1}$ and $\alpha_{\mu_i}$ ($i=1,2,\cdots,6$) 
are non-vanishing. These seven EFT 
parameters vanish in the scalar-tensor 
limit $\alpha_g \to 0$.
We note that the $\Lambda$CDM model can be 
recovered by taking the limit $g_{\rm eff} 
\to 0$, ${\cal G} \to 0$, $M^2 \to \Mpl^2$, 
$m_{\rm c} \to \hat{m}_{\rm c}={\rm constant}$, 
$\tilde{c} \to 0$, and 
$\Lambda(t)=\Lambda={\rm constant}$, 
with all the $\alpha$-basis parameters 
in Eq.~(\ref{aldef1}) vanishing. 
The strict $\Lambda$CDM model does not have DE perturbations. However, it may be possible 
to accommodate scenarios with an exact 
$\Lambda$CDM background, while featuring 
DE perturbations.\footnote{In the context of 
minimal theories of massive gravity \cite{DeFelice:2015hla,Aoki:2018brq}, 
it is possible to modify the evolution of 
perturbations by keeping the $\Lambda$CDM background. This is possible in other modified gravities~\cite{delaCruz-Dombriz:2006kob} 
and DE models~\cite{BeltranJimenez:2008enx,BeltranJimenez:2009arq}.}
This is achieved by e.g., allowing for  
a non-vanishing $\alpha_B$ and/or $\alpha_K$. Note that we choose to separate $\tilde{\alpha}_{m_{1}}$ and $\Omega_{\rm c}$ in the coefficient of the operator $(\delta N/\bar{N})(\delta n/\bar{n})$, which 
allows us to take a smooth, non-interacting limit.

It is informative to mention the relation 
between the EFT parameters and the couplings 
of concrete theories. 
In the subclass of non-shift-symmetric 
Horndeski theories with the action (\ref{Horndeski3}), the dictionaries 
between $M_2^4$, $\bar{M}_1^3$, $m_{\rm c}$, 
$m_1^4$, $m_2^4$ and $G_2$, $g_3$, $G_4$, 
$f_1$, $f_2$ are given in 
Eqs.~(\ref{M2def})-(\ref{m2def2}) 
of Appendix~\ref{Hocorres}, where 
$G_3=g_3+2Xg_{3,X}$.
We note that $m_{\rm c}$ corresponds 
to $f_1(\phi,X,Z)$ itself, while $m_1^4$ 
depends on $f_{1,X}$ and $f_{1,Z}$. 
The momentum transfer 
induced by $m_2^4$ is dependent on 
$f_{1,Z}$ and $f_{2,Z}$, while 
$\bar{m}_1^2=0$ due to the absence of 
the $\tilde{E}$ dependence in $f_1$ and $f_2$.
In the subclass of GP theories with 
the action (\ref{eq:GP2}), 
the relations between $M_2^4$, 
$\bar{M}_1^3$, $\gamma_1$, $m_{\rm c}$, 
$m_1^4$, $m_2^4$, $\bar{m}_1^2$ and $G_2$, $g_3$, $f_1$, $f_2$ are given in 
Eqs.~(\ref{M2d})-(\ref{bm1}) 
of Appendix~\ref{sec:GP}, 
where $G_3=g_3+2\tilde{X} g_{3,\tilde{X}}$.
They are analogous to those in scalar-tensor 
theories, but the difference is that 
the $\tilde{E}$ dependence in 
$f_1$ and $f_2$ leads to 
$\bar{m}_1^2 \neq 0$ in 
vector-tensor theories~\cite{Pookkillath:2024ycd}.

\section{Perturbation equations of motion} \label{sec:perturbation_EOMs}

In this section, we will derive the linear perturbation equations of motion for tensor and scalar perturbations. In the ADM line element (\ref{eq:3+1metric}), 
we incorporate the perturbed fields 
in the form 
\ba
N &=& \bar{N}(t) \left( 1+\alpha \right)\,,
\label{Ndef} \\
N_i &=& \bar{N}(t) \left( 
\partial_i \chi+S_i \right)\,,
\label{Nidef} \\
h_{ij} &=& a^2(t) \left[ (1+2\zeta) 
\delta_{ij}+2 \partial_i \partial_j E+\tilde{h}_{ij}
+2 \partial_j F_i \right]\,,
\label{hijdef}
\ea
where $\alpha$, $\chi$, $\zeta$, and 
$E$ are scalar perturbations, $S_i$ 
and $F_i$ are vector perturbations 
obeying the conditions 
$\partial^i S_i=0$ and 
$\partial^i F_i=0$, and $\tilde{h}_{ij}$ 
is the tensor perturbation 
satisfying the conditions 
$\partial^i \tilde{h}_{ij}=0$ 
and ${\tilde{h}_{i}}^i=0$. 
Since we focus on the dynamics of 
linear perturbations, we can use the 
partial derivative $\partial_i
=\partial/\partial x^i$ instead of 
the covariant derivative on the right-hand 
sides of Eqs.~(\ref{Nidef})-(\ref{hijdef}).
We recall that our EFT is formulated under 
the irrotational ansatz, so the contribution 
of vector-field perturbations 
is ignored in the following discussion. 
In Appendix \ref{vecsec}, we will discuss the propagation of intrinsic vector modes 
in the decoupling limit and obtain the relevant stability conditions.

In the configuration 
$A_{\mu}=(A_0, \partial_i A)$,  
the gauge condition $A=0$ is 
chosen to fix the residual gauge freedom 
of the combined $U(1)$ and time diffeomorphisms
given in Eq.~(\ref{shift_sym2}).
We also choose the comoving gauge $\phi^i=x^i$ for the DM fluid to fix the spatial diffeomorphism. 
In this case, we do not have additional gauge DOFs for scalar metric perturbations.
Since $\alpha$ and $\chi$ can be eliminated by the Hamiltonian and momentum constraints, 
there are two dynamical perturbed fields $\zeta$ and $E$.
As we will see later, the perfect-fluid 
matter also gives rise to the propagation 
of one additional scalar mode that corresponds to the density perturbations or, equivalently, the longitudinal phonons of the fluid.

In the following, we will derive 
the perturbation equations of motion 
for tensor and scalar perturbations, in turn. 
We will work both in real and Fourier space from now on and we will not distinguish between the real space and Fourier space variables. There will be no risk of confusion since the context will make it clear. Finally, from now on, spatial indices will be raised and lowered with the Euclidean metric $\delta^{ij}$, 
so for instance, $(\partial_i\zeta)^2=\delta^{ij}\partial_i\zeta
\partial_j\zeta$,
$\partial_i \alpha \partial_i \zeta
=\delta^{ij} \partial_i \alpha \partial_j \zeta$, 
and $\nabla^2\zeta=\delta^{ij}\partial_i\partial_j\zeta$.

\subsection{Tensor perturbations}

For intrinsic tensor perturbations, 
we set $N=\bar{N}(t)$, $N_i=0$, 
and $h_{ij}=a^2(t) (\delta_{ij}+\tilde{h}_{ij})$, where 
$\tilde{h}_{ij}$ is the perturbed part.
To satisfy the transverse and traceless conditions 
for $\tilde{h}_{ij}$, 
we consider the propagation of gravitational 
waves along the $z$ direction and choose 
\be
\tilde{h}_{11}=-\tilde{h}_{22}
=h_+(t,z)\,,
\qquad 
\tilde{h}_{12}=\tilde{h}_{21}
=h_{\times}(t,z)\,,
\ee
where $h_+$ and $h_{\times}$ are functions of $t$ and $z$. In Fourier space, up to second order in perturbations, we have 
\be
\delta K_{\mu \nu} \delta K^{\mu \nu}
=\sum_{\lambda=+,\times} \frac{1}{2}\dot{h}_{\lambda}^2\,,
\qquad 
\delta K^2=0\,,
\qquad 
{}^{(3)}R=
-\sum_{\lambda=+,\times} 
\frac{k^{2}}{2a^{2}} h_{\lambda}^2\,.
\ee
Expanding the action (\ref{Sfull}) up to 
quadratic order in tensor perturbations, 
the second-order action in Fourier space yields
\be
{\cal S}_{\rm T}^{(2)} =  
\sum_{\lambda = +, \times} 
\int \frac{\rd t \rd^3 k}{(2\pi)^3}
\bar{N}a^{3}\frac{M^2}{4} \left( \vert \dot{h}_{\lambda}\vert^2
-\frac{k^{2}}{a^{2}} \vert h_{\lambda} \vert^2
\right) \,. 
\label{ST}
\ee
This shows that the no-ghost condition 
and the squared propagation speed for 
tensor modes are given, respectively, by 
\begin{tcolorbox}[colback=black!5!white,colframe=black!75!white,title=Tensor mode stability]
\be
M^2 >0 \,,\qquad 
c_{\rm T}^{2} = 1 \,, \label{eq:NGNLtensor_mode}
\ee
\end{tcolorbox}
\noindent
whose results are expected as we have 
focused on theories with the luminal 
speed of gravitational waves. 
In the presence of additional EFT 
operators, $c_{\rm T}^2$ can deviate 
from 1 in general.
Varying the action (\ref{ST}) 
with respect to $h_{\lambda}$, we obtain 
\be
\ddot{h}_{\lambda}+H\left( 3+\alpha_M 
\right) \dot{h}_{\lambda}
+\frac{k^2}{a^2}h_{\lambda}=0\,,
\label{teneq}
\ee
where 
\be
\alpha_M:=\frac{2\dot{M}}{HM}\,.
\ee
The time variation of the effective Planck 
mass squared $M^2(t)=M_*^2 
f(t)$ affects the propagation of gravitational waves through the friction term $H \alpha_{\rm M} \dot{h}_{\lambda}$ in Eq.~(\ref{teneq}). 
This leads to a difference between the gravitational wave and electromagnetic luminosity distances. Assuming that $\alpha_M$ is constant and using the GWTC-3 catalog from the LIGO-Virgo-KAGRA Collaboration \cite{KAGRA:2021vkt}, Ref.~\cite{Chen:2024pln} 
obtained the bound $\alpha_M=0.5^{+3.5}_{-2.6}$ 
and hence the vanishing $\alpha_M$ is consistent 
with the data
(see also Ref.~\cite{Arai:2017hxj} for an earlier work). 
On the other hand, $M^2$ also determines the effective gravitational coupling of gravitational waves to matter, so it affects the emission of gravitational waves of e.g., binary systems. 
We can then use the constraint obtained in Ref.~\cite{BeltranJimenez:2015sgd} from the period variation 
of the Hulse-Taylor pulsar. In theories with 
$c_{\rm T}=1$ as the EFT of coupled DE and DM constructed here, the bound translates to 
$0.995 \lesssim M_{\text{Pl}}^2/M^2 \lesssim 1$. 

\subsection{Scalar perturbations}

The scalar metric perturbations $\alpha$, 
$\chi$, $\zeta$, and $E$ appear in 
$N$, $N_{i}$, and $h_{ij}$, as
\be
N = \bar{N}(t)(1 + \alpha)\,,\qquad 
N_{i} =\bar{N}(t) \partial_{i} \chi \,,\qquad
h_{ij} = a^{2}(t) \left[(1 + 2 \zeta) \delta_{ij} 
+ 2 \partial_{i} \partial_{j} E \right] \,,
\ee   
so that $\delta N=\bar{N}(t)\alpha$.
In real space, the intrinsic curvature 
is given by 
\be
\Rs = -4\frac{\nabla^2 \zeta}{a^2} 
+ \frac{1}{a^2}\left[ 6(\partial_{i}\zeta)^{2} 
+16 \zeta \nabla^2 \zeta
+ 8(\nabla^2\zeta) 
(\nabla^2 E)
+ 4\delta^{ij}\partial_{i} \zeta \partial_{j}
(\nabla^2 E) \right]
+{\cal O}(\epsilon^3)\,,
\label{Rs}
\ee
where $\epsilon^n$ represents the $n$-th 
order in perturbations. 
In Eq.~(\ref{Rs}), the first term represents 
the first-order contribution and terms 
inside the square bracket correspond to the second-order contribution.

We recall that, under the gauge choice 
$\phi^i=x^i$, the CDM number density 
is given by $n=\sqrt{{\rm det}\,g^{ij}}$.
The perturbed number density $\delta n$ 
relative to the background value $\bar{n}$ 
is expressed as
\be
\frac{\delta n}{\bar{n}}=
-3\zeta-\nabla^2 E+\frac{1}{2} 
\left[ 15\zeta^2+3(\nabla^2 E)^2
+10 (\nabla^2 E) \zeta-\frac{(\partial_i \chi)^2}
{a^2} \right]+{\cal O} (\epsilon^3) \,,
\ee
so that $\delta n=-\bar{n}(3\zeta+\nabla^2 E)$ 
at linear order. We also find
\be
\delta_2 \left( \frac{\sqrt{h}}{a^{3}} 
\frac{n}{\bar{n}} \right)
= -\frac{(\partial_i \chi)^2}{2a^2}\,.
\ee
From the definition of the extrinsic curvature 
in Eq.~(\ref{Kij}), we have
\ba
\delta K
&=& 3 \left( \dot{\zeta}- H \alpha 
\right)+\nabla^2 \dot{E}
-\frac{\nabla^2 \chi}{a^2}
+{\cal O}(\epsilon^2)\,,\\
\delta K_{\mu \nu} \delta K^{\mu \nu} 
&=& 2 \left(\dot{\zeta}-H \alpha \right)^{2} 
+\left( \dot{\zeta}-H \alpha 
+\nabla^2 \dot{E}
-\frac{\nabla^2 \chi}{a^2}
\right)^2+{\cal O}(\epsilon^3) \,.
\ea
Substituting the above relations into 
Eq.~(\ref{eq:EQ_EFT_alpha_action}) and 
performing the integration 
by parts, the resulting second-order action 
of scalar perturbations 
in real space yields
\ba
{\cal S}_{\rm S}^{(2)} &=& \int 
\rd^4x \bar{N} a^{3} 
\frac{M^{2}}{2} \biggl\{ 2 \left( \dot{\zeta} 
-H \alpha \right)^2+\left( \dot{\zeta} 
-H \alpha+\nabla^2 \dot{E}-\frac{1}{a^2}\nabla^2\chi 
\right)^2+\frac{2}{a^2}(\partial_i\zeta)^2
+\frac{4}{a^2} \partial_i \alpha 
\partial_i\zeta
\nn
& &~ 
+\left[ 4\tilde{\alpha}_B H \alpha 
-\alpha_{\mu_3} H \big(3\zeta+\nabla^2E\big)
-\alpha_{\mu_4} \frac{1}{a^2}\nabla^2 \chi \right] 
\left(3\dot{\zeta} 
-3H \alpha+\nabla^2 \dot{E}-\frac{1}{a^2}\nabla^2\chi
\right)
+\tilde{\alpha}_K H^2 \alpha^2 
+\frac{\tilde{\alpha}_{\bar{m}_1}}{a^4}
(\nabla^2\chi)^2
\nonumber \\
& &~
+\left(\alpha_{m_2}+3\Omega_{\rm c} \right)\frac{1}{a^2}H^2(\partial_i\chi)^2 
+\alpha_{\mu_1} H^2 
\left( 3\zeta+\nabla^2 E \right)^2 
+\left( \alpha_{\mu_2}-1 \right)
\left(3\dot{\zeta} 
-3H \alpha+\nabla^2 \dot{E}-\frac{1}{a^2}\nabla^2\chi
\right)^2
\nonumber \\
& &
~+\alpha_{\mu_5} \frac{1}{a^2}H \partial_i\alpha\partial_i \chi
-H \big(3\zeta+\nabla^2E\big) \left[
(\tilde{\alpha}_{m_1}-6\Omega_{\rm c})H\alpha 
-\alpha_{\mu_6} \frac{1}{a^2} \nabla^2\chi 
\right] \biggr\} 
+\tilde{\cal S}_{\rm m}^{(2)}\,.
\label{Ssfinal0}
\ea

To compute the second-order contribution to 
the Schutz-Sorkin action (\ref{Smatter}), 
we decompose the temporal and spatial components of $J^{\mu}$ into the background and perturbed parts, as 
\be
J^0={\cal N}_0+\delta J\,,\qquad
J^i=\frac{\delta^{ik} \partial_k \delta j}{a^2(t)}\,,
\ee
where ${\cal N}_0=\bar{n}_{\rm m} a^3$ is the conserved background fluid number, 
and $\delta J$, $\delta j$ are scalar perturbations. Varying (\ref{Smatter}) with respect to 
$J^{\mu}$, we find that the 
four-velocity $(u_{\rm m})^{\mu}
=J^{\mu}/(n_{\rm m}\sqrt{-g})$ is related 
to the Lagrange multiplier $\ell$, 
as $(u_{\rm m})_{\mu}
=\partial_{\mu} \ell/\rho_{{\rm m}, n_{\rm m}}$.
We write the spatial component of 
$(u_{\rm m})_{\mu}$, as 
$(u_{\rm m})_i=-\partial_i v_{\rm m}$, 
where $v_{\rm m}$ is the velocity potential.
Then, we can express $\ell$ in the form 
\be
\ell=-\int^t \bar{\rho}_{{\rm m}, n_{\rm m}}(\tilde{t})
\bar{N}(\tilde{t}) \rd \tilde{t} 
-\bar{\rho}_{{\rm m}, n_{\rm m}} v_{\rm m}\,.
\label{ell}
\ee
The matter density perturbation is defined by 
\be
\delta \rho_{\rm m}:=\frac{\bar{\rho}_{{\rm m}, 
n_{\rm m}}}{a^3} \left[ \delta J -{\cal N}_0  
\left( 3\zeta+\nabla^2 E \right) \right]\,.
\label{delmrho}
\ee
From the above definition, the perturbation 
of $n_{\rm m}=\sqrt{g_{\mu \nu}J^{\mu}J^{\nu}/g}$ 
can be expressed as
\be
\delta n_{\rm m}=\frac{\delta \rho_m}
{\bar{\rho}_{{\rm m},n_{\rm m}}}
-\frac{({\cal N}_0 
\partial_i \chi+\partial_i \delta j/\bar{N})^2}
{2{\cal N}_0 a^5}-\frac{(3\zeta+\nabla^2 E)\delta \rho_m}
{\bar{\rho}_{{\rm m},n_{\rm m}}}
-\frac{{\cal N}_0 (\zeta+\nabla^2 E)
(3 \zeta-\nabla^2 E)}{2a^3}+{\cal O}(\epsilon^3)\,.
\label{dnm}
\ee
We use Eq.~(\ref{delmrho}) to 
express $\delta J$ in terms of 
$\delta \rho_{\rm m}$, $\zeta$, $E$ 
and then expand the matter action (\ref{Smatter}) 
up to quadratic order on account of 
Eqs.~(\ref{ell}) and (\ref{dnm}).
Varying the resulting second-order action 
with respect to $\delta j$, it follows that 
\be
\partial_{i} \delta j=
-\left( \partial_{i}\chi
+\partial_{i} v_{\rm m} \right) 
{\cal N}_0 \bar{N}\,.
\ee
After the elimination of the 
$\partial_{i} \delta j$ term, 
the second-order matter action in real space 
takes the following 
form~\cite{Kase:2018aps} 
\ba
{\cal S}_{\rm m}^{(2)}
&=&
\int \rd^4 x\,\bar{N} a^3 
\Bigg\{ \left(\dot{v}_{\rm m}
-3H c_{\rm m}^2 v_{\rm m}
-\alpha \right) \delta \rho_{\rm m}
-\frac{c_{\rm m}^2}{2(\bar{\rho}_{\rm m}
+\bar{p}_{\rm m})} \delta \rho_{\rm m}^2
+\frac{\bar{p}_{\rm m}}{2}(\zeta+\nabla^2 E)
(3\zeta-\nabla^2 E)
\notag\\
\hspace{-0.3cm}
&&
-\frac{\bar{\rho}_{\rm m}+\bar{p}_{\rm m}}{2a^2} 
\Big[ (\partial_i v_{\rm m})^2
+2\partial_i v_{\rm m} \partial_i \chi 
\Big]
+ (3\zeta+\nabla^2 E) \Big[ 
(\bar{\rho}_{\rm m}+\bar{p}_{\rm m})
(\dot{v}_{\rm m}-3H c_{\rm m}^2 v_{\rm m}) 
-\bar{\rho}_{\rm m} \alpha
\Big]
 \Bigg\}\,,
\label{SMS2}
\ea
where $c_{\rm m}^2$ is the squared 
matter sound speed defined by 
\be
c_{\rm m}^2:=\frac{\bar{n}_{\rm m}
\bar{\rho}_{{\rm m},n_{\rm m}n_{\rm m}}}
{\bar{\rho}_{{\rm m},n_{\rm m}}}\,.
\ee
The other second-order matter contribution (\ref{SmNL2}) is given by 
\be
{\cal S}_{\rm m}^{{\rm NL}(2)}
=\int \rd^4 x \,\bar{N} a^3 
\left[ \bar{\rho}_{\rm m} 
\left(3\zeta+\nabla^2 E 
\right) \alpha-\frac{\bar{p}_{\rm m}}{2}
\left(\zeta+\nabla^2 E \right)
\left(3\zeta-\nabla^2 E \right) 
\right]\,.
\ee
Then, the second-order matter 
action $\tilde{{\cal S}}_{\rm m}^{(2)}
={\cal S}_{\rm m}^{(2)}
+{\cal S}_{\rm m}^{{\rm NL}(2)}$ 
reduces to
\begin{align}
\hspace{-0.5cm}
\tilde{{\cal S}}_{\rm m}^{(2)}
=&
\int \,\rd^4x 
\bar{N} a^{3} \biggl\{ \left( 
\dot{v}_{\rm m}-3H c_{\rm m}^2 v_{\rm m}
-\alpha \right) \delta \rho_{\rm m}
-\frac{c_{\rm m}^2}
{2(\bar{\rho}_{\rm m}
+\bar{p}_{\rm m})}\delta \rho_{\rm m}^2
-\frac{\bar{\rho}_{\rm m}
+\bar{p}_{\rm m}}{2a^2}  
\Big[ (\partial_i v_{\rm m})^2 
+2\partial_i v_{\rm m} \partial_i \chi
\Big] 
\nonumber \\
\hspace{-0.5cm}
&
+\left( \bar{\rho}_{\rm m}
+\bar{p}_{\rm m}\right)
\left(3\zeta+\nabla^2 E 
\right) \left( 
\dot{v}_{\rm m}-3H c_{\rm m}^2 v_{\rm m}\right) 
\biggr\}\,.
\label{Smfinal}
\end{align}

We define the CDM density, as  
$\rho_{\rm c}=m_{\rm c}(t)n$, where 
$m_{\rm c}(t)=\hat{m}_{\rm c}
+\Delta m_{\rm c}(t)$. 
Since the CDM density perturbation 
is given by 
$\delta \rho_{\rm c}=m_{\rm c}(t) 
\delta n$ in the unitary gauge, 
we introduce the corresponding 
CDM density contrast, as\footnote{An alternative 
definitions of the CDM density and its contrast 
are $\hat{\rho}_{\rm c}=\hat{m}_{\rm c}n$ 
and $\hat{\delta}_{\rm c}=
\hat{\delta \rho}_{\rm c}/
\bar{\hat{\rho}}_{\rm c}$ 
with $\hat{\delta \rho}_{\rm c}
=\hat{m}_{\rm c} \delta n$, 
giving the same relation 
$\hat{\delta}_{\rm c}=\delta n/\bar{n}$ 
as Eq.~(\ref{delc}).} 
\be
\delta_{\rm c}:=
\frac{\delta \rho_{\rm c}}{\bar{\rho}_{\rm c}}
=\frac{\delta n}{\bar{n}}=-\big(3\zeta+\nabla^2 E\big)\,.
\label{delc}
\ee
We also introduce the matter 
density contrast
\be
\delta_{\rm m}:=\frac{\delta 
\rho_{\rm m}}{\bar{\rho}_{\rm m}}\,.
\ee
Substituting 
$\nabla^2E=-(\delta_{\rm c}+3\zeta)$ and 
$\delta \rho_{\rm m}=\bar{\rho}_{\rm m}
\delta_{\rm m}$ into 
Eqs.~(\ref{Ssfinal0}) and 
(\ref{Smfinal}), the total second-order 
action of scalar perturbations 
is expressed as
\begin{tcolorbox}[colback=black!5!white,colframe=black!75!white,title=Second-order scalar action with matter in metric variables]
\ba
{\cal S}_{\rm S}^{(2)} &=& \int 
\rd^4x \bar{N} a^{3} 
\frac{M^{2}}{2} \Biggl\{ 2 \left( \dot{\zeta} 
-H \alpha \right)^2+\left( 2\dot{\zeta}
+\dot{\delta}_{\rm c} 
+H \alpha+\frac{1}{a^2}\nabla^2\chi 
\right)^2+\frac{2}{a^2}(\partial_i\zeta)^2
+\tilde{\alpha}_K H^2 \alpha^2
\nn
& &~ 
-\left( 4\tilde{\alpha}_B H \alpha 
+\alpha_{\mu_3} H \deltac
-\alpha_{\mu_4} \frac{1}{a^2}\nabla^2 \chi \right) 
\left(\dot{\delta}_{\text{c}} 
+3H \alpha+\frac{1}{a^2}\nabla^2\chi
\right)+\left(\alpha_{m_2}+3\Omega_{\rm c} \right)\frac{H^2}{a^2}(\partial_i\chi)^2 
\nonumber \\
& &~ 
+\frac{\tilde{\alpha}_{\bar{m}_1}}{a^4}
(\nabla^2\chi)^2
+\alpha_{\mu_1} H^2 
\deltac^2 +\alpha_{\mu_5} \frac{H}{a^2} 
\partial_i \alpha \partial_i \chi 
+\left( \alpha_{\mu_2}-1 \right)
\left(\dot{\delta}_{\text{c}} 
+3H \alpha+\frac{1}{a^2}\nabla^2\chi
\right)^2 
\nonumber \\
& &
+H \deltac \left[
(\tilde{\alpha}_{m_1}-6\Omega_{\rm c})H\alpha 
- \frac{\alpha_{\mu_6}}{a^2} \nabla^2\chi 
\right]+\frac{4}{a^2} \partial_i\alpha \partial_i\zeta 
+\frac{2}{M^2}
\Big[
\big(\dot{v}_{\rm m}-3H c_{\rm m}^2 v_{\rm m}
-\alpha \big) \rho_{\rm m} \delta_{\rm m}  
\nonumber \\
\hspace{-0.5cm}
& &
-\frac{c_{\rm m}^2}
{2(\bar{\rho}_{\rm m}
+\bar{p}_{\rm m})}\rho_{\rm m}^2
\delta_{\rm m}^2
-\frac{\bar{\rho}_{\rm m}
+\bar{p}_{\rm m}}{2a^2} 
\Big( (\partial_i v_{\rm m})^2 
+2\partial_i v_{\rm m} \partial_i \chi
\Big) 
-\left( \bar{\rho}_{\rm m}
+\bar{p}_{\rm m}\right)
\deltac \left( 
\dot{v}_{\rm m}-3H c_{\rm m}^2 v_{\rm m}\right) 
\Big] \Biggr\}.
\label{Ssfinal}
\ea
\end{tcolorbox}
Varying (\ref{Ssfinal}) with respect to 
the non-dynamical perturbations 
$\alpha$, $\chi$, and $v_{\rm m}$ and moving to Fourier space, we obtain the following constraint equations
\ba
& & \frac{4k^2}{a^2}\zeta
-2 (6+12\tilde{\alpha}_B -\tilde{\alpha}_K
-9 \alpha_{\mu_2}) H ^2 \alpha
+\frac{k^2}{a^2} (4+4\tilde{\alpha}_{B}
-6\alpha_{\mu_2}-3\alpha_{\mu_4}
+\alpha_{\mu_5}) H \chi
\nonumber \\
& &
-2 (2+2\tilde{\alpha}_B-3 \alpha_{\mu_2})H 
\dot{\delta}_{\rm c}  
-(6\Omega_{\rm c}-\tilde{\alpha}_{m_1}
+3 \alpha_{\mu_3})H^2\delta_{\rm c}
=\frac{2}{M^2}\bar{\rho}_m \delta_{\rm m}\,, 
\label{pereq1} \\
& &
H \left( 4+4\tilde{\alpha}_B-6 
\alpha_{\mu_2}-3\alpha_{\mu_4}+\alpha_{\mu_5} \right) \alpha
+2\left[ H^2 (3\Omega_{\rm c}
+\alpha_{m_2})
+\frac{k^2}{a^2} (\alpha_{\mu_2} 
+\alpha_{\mu_4}+\tilde{\alpha}_{\bar{m}_1})
\right] \chi\nonumber \\
& &
-\left( 2\alpha_{\mu_2}+\alpha_{\mu_4} \right) \dot{\delta}_{\rm c}
+\left( \alpha_{\mu_3}
+\alpha_{\mu_6} \right)
H \delta_{\rm c}
-4\dot{\zeta}
=\frac{2}{M^2}(1+w_{\rm m})
\bar{\rho}_{\rm m}
v_{\rm m}\,,
\label{pereq2} \\
& &
\dot{\delta}_{\rm m}
+3 (c_{\rm m}^2-w_{\rm m}) H 
\delta_{\rm m}
-(1+w_{\rm m}) \left[ \dot{\delta}_{\rm c}
-\frac{k^2}{a^2} (v_{\rm m}+\chi) 
\right]=0\,,
\label{pereq3}
\ea
where we have introduced the matter equation of state parameter
\be
w_{\rm m}:=\frac{\bar{p}_{\rm m}}
{\bar{\rho}_{\rm m}}\,.
\ee
On the other hand, the variation of (\ref{Ssfinal}) with respect to the dynamical perturbations 
$\delta_{\rm c}$, $\zeta$, and 
$\delta_{\rm m}$ leads to the corresponding 
evolution equations, which read as follows 
in Fourier space: 
\ba
& &
2\alpha_{\mu_2} \ddot{\delta}_{\rm c}
+2[ \dot{\alpha}_{\mu_2}
+H(3+\alpha_M) \alpha_{\mu_2}]
\dot{\delta}_{\rm c}
+4\ddot{\zeta}+4( 3+\alpha_M) 
H \dot{\zeta}+2(3\alpha_{\mu_2} 
-2\tilde{\alpha}_B-2) H\dot{\alpha}
-\frac{k^2}{a^2} (2\alpha_{\mu_2}
+\alpha_{\mu_4}) \dot{\chi}
\nonumber \\
& & +\{ H^2 [ 6\Omega_{\rm c}-12+2\alpha_M 
(3 \alpha_{\mu_2}-2\tilde{\alpha}_B-2)
-\tilde{\alpha}_{m_1}+18 \alpha_{\mu_2} 
+3 \alpha_{\mu_3}-12 \tilde{\alpha}_B ]
+6 (H \dot{\alpha}_{\mu_2}+ \dot{H} \alpha_{\mu_2})
\nonumber \\
& &
-4[\dot{H}(1+\tilde{\alpha}_B)
+H\dot{\tilde{\alpha}}_B]\} \alpha 
-\frac{k^2}{a^2} \{ H [ (1+\alpha_M)
(2\alpha_{\mu_2}+\alpha_{\mu_4}) 
+\alpha_{\mu_3}+\alpha_{\mu_6} ]
+2\dot{\alpha}_{\mu_2}
+\dot{\alpha}_{\mu_4}\}\chi
\nonumber \\
& &
-\{ H^2 [2 \alpha_{\mu_1}+(3+\alpha_M)
\alpha_{\mu_3}]+H \dot{\alpha}_{\mu_3} 
+\dot{H} \alpha_{\mu_3} \} \delta_{\rm c}
+\frac{2}{M^2}\bar{\rho}_{\rm m}(1+w_{\rm m})
(\dot{v}_{\rm m}-3H c_{\rm m}^2 v_{\rm m})=0\,,
\label{pereq4}\\
& &
k^2 \left[ \alpha+\zeta+\dot{\chi}
+H (1+\alpha_M) \chi \right]-a^2
\left[ 3\ddot{\zeta}+\ddot{\delta}_{\rm c}
+H(3+\alpha_M)( 3\dot{\zeta}
+\dot{\delta}_{\rm c})\right]
=0\,,\label{pereq5}\\
& & 
\dot{v}_{\rm m}-3H c_{\rm m}^2 v_{\rm m}
-\alpha-\frac{c_{\rm m}^2}{1+w_{\rm m}}
\delta_{\rm m}=0\,.
\label{pereq6}
\ea
The dynamics of linear scalar perturbations 
are fully determined by integrating
Eqs.~(\ref{pereq1})-(\ref{pereq3}) and 
(\ref{pereq4})-(\ref{pereq6}) for given 
initial conditions. 
Solving Eqs.~(\ref{pereq1})-(\ref{pereq3}) 
for the non-dynamical fields
$\alpha$, $\chi$, $v_{\rm m}$ and substituting 
them into Eqs.~(\ref{pereq4})-(\ref{pereq6}), 
we obtain coupled second-order differential 
equations for the dynamical perturbations 
$\delta_{\rm c}$, $\zeta$, and $\delta_{\rm m}$. 
An alternative procedure is to remove the three 
non-dynamical perturbations from the second-order action (\ref{Ssfinal}) by using their equations of motion. 
Indeed, the latter is convenient for deriving the linear stability conditions of three dynamical perturbations. 
We will address this issue in Sec.~\ref{stasec}.

\section{Stability conditions for scalar 
perturbations and the CDM gravitational coupling}
\label{stasec}

To confront the EFT of coupled DE and DM with 
the observation of galaxy clusterings and weak 
lensing, we are mostly interested in the behavior of perturbations for the modes deep inside the DE sound horizon. In this section, we derive the linear stability conditions of scalar perturbations by taking the small-scale limit. Then, we compute the effective CDM gravitational coupling $G_{\rm eff}$ under a so-called 
quasi-static approximation. Then, we will apply the result of $G_{\rm eff}$ 
to the case in which the three EFT parameters 
$\alpha_{m_1}$, $\alpha_{m_2}$, and 
$\alpha_{\bar{m}_1}$ are vanishing. 
Let us caution that the DE sound horizon can be 
well inside the Hubble horizon, so the scales relevant for observational probes could 
lie inside the Hubble horizon but outside the DE 
sound horizon. Then, one needs to be more 
careful if this situation occurs.

\subsection{Linear stability conditions}

Taking the limit $k \to \infty$ in Fourier space,  
the EFT parameters $\tilde{\alpha}_{\bar{m}_1}$, 
${\alpha}_{\mu_i}$ ($i=1,2,\cdots,6$), 
and ${\cal G}$, which are defined in 
Eqs.~(\ref{aldef1}) and (\ref{aldef2}), 
are vanishing. However, the next-to-leading 
order terms in these EFT parameters 
can affect the stability conditions 
discussed below. 
Hence, we retain all the higher-order 
$k$-dependent contributions to the 
EFT parameters and finally expand 
quantities associated with the stability of scalar perturbations with respect to 
large values of $k$.

We solve Eqs.~(\ref{pereq1})-(\ref{pereq3}) for $\alpha$, $\chi$, $v_{\rm m}$, and substitute them into Eq.~(\ref{Ssfinal}) in Fourier space. 
After doing some integration by parts, the second-order 
action can be expressed in the form 
\be
{\cal S}^{(2)}=\int 
\frac{\rd t \rd^3 k}{(2\pi)^3}\,
\bar{N} a^{3} \left(
 \dot{\vec{\mathcal{X}}}_{\bm k}^{t}{\bm K}
\dot{\vec{\mathcal{X}}}_{-\bm k}
-\vec{\mathcal{X}}_{\bm k}^{t}\tilde{{\bm G}}
\vec{\mathcal{X}}_{-\bm k} 
-\frac{k}{a}\vec{\mathcal{X}}^{t}_{\bm k}{\bm B}
\dot{\vec{\mathcal{X}}}_{-\bm k}
\right) \,,
\label{S2c}
\ee
where we have defined
\be
\vec{\mathcal{X}}^{t}_{\bm k}
=\left[ \frac{\delta_{\rm c}({\bm k})}{k}, 
\zeta({\bm k}), \frac{\delta_{\rm m}({\bm k})}{k} \right] \,.
\label{Xt}
\ee
The normalization of the two dynamical fields 
$\delta_{\rm c}$ and $\delta_{\rm m}$ 
in Eq.~(\ref{Xt}) has been introduced 
to have the canonical 
$k^0$ behaviour in the diagonal 
components of the kinetic matrix ${\bm K}$. 
In the quadratic action \eqref{S2c}, the matrices $\bm{K}$, $\tilde{\bm{G}}$ are  
symmetric $3 \times 3$ matrices, and 
we have performed the integration by parts 
to define $\bm{B}$ as an antisymmetric $3 \times 3$ matrix. In the following, we will make use 
of the following decomposition:
\be
\tilde{\bm{G}}=\frac{k^2}{a^2}\bm{G}+\bm{M}\,,
\ee
defined perturbatively in the large $k$ limit so that ${\bm G}$ corresponds to the terms in the expansion of $\tilde{{\bm G}}$ with strictly positive powers of $k$ (i.e., the local part), while $\bm{M}$ will include the non-positive powers of $k$, i.e., the leading order $k^0$ and the non-local inverse powers of $k$.

Taking the limit $k \to \infty$, the 
non-vanishing components of ${\bm K}$ 
are given by 
\be
K_{11}=\frac{a^2 M^2 H^2}{2} \left( 
3\Omega_{\rm c}+\alpha_{m_2}
-\alpha_{\bar{m}_1}^2 
\right),\qquad 
K_{22}=\frac{M^2(6\alpha_B^2
+\tilde{\alpha}_K)}{2(1+\alpha_B)^2},
\qquad
K_{33}=\frac{a^2 \bar{\rho}_{\rm m}}
{2(1+w_{\rm m})}.
\label{Kcom}
\ee
The off-diagonal components have the 
following scale dependence:
\be
K_{12}=K_{21}={\cal O}(k^{-1})\,,\qquad
K_{23}=K_{32}={\cal O}(k^{-1})\,,\qquad
K_{13}=K_{31}={\cal O}(k^{-2})\,,
\ee
which are suppressed in comparison to 
the diagonal components. 
In the small-scale limit, the scalar ghosts 
are absent 
under the following two conditions:
\begin{tcolorbox}[colback=black!5!white,colframe=black!75!white,title=No-ghost conditions for DM and DE]
\ba
q_{\rm c}: &=& 
\frac{K_{11}}{a^2}=
\frac{M^2 H^2}{2} \left( 
3\Omega_{\rm c}+\alpha_{m_2}
-\alpha_{\bar{m}_1}^2 
\right)>0\,,\label{snoghost1}\\
q_{\rm s}: &=& K_{22}=
\frac{M^2(6\alpha_B^2
+\tilde{\alpha}_K)}{2(1+\alpha_B)^2}>0\,,
\label{snoghost2}
\ea
\end{tcolorbox}
\noindent
supplemented with the condition $\bar{\rho}_{\rm m}(1+w_{\rm m})>0$ that guarantees the absence of ghosts in the matter sector. Using the tensor no-ghost 
condition $M^2>0$, the inequalities 
(\ref{snoghost1}) and (\ref{snoghost2}) 
reduce, respectively, to 
$3\Omega_{\rm c}+\alpha_{m_2}
-\alpha_{\bar{m}_1}^2>0$ and 
$6\alpha_B^2+\tilde{\alpha}_K>0$. 
The no-ghost conditions of CDM, DE, 
and the matter fluid correspond to 
$q_{\rm c}>0$, $q_{\rm s}>0$, and 
$\bar{\rho}_{\rm m} (1+w_{\rm m})>0$, 
respectively. In the small-scale limit, 
the quantity $\tilde{\alpha}_K$ has 
the following behavior 
\be
\tilde{\alpha}_K \to \alpha_K
+6\Omega_{\tilde{c}}\,,
\ee
where the term $6\Omega_{\tilde{c}}$ does 
not generally vanish in scalar-tensor theories.
In vector-tensor theories, we have 
$\tilde{\alpha}_K \to \alpha_K$ due 
to the condition (\ref{cprimecon}). 
The momentum exchange between DE and CDM 
affects the CDM no-ghost condition through 
the EFT parameters $\alpha_{m_2}$ and 
$\alpha_{\bar{m}_1}$. 
The EFT parameter $\alpha_{m_1}$ associated 
with the energy exchange does not explicitly appear either in $q_{\rm c}$ or $q_{\rm s}$.
We note that the positivity of the kinetic matrix ${\bm K}$ is required to maintain
the hyperbolic character of the perturbation equations. Indeed, the sign flip of the 
determinant of ${\bm K}$ during the cosmological 
evolution signals a strong coupling problem 
at the point of the vanishing determinant.
Let us also notice that the obtained ghost-free conditions have been derived by taking the 
small-scale limit, so they do not necessarily guarantee the stability of theories at all scales.

In the small-scale limit, the diagonal 
components of ${\bm G}$ are given by 
\ba
& &
G_{11}=0,\label{G11} \\
& &
G_{22}=-\frac{M^2}{2(1+\alpha_B)^2}
\biggl[ 3\Omega_{\rm c}+3\Omega_{\rm m} (1+w_{\rm m})
+2(1+\alpha_B) 
\left( \alpha_B-\alpha_M-\epsilon_H \right)+\frac{2\dot{\alpha}_{B}}{H} 
\nonumber \\
& &\qquad \quad
-4\alpha_B^2 \alpha_g^2
+\alpha_{m_2}
-\alpha_{\bar{m}_1}^2
-4\alpha_g \alpha_B \alpha_{\bar{m}_1}
\biggr],\\
& &
G_{33}=\frac{a^2 \bar{\rho}_{\rm m}}
{2(1+w_{\rm m})}c_{\rm m}^2\,,
\ea
where
\be
\Omega_{\rm m}:=\frac{\bar{\rho}_{\rm m}}
{3M^2 H^2}\,,\qquad
\epsilon_H:=-\frac{\dot{H}}{H^2}\,.
\ee
The off-diagonal components of ${\bm G}$ 
have the following scale dependences 
\be
G_{12}=G_{21}={\cal O}(k^{-1})\,,\qquad
G_{23}=G_{32}={\cal O}(k^{-1})\,,\qquad
G_{13}=G_{31}={\cal O}(k^{-2})\,.
\ee
In the limit $k \to \infty$, 
the anti-symmetric matrix ${\bm B}$ 
possesses the following non-vanishing components 
\be
B_{12}=-B_{21}=\frac{aM^2H}{4(1+\alpha_B)}
\left( \alpha_{m_1}+2\alpha_{m_2}
-2\alpha_{\bar{m}_1}^2
-4\alpha_g \alpha_B \alpha_{\bar{m}_1}
\right)\,,
\label{B12}
\ee
while the other components of ${\bm B}$ 
have the $k$ dependences 
\be
B_{13}=-B_{31}={\cal O}(k^{-1})\,,
\qquad
B_{23}=-B_{32}={\cal O}(k^{-2})\,.
\ee
The squared propagation speed of the 
matter fluid is given by 
\be
c_{\rm m}^2=\frac{G_{33}}{K_{33}}\,,
\ee
which is not affected by the matrix 
${\bm B}$. So long as 
\be
c_{\rm m}^2>0\,,
\ee
the Laplacian instability is absent for 
the matter density contrast $\delta_{\rm m}$.
The presence of non-vanishing components 
(\ref{B12}) in ${\bm B}$ affects 
the propagation of the DM and DE perturbations. 
To obtain the latter two propagation speeds, we substitute the solutions ${\cal X}_j=\tilde{{\cal X}}_j e^{i (\omega t-k x)}$ into their 
equations of motion, where 
${\cal X}_1=\delta_{\rm c}/k$, 
${\cal X}_2=\zeta$, and
$\omega$ is an angular frequency. 
Note that $\tilde{{\cal X}}_j$ 
(with $j=1,2$) are dealt as constants 
under the WKB approximation.
Picking up the terms of 
orders $\omega^2$, $\omega k$, 
and $k^2$, we find
\ba
& & \omega^2 \tilde{{\cal X}}_1-\hat{c}_{\rm c}^2\frac{k^2}{a^2}
\tilde{{\cal X}}_1-i \omega \frac{k}{a} \frac{B_{12}}{K_{11}} 
\tilde{{\cal X}}_2 \simeq 0\,,
\label{disp1}\\
& & \omega^2 \tilde{{\cal X}}_2-\hat{c}_{\rm s}^2\frac{k^2}{a^2}
\tilde{{\cal X}}_2
+i \omega \frac{k}{a} 
\frac{B_{12}}{K_{22}} 
\tilde{{\cal X}}_1  \simeq 0\,,\label{disp2}
\ea
where 
\be
\hat{c}_{\rm c}^2=\frac{G_{11}}{K_{11}}\,,
\qquad
\hat{c}_{\rm s}^2=\frac{G_{22}}{K_{22}}\,.
\ee
On using Eqs.~(\ref{Kcom}) and (\ref{G11}), 
we have that $\hat{c}_{\rm c}^2=0$. 
Then, from Eq.~(\ref{disp1}), 
we obtain the following two solutions
\ba
& &
\omega=0\,,\label{branch1}\\
& &
\omega \tilde{\cal X}_1=i \frac{k}{a}
\frac{B_{12}}{K_{11}}
\tilde{{\cal X}}_2 \,,
\label{branch2}
\ea
The solution (\ref{branch1}) corresponds 
to the dispersion relation for the CDM 
density contrast. 
Since the CDM squared sound speed satisfies 
$c_{\rm CDM}^2=\omega^2 a^2/k^2$, 
we obtain 
\be
c_{\rm CDM}^2=0\,.
\ee
The other solution (\ref{branch2}) 
corresponds to the dispersion relation 
$\omega^2=c_{\rm s}^2\,k^2/a^2$
for the DE perturbation $\zeta$.
Substituting Eq.~(\ref{branch2})
into Eq.~(\ref{disp2}), the DE 
squared sound speed is given by 
\begin{tcolorbox}[colback=black!5!white,colframe=black!75!white,title=Speed of propagation for DE]
\be
c_{\rm s}^2=\hat{c}_{\rm s}^2
+\Delta c_{\rm s}^2\,,
\label{cst}
\ee
\end{tcolorbox}
\noindent
with 
\ba
\hat{c}_{\rm s}^2 &=&
\frac{G_{22}}{K_{22}}=
-\frac{1}{(6\alpha_B^2+\tilde{\alpha}_K)}
\biggl[ 3\Omega_{\rm c}+3\Omega_{\rm m}(1+w_{\rm m})
+2(1+\alpha_B) \left( \alpha_B-\alpha_M-\epsilon_H \right)+\frac{2\dot{\alpha}_{B}}{H} \nonumber \\
& &\qquad \quad
-4\alpha_B^2 \alpha_g^2+\alpha_{m_2}
-\alpha_{\bar{m}_1}^2
-4\alpha_g \alpha_B \alpha_{\bar{m}_1} \biggr]\,,\label{hatcs} \\
\Delta c_{\rm s}^2 &=& \frac{B_{12}^2}
{K_{11} K_{22}}=
\frac{(\alpha_{m_1}+2\alpha_{m_2}
-2\alpha_{\bar{m}_1}^2-4\alpha_g \alpha_B \alpha_{\bar{m}_1})^2}{4(3\Omega_{\rm c}+\alpha_{m_2}
-\alpha_{\bar{m}_1}^2)(6\alpha_B^2
+\tilde{\alpha}_K)}\,.
\ea
To avoid Laplacian instabilities in 
the DE sector, we require that the corresponding squared propagation speed remains positive, i.e., 
\be
c_{\rm s}^2>0\,.
\label{Lap}
\ee
Under the two no-ghost conditions 
$K_{11}>0$ and $K_{22}>0$, we have that 
$\Delta c_{\rm s}^2>0$. Then, 
so long as the inequality 
$\hat{c}_{\rm s}^2>0$ holds, the condition 
(\ref{Lap}) is satisfied.
We observe that $\hat{c}_{\rm s}^2$ is affected by the momentum transfer (weighed by 
$\alpha_{m_2}$ and $\alpha_{\bar{m}_1}$). 
The energy transfer associated with the 
time-dependent CDM mass appears as the 
density parameter $\Omega_{\rm c}$ 
in $\hat{c}_{\rm s}^2$, while the 
other EFT parameter $\alpha_{m_1}$ 
is not present in $\hat{c}_{\rm s}^2$.
On the other hand, $\Delta c_{\rm s}^2$ 
contains all of the functions 
$\Omega_{\rm c}$, 
$\alpha_{m_1}$, $\alpha_{m_2}$, and $\alpha_{\bar{m}_1}$
In vector-tensor theories, there are two 
terms $-4\alpha_B^2 \alpha_g^2$ and 
$-4\alpha_g \alpha_B \alpha_{\bar{m}_1}$
that contribute to $c_{\rm s}^2$. 
The latter contribution arises from 
the momentum transfer associated with the EFT parameter $\alpha_{\bar{m}_1}$. 
These two terms vanish in scalar-tensor theories.

Before closing this subsection, it would be worth mentioning stability conditions for the intrinsic vector mode in vector-tensor theories. 
We leave their derivation in Appendix~\ref{vecsec} and borrow the results from \eqref{GF_vector}. By using the $\alpha$-basis parameters, the stability conditions \eqref{GF_vector} are written as 
\begin{tcolorbox}[colback=black!5!white,colframe=black!75!white,title=Vector mode stability]
\begin{align}
\gamma_1>0\,, \qquad 3\Omega_{\rm c} +\alpha_{m_2} - \alpha^2_{\bar{m}_1}>0 \,. \label{eq:Vector_stability}
\end{align}
\end{tcolorbox}
\noindent
The latter condition is the same as the ghost-free condition of the scalar mode $q_{\rm c}>0$. Hence, the stability of the vector modes does not give rise to additional conditions on the $\alpha$-basis parameters.

\subsection{Effective gravitational 
coupling for CDM}
\label{Geffsec}

Given that we have obtained the closed-form action (\ref{S2c}) for the dynamical perturbations, we can study the CDM gravitational coupling $G_{\rm eff}$ 
relevant to the growth of large-scale structures. 
For this purpose, we neglect the contribution 
of matter fluids (baryons and radiation) 
and set $\bar{\rho}_{\rm m}=0=\bar{p}_{\rm m}$ and $\delta_{\rm m}=0=v_{\rm m}$. 
Note that baryons also give contributions to 
the CDM gravitational coupling, but 
the main aim here is to see how the direct 
DE and CDM couplings affect $G_{\rm eff}$.
We will focus on the dynamics of perturbations 
on small scales in the sense that we will make 
clear below.

Varying the action (\ref{S2c}) with respect to the two dynamical fields ${\cal X}_1=
\delta_{\rm c}/k$ and ${\cal X}_2=\zeta$ and 
converting to the perturbation equations
for $\delta_{\rm c}$ and $\zeta$, it follows that 
\ba
& &
\ddot{\delta}_{\rm c}+\left( 3H
+\frac{\dot{K}_{11}}{K_{11}} \right)
\dot{\delta}_{\rm c}
+\left( \frac{G_{11}}{K_{11}}\frac{k^2}{a^2} 
+\frac{M_{11}}{K_{11}} \right) 
\delta_{\rm c}+\frac{K_{12}}{K_{11}} 
k\ddot{\zeta}-\left( \frac{k}{a} 
\frac{B_{12}}{K_{11}}-\frac{\dot{K}_{12}
+3H K_{12}}{K_{11}}
\right)k \dot{\zeta} \nonumber \\
& &+\left[ \frac{G_{12}}{K_{11}} 
\frac{k^2}{a^2}-\frac{k(\dot{B}_{12}
+2H B_{12})}{2aK_{11}}+\frac{M_{12}}
{K_{11}} \right] k \zeta=0\,,\label{perde1} \\
& &
\ddot{\zeta}+\left( 3H
+\frac{\dot{K}_{22}}{K_{22}} \right)
\dot{\zeta}
+\left( \frac{G_{22}}{K_{22}}\frac{k^2}{a^2} 
+\frac{M_{22}}{K_{22}} \right) 
\zeta+\frac{K_{12}}{K_{22}} 
\frac{\ddot{\delta}_{\rm c}}{k}
+\left( \frac{k}{a} 
\frac{B_{12}}{K_{22}}+\frac{\dot{K}_{12}
+3H K_{12}}{K_{22}}
\right) \frac{\dot{\delta}_{\rm c}}
{k} \nonumber \\
& &+\left[ \frac{G_{12}}{K_{22}} 
\frac{k^2}{a^2}+\frac{k(\dot{B}_{12}
+2H B_{12})}{2aK_{22}}+\frac{M_{12}}
{K_{22}} \right] \frac{\delta_{\rm c}}{k}=0\,.
\label{perde2}
\ea
We recall that $G_{11} \to 0$ in the 
small-scale limit, by reflecting 
the fact that the CDM sound speed vanishes.
In the same limit, the other matrix components in Eqs.~(\ref{perde1}) and (\ref{perde2}) have the scale-dependence $K_{11} \propto k^0$, 
$M_{11} \propto k^0$, 
$K_{12} \propto k^{-1}$, 
$B_{12} \propto k^0$, 
$G_{12} \propto k^{-1}$, 
$M_{12} \propto k^0$, $K_{22} \propto k^0$, 
$G_{22} \propto k^0$, 
and $M_{22} \propto k^0$.
Except for $M_{11}$, we can neglect the 
mass matrix components $M_{12}$, $M_{22}$ 
as well as the terms containing $K_{12}$. 

The detailed analysis of Eqs.~\eqref{perde1} and \eqref{perde2} governing the evolution for the density contrast and metric perturbations is in general cumbersome and it will depend on the specific form of coupling functions. 
We can however gain some insights by considering some limits. In particular, we are interested in elucidating under which circumstances the growth of structures can be suppressed with respect to the $\Lambda$CDM model on sub-horizon scales. 
Thus, we will focus on sufficiently sub-horizon modes in the sense explained below.

Since the CDM component has a vanishing propagation speed, we expect to have {\it slow} modes, i.e., 
those that do not oscillate and whose time scale is set 
by $H$. The existence of this slow mode stems from 
the fact that $G_{11}$ vanishes at 
leading order in the small-scale limit. 
This can induce a slow mode for the perturbation $\zeta$ as well. Since $G_{22}$ is not zero, 
we expect to have a {\it fast} oscillating mode 
for $\zeta$, which is a consequence of the DE sector 
having pressure. If we assume that the attractor solution for $\zeta$ is dominated by the slow mode, we can neglect temporal derivatives over $(k^2/a^2)\zeta$, since they will be suppressed by a factor $a^2H^2/k^2\ll 1$ in the sub-Hubble limit. 

A cautionary remark is that we are implicitly assuming that there are no additional hierarchies introduced by the scales present in the coupling functions of the EFT action. 
For instance, it may occur that $M_{22}$ dominates over $G_{22}k^2/a^2$ across a wide range of 
sub-Hubble modes relevant to the linear regime. In other words, the sub-horizon regime for the DE component could be substantially below the Hubble horizon. We will assume that this does not happen and that the DE horizon parametrically coincides with the Hubble horizon. Under the above approximation scheme, 
which we will refer to as the quasi-static approximation, Eq.~(\ref{perde2}) yields 
\be
\hat{c}_{\rm s}^2 \frac{k^2}{a^2} \zeta
+\frac{b_{12}}{q_{\rm s}} \dot{\delta}_{\rm c}
+\frac{2g_{12}+(\dot{b}_{12}+3H b_{12})}
{2q_{\rm s}} \delta_{\rm c} 
\simeq 0\,,
\label{perq1}
\ee
where $q_{\rm s}$ and $\hat{c}_{\rm s}^2$ 
are defined, respectively, 
in Eqs.~(\ref{snoghost2}) and (\ref{hatcs}), and 
\be
b_{12}:=\frac{B_{12}}{a}\,,\qquad 
g_{12}:=\frac{k G_{12}}{a^2}\,.
\ee
Using the quantity $q_{\rm c}$ defined 
in Eq.~(\ref{snoghost1}), 
we can approximate Eq.~(\ref{perde1}) to give 
\be
\ddot{\delta}_{\rm c}+\left( 5H 
+\frac{\dot{q}_{\rm c}}{q_{\rm c}} \right)
\dot{\delta}_{\rm c}
-\frac{\mu_{11}}{q_{\rm c}} \delta_{\rm c}
-\frac{b_{12}}{q_{\rm c}} \frac{k^2}{a^2} 
\dot{\zeta}+\frac{2g_{12}-(\dot{b}_{12}+3H b_{12})}
{2q_{\rm c}}\frac{k^2}{a^2}\zeta \simeq 0\,,
\label{perq2}
\ee
where 
\be
\mu_{11}:=-\frac{M_{11}}{a^2}\,.
\ee
We solve Eq.~(\ref{perq1}) for $\zeta$ 
and take its time derivative. Substituting them 
into Eq.~(\ref{perq2}), we obtain 
the closed-form differential equation 
for $\delta_{\rm c}$, as
\be
\ddot{\delta}_{\rm c}+{\cal C}
\dot{\delta}_{\rm c}
-4\pi G_{\rm eff} 
\bar{\rho}_{\rm c}\delta_{\rm c} 
\simeq 0\,,
\label{delcqu}
\ee
where 
\ba
{\cal C} &=&
\frac{(5H q_{\rm c}+\dot{q}_{\rm c})\nu_{\rm s}^2
+(2 b_{12} \dot{b}_{12}+
5H b_{12}^2)\nu_{\rm s}
-b_{12}^2 \dot{\nu}_{\rm s}}
{q_{\rm c}\nu_{\rm s}^2+b_{12}^2 \nu_s}\,,\label{cC} 
\ea
\begin{tcolorbox}[colback=black!5!white,colframe=black!75!white,title=Effective gravitational coupling of CDM]
\begin{align}
G_{\rm eff} &= \frac{1}
{16 \pi \bar{\rho}_{\rm c}
(q_{\rm c}\nu_{\rm s}^2
+b_{12}^2 \nu_{\rm s})}
\biggl\{ 4 \mu_{11} \nu_{\rm s}^2 
+ 4 g_{12}^2 \nu_{\rm s} - 
\Big[3H^2(7 - 2\epsilon_H) b_{12}^2 + \dot{b}_{12}^2 \Big] \nu_{\rm s} \nonumber \\
& 
- 2 (\ddot{b}_{12}+8H \dot{b}_{12}
+2\dot{g}_{12}+4H g_{12})
b_{12}\nu_{\rm s} + 2( \dot{b}_{12}+3 H b_{12} 
+ 2 g_{12})b_{12}\dot{\nu}_{\rm s}
\biggr\} \,,
\label{Geff}
\end{align}
\end{tcolorbox}
\noindent
with 
\be
\nu_{\rm s}:=q_{\rm s} 
\hat{c}_{\rm s}^2\,.
\ee
The result (\ref{Geff}) corresponds to the effective gravitational coupling of CDM 
derived under the quasi-static approximation 
for perturbations deep inside the DE 
sound horizon. In the $\Lambda$CDM model, 
$G_{\rm eff}$ is equivalent to the 
Newton gravitational constant 
$G_N=1/(8\pi \Mpl^2)$. 
The presence of DE and DM couplings as well 
as the modification of gravity from General 
Relativity lead to the deviation of $G_{\rm eff}$ 
from $G_N$. In Secs.~\ref{nointsec} 
and \ref{intsec}, we will estimate $G_{\rm eff}$ for 
several different cases. 
For the moment, let us point out that, in the regime where the last term on the left-hand side 
of Eq.~\eqref{delcqu} can be neglected, i.e., 
$4\pi G_{\rm eff} \bar{\rho}_{\rm c}\ll H^2$, 
Eq.~(\ref{delcqu}) admits a constant mode as a solution for the density contrast. If the friction term ${\cal C}$ remains positive, 
this constant mode will in turn be the dominant solution. In this situation, the evolution of the density contrast is oblivious to the details of the interactions and we can make relatively universal predictions.\footnote{For instance, this 
situation occurs in the scenarios studied in 
Refs.~\cite{Asghari:2019qld,Figueruelo:2021elm,BeltranJimenez:2020qdu}.} The question is whether such a regime is possible. In any case, we will show later that the energy and momentum transfers associated with 
the nonvanishing matrix component $B_{12}$ admit scenarios where the growth of structures can be suppressed.

We will finish this subsection with a discussion on the difference between the two gauge-invariant gravitational potentials defined by 
\be
\Psi=\alpha+\left( \chi-a^2 \dot{E} \right)^{\cdot}\,,
\qquad 
\Phi=-\zeta-H \left( \chi-a^2 \dot{E} \right)\,.
\ee
Then, we can express Eq.~(\ref{pereq5}) 
in the form 
\be
\Psi-\Phi
=\alpha_M \left( \Phi+\zeta \right)\,.
\label{PsiPhi}
\ee
When $\alpha_M=0$, i.e., 
for constant $M^2=M_*^2 f$, we have that $\Psi=\Phi$. 
The gravitational slip between $\Psi$ and 
$\Phi$ is induced by the time variation 
of the effective Planck mass 
($\alpha_M \neq 0$).
We stress that the $\alpha$-basis 
EFT parameters associated with the energy or momentum exchange do not appear 
in Eq.~(\ref{PsiPhi}).
Moreover, the relation (\ref{PsiPhi}) is 
valid without using the quasi-static 
approximation for perturbations deep 
inside the DE sound horizon. 
We could relate the non-vanishing of the gravitational slip with the presence of DE anisotropic stresses governed by $\alpha_M$.

\subsection{Theories with 
$\alpha_{m_1}=0$,  
$\alpha_{m_2}=0$, 
$\alpha_{\bar{m}_1}=0$}
\label{nointsec}

Before proceeding to the discussion of the 
effect of energy and momentum transfers 
induced by the EFT parameters 
$\alpha_{m_1}$, $\alpha_{m_2}$, and 
$\alpha_{\bar{m}_1}$, 
we will first consider theories in which these three EFT parameters are vanishing:
\be
\alpha_{m_1}=0\,,\qquad 
\alpha_{m_2}=0\,,\qquad 
\alpha_{\bar{m}_1}=0\,.
\ee
We note that there is yet another energy 
transfer induced by the time variation of 
the effective CDM mass $m_{\rm c}$. 
As we will see below, this effect manifests 
itself for both ${\cal C}$ and $G_{\rm eff}$.
Eqs.~(\ref{cC}) and 
(\ref{Geff}) reduce, respectively, to 
\be
{\cal C} = 5H+
\frac{\dot{q}_{\rm c}}{q_{\rm c}}\,,\qquad 
G_{\rm eff} = 
\frac{\mu_{11} \nu_{\rm s}
+g_{12}^2}{4\pi \bar{\rho}_{\rm c} 
q_{\rm c} \nu_{\rm s}}\,.
\label{Geffun}
\ee
In this uncoupled limit, we have 
$q_{\rm c} \to 3M^2 H^2 
\Omega_{\rm c}/2=m_{\rm c} \bar{n}/2$ 
and hence 
\be
{\cal C}=H \left( 2+\alpha_{m_{\rm c}} 
\right)\,,
\ee
where 
\be
\alpha_{m_{\rm c}}:=
\frac{\dot{m}_{\rm c}}
{Hm_{\rm c}}\,.
\ee
The time variation of $m_{\rm c}$ modifies 
the standard friction term 
$2H\dot{\delta}_{\rm c}$ in General Relativity.

The effective gravitational coupling 
in Eq.~(\ref{Geffun}) yields 
\be
G_{\rm eff}=\frac{1}{8\pi M^2} 
\left[1+\frac{2(\alpha_B-\alpha_M
+\alpha_{m_{\rm c}})^2}{V_{\rm s}} \right]\,,
\label{Geff2}
\ee
where 
\ba
V_{\rm s}: 
&=&
2(2 \alpha_g^2 - 1) \alpha_B^2 + 2(\alpha_M 
- \epsilon_B + \epsilon_H 
- 1)\alpha_B + 2\alpha_M 
- 3\Omega_{\rm c} + 2\epsilon_H 
\nonumber \\
&=& \frac{2}{M^2}(1+\alpha_B)^2 \nu_{\rm s}\,,
\ea
with $\epsilon_B:= \dot{\alpha}_B/(H \alpha_B)$.
To ensure the stability of scalar perturbations, we consider the case in which 
the conditions $q_{\rm s}>0$ and 
$\hat{c}_{\rm s}^2>0$ are satisfied. 
Since $\nu_{\rm s}>0$ in this case, 
we have $2(\alpha_B-\alpha_M
+\alpha_{m_{\rm c}})^2/V_{\rm s}>0$ 
in Eq.~(\ref{Geff2}). 
Thus, the CDM gravitational coupling is enhanced by the non-vanishing EFT parameters 
$\alpha_B$, $\alpha_M$, and 
$\alpha_{m_{\rm c}}$.
For $\alpha_{m_{\rm c}}=0$, the result 
(\ref{Geff2}) agrees with those derived 
for the subclasses of Horndeski theories~\cite{DeFelice:2011hq,Tsujikawa:2015mga,Kase:2018aps} and GP theories~\cite{DeFelice:2016uil,Amendola:2017orw,Aoki:2021wew} 
with the luminal speed of tensor perturbations. 
For $\alpha_B \neq 0$ or $\alpha_M \neq 0$, 
the gravitational interaction is enhanced by 
indirect couplings between DE and CDM 
mediated by gravity. 
This increased gravitational attraction is 
attributed to the positive term 
$4 g_{12}^2 \nu_{\rm s}$ in $G_{\rm eff}$.

In Eq.~(\ref{Geff2}), there is yet another contribution from the time-varying mass 
$m_{\rm c}(t)$ that contributes to the 
enhancement of $G_{\rm eff}$. 
The time dependence of $m_{\rm c}(t)$ can be
acquired by the energy transfer between DM and 
DE, e.g., through the interaction 
$f_1(\phi)\hat{m}_{\rm c}n$ in 
non-shift-symmetric scalar-tensor theories. 
Indeed, as shown in Refs.~\cite{Amendola:2020ldb,Kase:2019veo,Kase:2020hst} this type of energy transfer 
enhances the gravitational attraction. 
In our scheme of coupled DE and DM based on the 
unitary gauge, the parameter 
$\alpha_{m_{\rm c}}$
does not explicitly appear in the second-order 
action of scalar perturbations. 
However, this effect manifests itself in the effective CDM gravitational coupling. 
In vector-tensor theories and shift-symmetric 
scalar-tensor theories, we recall 
that the consistency condition (\ref{consistency3Mr}) holds. 
This shows that, so long as $\alpha_{m_1}=0$, we have $\alpha_{m_{\rm c}}=0$, in which case 
there is no energy transfer arising from 
the time-varying mass $m_{\rm c}$. 
In non-shift-symmetric scalar-tensor theories, 
the enhancement of $G_{\rm eff}$ can occur through the non-vanishing EFT parameter 
$\alpha_{m_{\rm c}}$ even for 
$\alpha_{m_1}=0$.

\section{Effects of DE and DM couplings on gravitational clustering}
\label{intsec}

In this section, we apply our general results of 
the approximate perturbation equation for $\delta_{\rm c}$ obtained in Sec.~\ref{Geffsec} 
to several concrete models with the energy and transfers. We also discuss the validity of the 
quasi-static approximation exploited for 
the derivation of Eq.~(\ref{delcqu}). 
In Sec.~\ref{nointsec} we discussed the case in 
which all of $\alpha_{m_1}$, $\alpha_{m_2}$, 
and $\alpha_{\bar{m}_1}$ are vanishing, but now 
we will study the cases in which at least one of the EFT parameters $\alpha_{m_1}$, $\alpha_{m_2}$, 
and $\alpha_{\bar{m}_1}$ are non-zero. 
In the latter cases, the mixing matrix components 
\be
b_{12}=-b_{21}=\frac{M^2H}{4(1+\alpha_B)}
\left( \alpha_{m_1}+2\alpha_{m_2}
-2\alpha_{\bar{m}_1}^2
-4\alpha_g \alpha_B \alpha_{\bar{m}_1} 
\right)
\label{b12D}
\ee
are non-vanishing in general.
This modifies the effective CDM gravitational 
coupling in two ways. First, there is 
an overall modification of $G_{\rm eff}$ 
induced by the $b_{12}^2 \nu_{\rm s}$ term 
in the denominator of Eq.~(\ref{Geff}).
This term, which can also be 
expressed as $q_{\rm c} q_{\rm s}\nu_{\rm s} 
\Delta c_{\rm s}^2$, is always positive 
under the linear stability conditions, so that 
$G_{\rm eff}$ is suppressed compared to 
the uncoupled case with $b_{12}=0$. 
Second, the numerator of 
Eq.~(\ref{Geff}) is affected by the 
presence of the $b_{12}$ 
term and its time derivatives. 
In particular, the contribution 
\be
-\Big[3H^2(7 - 2\epsilon_H) 
b_{12}^2 + \dot{b}_{12}^2\Big] 
\nu_{\rm s}
\label{geff1}
\ee
is always negative for 
$\epsilon_H \le 7/2$ (which is the case 
for the matter- and DE-dominated epochs). 
This repulsive interaction, which arises 
from the non-vanishing DE pressure present 
in the term  
$\nu_{\rm s}=q_{\rm s} \hat{c}_{\rm s}^2$, 
works to counteract the attractive 
gravitational force for CDM.
In the second line of Eq.~(\ref{Geff}), 
there are also other contributions to 
$G_{\rm eff}$ arising from $b_{12}$. 
Since the signs of these terms depend on the 
type of energy and momentum transfers, 
we need to specify the form of interactions 
to see whether $G_{\rm eff}$ is suppressed 
or not in comparison to the uncoupled 
case with $b_{12}=0$. 

As studied in the context of scalar-tensor 
theories~\cite{Pourtsidou:2013nha,Boehmer:2015sha,Skordis:2015yra,Koivisto:2015qua,Pourtsidou:2016ico,Dutta:2017kch,Linton:2017ged,Kase:2019veo,Kase:2019mox,Chamings:2019kcl,Kase:2020hst,Linton:2021cgd} and vector-tensor theories~\cite{DeFelice:2020icf,Pookkillath:2024ycd}, 
the momentum transfer can lead to the 
reduction of the effective 
CDM gravitational coupling. 
The energy transfer arising from the time 
variation of $m_{\rm c}$ enhances 
$G_{\rm eff}$ in general, but the 
coexistence of the momentum transfer 
can work to suppress $G_{\rm eff}$ 
at low redshift~\cite{Amendola:2020ldb,Liu:2023mwx}. 
Our general expression of 
$G_{\rm eff}$ given in Eq.~(\ref{Geff}) accommodates particular energy and momentum exchanges studied in the literature. 
Moreover, there is a new type of the energy 
transfer associated with the EFT 
parameter $\alpha_{m_1}$. 
We also note that the change in the 
friction term ${\cal C}\dot{\delta}_{\rm c}$ 
affects the evolution of $\delta_{\rm c}$, 
so we need to check its behavior besides 
$G_{\rm eff}$. The crucial point is that the 
gravitational interaction weaker than 
in the $\Lambda$CDM model can be realized 
by the direct interactions between CDM and DE 
in our EFT scheme.

To be concrete, we estimate the effective 
CDM gravitational coupling for several examples 
of the DE and DM interactions.
We will focus on the case in which the condition 
\be
\alpha_{M}=0
\label{alMcon}
\ee
is satisfied. Then, we obtain $\Psi=\Phi$ from 
Eq.~(\ref{PsiPhi}), so that there is 
no gravitational slip. 
Under the condition (\ref{alMcon}), we have 
$M^2=\Mpl^2=1/(8\pi G_N)$, where $G_N$ 
is the Newton gravitational constant. 
We will express $M^2$ 
by using $G_N$ in the following discussion.

\subsection{Energy transfer by $\alpha_{m_1}$}
\label{model1sec}

Let us first study theories given by 
the EFT functions
\be
\alpha_{B}=0\,,\qquad
\alpha_{m_1} \neq 0\,,\qquad 
\alpha_{m_2} = 0\,,\qquad 
\alpha_{\bar{m}_1} = 0\,.
\ee
Since the cubic-order Lagrangians 
$G_3(X) \square \phi$ and 
$G_3(\tilde{X}) \nabla_{\mu}A^{\mu}$ lead to 
$\alpha_{B} \neq 0$, we are now considering theories without those contributions. 
In the subclass of GP theories with the luminal speed of gravitational waves, 
we need the contribution 
$G_3(\tilde{X}) \nabla_{\mu}A^{\mu}$ to 
realize cosmic acceleration. 
In this sense, the discussion below is 
relevant to the subclass of scalar-tensor theories without 
the cubic-order Lagrangian $G_3(X) \square \phi$.
The EFT parameter $\alpha_{m_1}$ arises from 
the $X,Z$ dependence in $f_1$ 
in scalar-tensor theories.
In shift-symmetric scalar-tensor theories, $\alpha_{m_1}$ is related to 
$\alpha_{m_{\rm c}}$ via the consistency 
condition (\ref{consistency3Mr}). 
This is not the case in non-shift-symmetric 
scalar-tensor theories, where 
$\alpha_{m_{\rm c}}$
is independent of $\alpha_{m_1}$. 
In the following, we will estimate 
$G_{\rm eff}$ without 
imposing this consistency condition.

The effective gravitational coupling (\ref{Geff}) 
can be expressed in the form 
\be
G_{\rm eff}=G_N 
\left( 1+ {\cal G}_{\rm c} \right)\,,
\label{Geff1}
\ee
where the sign of ${\cal G}_{\rm c}$ characterizes 
whether the strong or weak gravitational 
interaction 
is realized or not.
The general expression of ${\cal G}_{\rm c}$ is 
cumbersome, so we will specify the model 
in which there is an attractor solution 
with cosmic acceleration characterized by  
$\epsilon_H={\rm constant} \neq 0$ and 
$\Omega_{\rm c}=0$. For example, the coupled 
quintessence given by the action 
\be
{\cal S}=\int \rd^4 x \sqrt{-g} \left[ 
X-V_0 e^{-\lambda \phi/\Mpl}-f_1 (\phi,X,Z)n 
\right]+{\cal S}_{\rm m}\,,
\label{model1}
\ee
where $V_0$ and $\lambda$ are constants, 
allows such a possibility. The presence of the 
exponential potential with $\lambda \neq 0$, 
which breaks the shift symmetry, leads to 
a tilt from the exact de Sitter solution 
at the attractor point. 
To estimate $G_{\rm eff}$ around the 
attractor solution, we expand ${\cal G}_{\rm c}$ 
around $\Omega_{\rm c}=0$ and 
$\dot{\epsilon}_H=0$, which leads to 
\be
{\cal G}_{\rm c}=-\frac{2(6H^2\alpha_{m_1}
+5H \dot{\alpha}_{m_1}+\ddot{\alpha}_{m_1})
+4H \epsilon_H[(\epsilon_H-4)H 
\alpha_{m_1}-\dot{\alpha}_{m_1}]}
{3H^2 \alpha_{m_1} \Omega_{\rm c}}
+{\cal O}(\Omega_{\rm c}^{0},
\dot{\epsilon}_H)\,,
\label{cG1}
\ee
where we have used $\alpha_g=0$ for 
scalar-tensor theories.
If the time variation of $\alpha_{m_1}$ 
is negligibly small and $\epsilon_H$ is suppressed relative to $\alpha_{m_1}$, the leading-order term 
is estimated to be
${\cal G}_{\rm c} \simeq -4/\Omega_{\rm c}$. 
Since ${\cal G}_{\rm c}$ is negative around 
the attractor with cosmic acceleration, 
$G_{\rm eff}$ is suppressed at low redshift.
Expanding the friction coefficient ${\cal C}$ 
in Eq.~(\ref{cC}) around $\Omega_{\rm c}=0$ and 
$\dot{\epsilon}_H=0$ and assuming that the time 
variations of $\alpha_{m_1}$ are negligible, 
it follows that
\be
{\cal C}=(5-2\epsilon_H)H
+{\cal O}(\Omega_{\rm c}^{0},
\dot{\epsilon}_H)\,,
\label{C1}
\ee
whose leading term is close to $5H$ for 
$\epsilon_H \ll 1$.
Taking the limits $\Omega_{\rm c} \to 1$, 
$\epsilon_H \to 3/2$, and 
$\dot{\alpha}_{m_1} \to 0$ 
in the deep matter-dominated epoch, 
we obtain the standard friction term 
${\cal C} \to 2H$.
So long as ${\cal C}$ continuously increases 
from $2H$ to the asymptotic value 
$(5-2\epsilon_H)H$, the friction term 
${\cal C}\dot{\delta}_{\rm c}$ works to suppress
the growth of $\delta_{\rm c}$ as well.

The leading-order terms in Eqs.~(\ref{cG1}) 
and (\ref{C1}) do not contain the dependence of $\alpha_{m_{\rm c}}$, so that the energy transfer arising from the time-dependent mass 
$m_{\rm c}$ is unimportant around the attractor. 
However, as shown in Refs.~\cite{Amendola:2020ldb,Liu:2023mwx}, 
the latter energy transfer can not be generally 
neglected at high redshift. 
We leave the detailed study about 
the evolution of perturbations at 
intermediate redshift for future work. 
The main message here is that the DE-DM interaction 
arising from $\alpha_{m_1}$ allows the possibility 
for realizing the growth rate of $\delta_{\rm c}$ 
smaller than in the $\Lambda$CDM model.

\subsection{Momentum transfer by 
$\alpha_{m_2}$}
\label{model2sec}

The next example is the EFT functions given by 
\be
\alpha_{B}=0\,,\qquad
\alpha_{m_2} \neq 0\,,\qquad 
\alpha_{m_1} = 0\,,\qquad 
\alpha_{\bar{m}_1} = 0\,.
\ee
The first condition $\alpha_{B}=0$ implies 
that we are now considering the subclass of 
scalar-tensor theories without the 
cubic-order Lagrangian $G_3(X) \square \phi$.
The EFT parameter $\alpha_{m_2}$ is 
associated with the $Z$ dependence in $f_1$ 
and $f_2$ in scalar-tensor theories.
In this case, the consistency condition 
(\ref{consistency3Mr}) imposes that
\be
\alpha_{m_{\rm c}}=0\,,
\label{almcon}
\ee
which is applied to  
shift-symmetric scalar-tensor theories.
The condition (\ref{almcon}) does not need to 
hold in non-shift-symmetric scalar-tensor 
theories, but we will consider the case where 
$\alpha_{m_{\rm c}}=0$ to extract the effect 
of the momentum transfer induced by $\alpha_{m_2}$.

The CDM effective gravitational coupling 
is given by the following simple form 
\ba
G_{\rm eff}=G_N 
\frac{3\Omega_{\rm c}(3\Omega_{\rm c}
+\alpha_{m_2}-2\epsilon_H)}
{3\Omega_{\rm c}(3\Omega_{\rm c}
+2\alpha_{m_2}-2\epsilon_H)
-2\epsilon_H \alpha_{m_2}}\,,
\label{Geffmo2}
\ea
where we have not used any expansion 
around the fixed point with cosmic acceleration. 
In the limit $\Omega_{\rm c} \to 0$, 
so long as $\epsilon_H \alpha_{m_2} 
\neq 0$, we have that $G_{\rm eff} \to 0$.
This property is consistent with 
those known for several coupled DE-DM 
models, e.g., quintessence coupled 
to DM with the interacting Lagrangian  
$f_2(Z)$~\cite{Kase:2019mox,Amendola:2020ldb}
and perfect fluid DE (described by the purely 
k-essence Lagrangian $K(X)$) coupled to DM 
with the function $f_2(Z)$~\cite{BeltranJimenez:2020qdu,BeltranJimenez:2021wbq}. 
Expanding the friction term ${\cal C}$ at an 
attractor point with $\Omega_{\rm c}=0$ and 
$\dot{\epsilon}_H=0$ and ignoring the time 
derivatives of $\alpha_{m_2}$, we obtain the 
same leading-order term as the one given in 
Eq.~(\ref{C1}). In the deep matter era characterized by $\Omega_{\rm c} \to 1$ 
and $\epsilon_H \to 3/2$, we have the asymptotic 
behavior ${\cal C} \to 2H$. As long as  
${\cal C}$ continuously increases toward the 
attractor, the friction term also suppresses 
the growth of $\delta_{\rm c}$ at the late 
cosmological epoch. 

We recall that the linear stability conditions 
are satisfied if
$q_{\rm s}>0$, $q_{\rm c}>0$, 
and $c_{\rm s}^2>0$, 
which translate, respectively, to
\ba
& &
\tilde{\alpha}_K>0\,,\qquad 
3\Omega_{\rm c}+\alpha_{m_2}>0\,,
\label{NGmo2}\\
& &
c_{\rm s}^2=\frac{2\alpha_{m_2}(\epsilon_H-3\Omega_{\text{c}})+3\Omega_{\text{c}}(2\epsilon_H-3\Omega_{\text{c}})}{\tilde{\alpha}_K(3\Omega_{\text{c}}+\alpha_{m_2})}>0\,.\label{csmo2}
\ea
Provided the two no-ghost conditions in 
Eq.~(\ref{NGmo2}) are satisfied, 
we have $c_{\rm s}^2>0$ for 
$2\alpha_{m_2}(\epsilon_H-3\Omega_{\text{c}})+3\Omega_{\text{c}}(2\epsilon_H-3\Omega_{\text{c}})>0$. 
In the limit $\Omega_{\text{c}} \to 0$, 
the squared sound speed reduces to 
$c_{\rm s}^2 \to 2\epsilon_H/\tilde{\alpha}_K$. 
So long as $\epsilon_H>0$ on the 
attractor with cosmic acceleration, 
the Laplacian instability 
is absent. If the attractor corresponds 
to a de Sitter fixed point with $\epsilon_H=0$, 
there may be a strong coupling problem 
associated with the vanishing value of $c_{\rm s}^2$. 
In the latter case, both the denominator and 
numerator of Eq.~(\ref{Geffmo2}) 
approach 0 toward the de Sitter solution, 
so that $G_{\rm eff}$ does not generally vanish 
unlike the case $\epsilon_H>0$.

For concreteness, we consider a momentum 
transfer model in the context of 
scalar-tensor theories \cite{Pourtsidou:2013nha,Boehmer:2015kta,Kase:2019veo,Amendola:2020ldb,Kase:2020hst}, 
which is characterized by the action 
\be
{\cal S}=\int {\rm d}^4 x \sqrt{-g} \left[ 
X-V_0 e^{-\lambda \phi/\Mpl} 
+\frac{\Mpl^2}{2}R-\hat{m}_{\rm c} n 
+\beta Z^2\right]+{\cal S}_{\rm m}\,,
\label{actionmo}
\ee
where $V_0$, $\lambda$, $\Mpl$, 
$\hat{m}_{\rm c}$, 
and $\beta$ are constants. Since 
$\hat{m}_{\rm c}$ is constant, there is 
no direct energy transfer between DE and DM.
This belongs to a subclass of Horndeski theories 
in which the coupling functions are given by
Eq.~(\ref{Gfun}) with $Q=0$. 
In this theory, the $\alpha$-basis parameters 
can be expressed as 
\be
\tilde{\alpha}_K=6\Omega_{\tilde{c}} 
=6 \left( 1+2\beta \right)x^2\,,
\qquad 
\alpha_{m_2}=12\beta x^2\,,
\label{alKmo2}
\ee
where $x:=\dot{\phi}/(\sqrt{6}H \Mpl)$, and 
$\alpha_K=0$. The ghost-free conditions 
in Eq.~(\ref{NGmo2}) are satisfied if
\be
1+2\beta>0\,,\qquad 
\Omega_{\rm c}+4\beta x^2>0\,.
\label{nogomo2}
\ee
In the background Eqs.~(\ref{back1}) 
and (\ref{back2}), we have 
$M_*^2f \to \Mpl^2$, 
$\Lambda \to 
V_0 e^{-\lambda \phi/\Mpl}$, 
$\tilde{c}/\bar{N}^2 \to 2\tilde{c}X \to \tilde{c} \dot{\phi}^2$, 
$\tilde{c} \to (1+2\beta)/2$, and $d \to 0$ 
in the current model, with $g_M=0$, 
$\bar{\rho}_{\rm m}=0$, and $\bar{p}_{\rm m}=0$. 
Subtracting 
Eq.~(\ref{back2}) from Eq.~(\ref{back1}), 
we obtain
\be
\epsilon_H=\frac{3}{2}\Omega_{\rm c}
+3 \left( 1+2\beta \right)x^2\,.
\label{epHmo2}
\ee
Substituting Eqs.~(\ref{alKmo2}) and (\ref{epHmo2}) into 
Eqs.~(\ref{Geffmo2}) and (\ref{csmo2}), we find
\be
G_{\rm eff}=G_N \frac{\Omega_{\rm c}}
{\Omega_{\rm c}+4\beta (1+2\beta)x^2}\,,
\qquad
c_{\rm s}^2=\frac{\Omega_{\rm c}
+4 \beta x^2 (1+2\beta)}{(1+2\beta)
(\Omega_{\rm c}+4 \beta x^2)}\,,
\ee
both of which agree with those obtained 
in Refs.~\cite{Kase:2019mox,Amendola:2020ldb}.
For $\beta>0$, the no-ghost conditions (\ref{nogomo2}) 
and the other condition $c_{\rm s}^2>0$ are 
automatically satisfied. 
As derived in Ref.~\cite{Kase:2019mox,Amendola:2020ldb}, 
the fixed point of the matter 
era is $(x,\Omega_{\rm c})=(0,1)$, 
whereas the fixed point of the  
late-time cosmic acceleration is 
$(x, \Omega_{\rm c})=(\lambda/[\sqrt{6}(1+2\beta)], 0)$. Then, $c_{\rm s}^2$ evolves 
from the initial value $1/(1+2\beta)$ and then 
finally approaches 1. We also find that 
$G_{\rm eff}$ starts from the initial value 
$G_N$ and evolves toward the asymptotic 
value $0$. The realization of weak gravity 
at low redshift is attributed to the fact that 
the scalar-field kinetic term $x^2$ dominates over $\Omega_{\rm c}$ around the fixed point with 
cosmic acceleration. 
In other words, neither $\epsilon_H$ nor $\alpha_{m_2}$ is vanishing at this fixed 
point, so that $G_{\rm eff} \to 0$ as $\Omega_{\rm c} \to 0$ in Eq.~(\ref{Geffmo2}).

The above discussion shows that, so long as 
$\epsilon_H$ and $\alpha_{m_2}$ are nonvanishing 
at the fixed point with cosmic acceleration, 
the weak cosmic growth rate of $\delta_{\rm c}$ 
can be realized at low redshift by the $Z$ 
dependence in $f_1$ or $f_2$.
We note that the term $f_{1,Z}$  
in $m_2^4$ is multiplied by the background 
CDM number density $\bar{n}$ 
proportional to $a^{-3}$, 
while this is not the case for $f_{2,Z}$.
In this sense, the $Z$ dependence in $f_2$ plays a more crucial role than its dependence in $f_1$. 
In perfect fluids models of coupled DE and DM, 
we need to be careful when using the 
quasi-static approximation at high 
redshift~\cite{BeltranJimenez:2020qdu,BeltranJimenez:2021wbq}.
In such cases, numerical studies are necessary 
to precisely trace the evolution of perturbations
for a wide range of scales.

The model given by the action (\ref{actionmo}) corresponds to the case in which $\tilde{c}$ 
does not vanish, while $\dot{d}=0$. 
In the EFT perspective, we may consider a model 
in which both $\tilde{c}$ and $\dot{d}$ are 
vanishing, with $M_*^2 f=\Mpl^2$. 
In such a case, the background reduces to 
pure $\Lambda$CDM, i.e., 
$\epsilon_H=3\Omega_{\text{c}}/2$.
Then, Eq.~(\ref{csmo2}) yields
$c_{\rm s}^2=-3\Omega_{\rm c} 
\alpha_{m_2}/[\tilde{\alpha}_K 
(3\Omega_{\text{c}}+\alpha_{m_2})]$.
Taking into account the ghost-free conditions 
(\ref{NGmo2}), the linearly stable region is 
characterized by $-3\Omega_{\text{c}}<\alpha_{m_2}<0$ 
and $\tilde{\alpha}_K>0$. As we approach the attractor with $\Omega_{\text{c}}\to0$, the window closes and only $\alpha_{m_2}=0$ is allowed. 
This is in contrast with the model 
(\ref{actionmo}), in which $\alpha_{m_2}$ 
approaches a positive constant for $\beta>0$.
Substituting the $\Lambda$CDM background solution
$\epsilon_H=3\Omega_{\text{c}}/2$ into 
Eq.~(\ref{Geffmo2}), it follows that 
$G_{\rm eff}=G_N$. In the same limit, 
we also have ${\cal C} \to 2H$. 
Hence, the CDM density contrast under the 
quasi-static approximation evolves according to 
\be
\ddot{\delta}_\text{c}+2H\dot{\delta}_\text{c}
-4\pi G_N \bar{\rho}_{\rm c} \delta_{\rm c}=0\,,
\ee
which is nothing but the standard equation for $\delta_{\rm c}$ in the $\Lambda$CDM model.
The evolution is thus insensitive to the EFT functions $\tilde{\alpha}_K$ and $\alpha_{m_2}$. 
This result shows how the above particular 
EFT scenario with the fixed $\Lambda$CDM background 
is difficult to realize the suppressed growth  
of $\delta_{\rm c}$. On the other hand, the 
dynamical DE scenario with concrete actions like 
(\ref{actionmo}) can naturally lead to 
the weak cosmic growth rate.

\subsection{Momentum transfer 
by $\alpha_{\bar{m}_1}$}
\label{model3sec}

The next example is the EFT functions given by 
\be
\alpha_{\bar{m}_1} \neq 0\,,\qquad
\alpha_{m_1} = 0\,,\qquad 
\alpha_{m_2} = 0\,,
\ee
From the consistency condition 
(\ref{consistency3Mr}), we have
\be
\alpha_{m_{\rm c}}=0\,,
\ee
which will be imposed in the following. 
The coupling $\bar{m}_1^2$ is present only for 
vector-tensor theories with the 
functions $f_1$ and $f_2$ depending 
on $\tilde{E}$. 
In GP theories, the background solution of the type 
$(A_0)^p \propto H^{-1}$, where $p$ is a positive constant, 
can be realized by the presence of the couplings 
$G_2 (\tilde{X})$ and $G_3 (\tilde{X})$ \cite{DeFelice:2020icf} (see also~\cite{BeltranJimenez:2016wxw}).
Since $\alpha_B \neq 0$ in such cases, we will include
the non-vanishing EFT parameter $\alpha_B$ 
in the following discussion. 
In this case, there is a de Sitter attractor
characterized by $\epsilon_H=0$ 
and $\Omega_{\rm c}=0$.\footnote{One could also 
have accelerated solutions with $\alpha_B=0$ from a minimum of the potential of the vector field. However, these solutions are prone to strong coupling problems owed to the vanishing of the propagation speed of the longitudinal mode that would require higher-order operators as 
in the ghost-condensate scenarios.}
We also assume that the time variations of 
$\alpha_{\bar{m}_1}$, $\alpha_B$, $\alpha_K$, 
and $\alpha_g$ are negligible 
at low redshift.
Expanding $G_{\rm eff}$ around the de Sitter 
self-accelerating solution with $\epsilon_H=0$ 
and $\Omega_{\rm c}=0$, it follows that 
$G_{\rm eff}$ is expressed in the form 
\ba
G_{\rm eff}=-G_N\frac{2\alpha_{\bar{m}_1}
(2\alpha_{\bar{m}_1}+5\alpha_B \alpha_g)}
{3 \alpha_B (\alpha_B+1)}
+{\cal O}(\epsilon_H,\Omega_{\rm c})\,.
\label{Geff3}
\ea
For $\alpha_g$ close to 0, the 
leading-order term of Eq.~(\ref{Geff3}) is 
$G_{\rm eff} \simeq -4G_N\alpha_{\bar{m}_1}^2
/[3 \alpha_B (\alpha_B+1)]$, and hence  
$G_{\rm eff}<0$ for $\alpha_B (\alpha_B+1)>0$.
Then, there is a possibility of realizing 
the repulsive gravitational interaction 
around the de-Sitter attractor. 
So long as the EFT parameter $\alpha_{\bar{m}_1}$ is subdominant at high 
redshift, $G_{\rm eff}$ can be positive 
initially and then decreases toward 
the asymptotic value (\ref{Geff3}). 
Applying a similar approximation to the one used 
for the derivation of $G_{\rm eff}$ around the de Sitter solution, we obtain the friction term 
${\cal C}=5H$ at leading order. Since 
the asymptotic value of ${\cal C}$ in the 
deep matter era is $2H$, the friction term 
works to suppress the growth of $\delta_{\rm c}$ 
for an increasing function ${\cal C}$.
Thus, there are possibilities for realizing the 
cosmic growth rate weaker than that in the 
$\Lambda$CDM model.
However, we need to caution that the ghost-free 
condition $q_{\rm c}>0$ imposes that 
$3\Omega_{\rm c}-\alpha_{\bar{m}_1}^2>0$. 
To ensure this inequality, we require that 
$\alpha_{\bar{m}_1}^2$ decreases faster than 
$3\Omega_{\rm c}$ at the late cosmological epoch. 
In comparison to the momentum transfer induced by $\alpha_{m_2}$, 
the fact that the term $-\alpha_{\bar{m}_1}^2$ 
in $q_{\rm c}$ is always negative implies that the realization of weak gravity without ghosts is more restrictive. 

\subsection{An explicit example}
\label{examplesec}

Finally, we will consider a simple model characterized by 
\be
\tilde{\alpha}_K=6\,,\qquad 
\alpha_{m_1}={\rm constant} \neq 0\,,\qquad 
\alpha_{m_2}={\rm constant} \neq 0\,,
\ee
while all the other 
$\alpha$-basis parameters are zero. 
We have chosen $\tilde{\alpha}_K=6$ because it nicely leads to a diagonal kinetic matrix ${\bm{K}}$ that simplifies the perturbation equations of motion. Furthermore, we will assume a $\Lambda$CDM 
background evolution, so that  
$\epsilon_H=3\Omega_\text{c}/2$ and 
$\dot{\Omega}_{\rm c}=3H \Omega_{\rm c} 
(\Omega_{\rm c}-1)$.

After eliminating non-dynamical perturbations, 
the dynamics of $\delta_{\rm c}$ and $\zeta$ 
are governed by the following equations
\begin{align}
&\deltacpp+2H \nu_{\text{eff}} \deltacp
-\frac{3}{2} H^2 \Omegac m_{\text{eff}}^2\, \deltac=\frac{k^2}{4a^2(3\Omegac+\alpham2)}\left[\frac{2(\alpham1+2\alpham2)}{H}\dot{\zeta}+\Big\{ 12\alpham2(1-\Omegac)+\alpham1 (3\Omegac+\alpham2)\Big\} \zeta\right]\,,\\
&\ddot{\zeta}+3H\dot{\zeta}-\frac{\alpham2 k^2}{6a^2}\zeta=-\frac{\alpham1+2\alpham2}{12}H\deltacp+\frac{6\alpham1(\Omegac-1)-\alpham2(6\Omegac-\alpham1)}{24}H^2\deltac\,,
\end{align}
where we have defined 
\begin{align}
\nu_{\text{eff}}:=&\frac{3(2-\alpham2)\Omegac+5\alpham2}{2(3\Omegac+\alpham2)}\,,\\
m_{\text{eff}}^2:=& \Big\{3\Omegac \big[ 6(4+4\alpham2-3\alpham1)\Omegac+\alpham1(8+\alpham1)-2\alpham2(8+5\alpham1) \big]+\alpham1\alpham2(20+\alpham1) \Big\}\nonumber \\
&/\big[24\Omegac (3\Omegac+\alpham2)\big]\,.
\end{align}
In the regime of small $\alpham1$ and $\alpham2$ 
close to 0, we can expand these 
expressions to obtain 
\begin{eqnarray}
\nu_{\text{eff}}&\simeq& 1-\frac{1}{2}\alpham2\left(1-\frac{1}{\Omegac}\right)\,,\\
m_{\text{eff}}^2&\simeq& 1-\frac{3}{4}\alpham1\left(1-\frac{4}{9\Omegac}\right)+\alpham2\left(1-\frac{1}{\Omegac}\right)\,,
\end{eqnarray}
at leading order.
For sufficiently super-Hubble modes ($k\to0$) 
and deep inside the matter-dominated 
epoch ($\Omegac \to 1$), the equation for the density contrast $\deltac$ decouples from $\zeta$. 
Furthermore, we have $\nu_{\text{eff}} \simeq 1$ 
and $m_{\text{eff}}^2 \simeq 
(\alpham1-4)(\alpham1-6)/24$ 
are nearly constant, so that we can obtain the following analytic solution for the density contrast:
\be
\deltac\big\vert_{k\to0} \simeq 
D_{1} t^{2/3-\alpham1/6}+
D_{2} t^{-1+\alpham1/6}\,.
\label{dck=0}
\ee
The mode $D_1$ gives the dominant solution for $\alpham1<5$, while for $\alpham1>5$ the dominant solution corresponds to the mode $D_2$. It is interesting to note that both modes give decaying solutions for $4<\alpham1<6$. Hence, for those values, there is no growing mode and, thus, no growth of structures. For $\alpham2=5$, both modes are degenerate. We can however see that the dominant mode grows slower than in the non-interacting case for $0<\alpham1<10$, 
so it realizes a weaker clustering.
It may look surprising that $\alpham2$ does not appear in this analysis of super-Hubble modes, 
but we should recall that $\alpham2$ represents 
the momentum interaction that is expected to be negligible on super-horizon scales because of 
the large-scale isotropy of the Universe.

Let us now turn our attention to smaller scales, 
i.e., below the corresponding sound horizon. 
The propagation speed for our choice of EFT functions 
is given by
\be
c_{\text{s}}^2=-\frac{12\alpha_{m_2}\Omega_{\text{c}}
-\alpha_{m_1}(\alpha_{m_1}+4\alpha_{m_2})}{24(3\Omega_{\text{c}}+\alpha_{m_2})}\,.
\ee
We are interested in the matter-dominated epoch wherein $\Omega_{\text{c}}\simeq1$ until $\Lambda$ starts dominating at late times. If we assume that 
$\alpham2 \ll 1$ (not necessarily $\alpham1$), the propagation speed deep in the matter-dominated 
epoch can be written as
\be
c_{\text{s}}^2\simeq -\frac{\alpham2}{6}(1+r),\quad\text{with}\quad r:=-\frac{\alpham1^2}{12\alpham2}\,.
\ee
Thus, Laplacian instabilities are avoided if we choose $\alpham2<0$. Furthermore, since today we have $\Omega_{\text{c},0}\simeq 0.3$, the assumption of small $\alpham2$ also guarantees that the ghost-free condition $3\Omega_{\text{c}}+\alpham2>0$ is satisfied until today. The ghost-free condition is automatically satisfied for $\alpham2>0$, in which case we need $12\alpha_{m_2}\Omega_{\text{c}}
-\alpha_{m_1}(\alpha_{m_1}+4\alpha_{m_2})<0$ to avoid Laplacian instabilities. We will see that the suppression of structures requires $\alpham2<0$, 
so we will assume that we are in this case from now on. Let us warn however that this regime cannot be sustained during the radiation-dominated epoch to avoid running into ghosts and/or Laplacian instabilities, so our choice of EFT functions must be understood as a proxy for the regime of interest. This regime must also be abandoned in the future evolution if the attractor corresponds to $\Omegac \to 0$. 
Let us notice in this respect that one of the original motivations for considering interacting models is in turn to obtain attractor solutions where the DM and DE density parameters are of the same 
order \cite{Wetterich:1994bg,Amendola:1999er,Amendola:1999qq}. 
In these cases, the regime with $\Omegac=0$ is never reached and the ghost-free condition may be less jeopardizing.

\begin{figure}[t!]
\centering    
\includegraphics[width=0.49\textwidth]{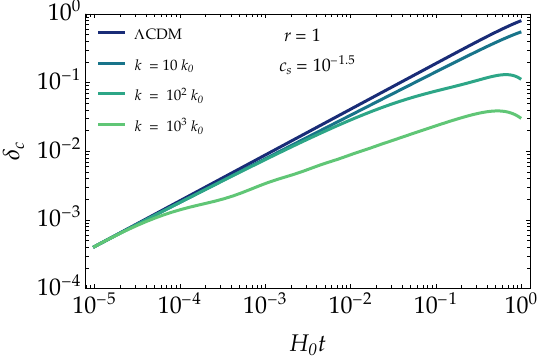}
\includegraphics[width=0.49\textwidth]{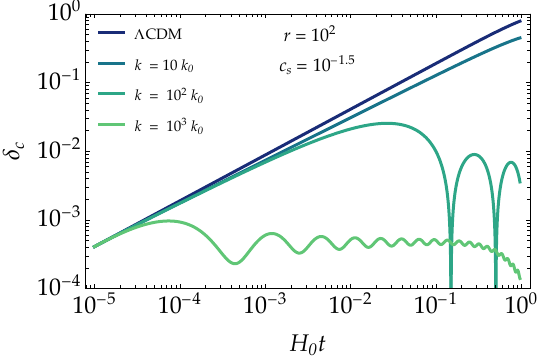}
\includegraphics[width=0.49\textwidth]{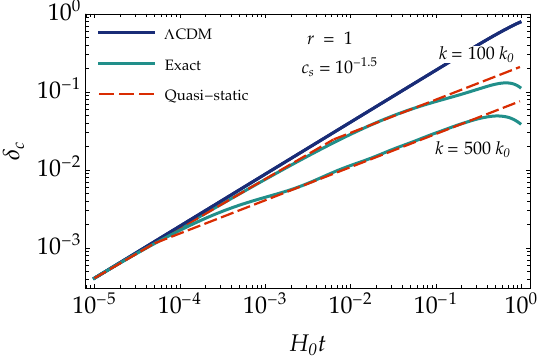}
\includegraphics[width=0.49\textwidth]{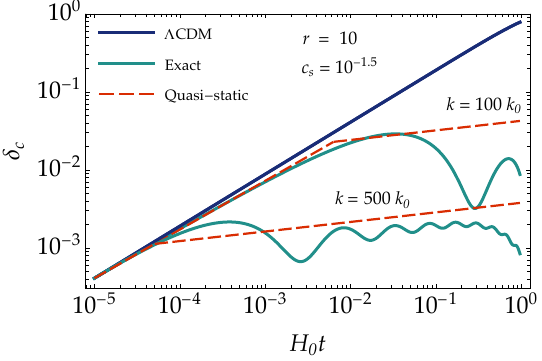}
\caption{Evolution of the DM density contrast for different Fourier modes (normalized to the present Hubble horizon $k_0=a_0H_0$) and EFT parameters. We have set initial conditions with $\deltac(t_{\text{ini}})=a(t_{\text{ini}})$, $\dot{\delta}_{\rm c}(t_{\text{ini}})=a(t_{\text{ini}})H(t_{\text{ini}})$, and $\zeta(t_{\text{ini}})=\dot{\zeta}(t_{\text{ini}})=0$. 
We have checked that $\zeta$ quickly reaches the attractor solution, so our results are not sensitive to the initial condition on $\zeta$ for the considered range of parameters. In all cases we have chosen the parameters $\alpham1$ and $\alpham2$, 
so that the sound horizon during matter domination is fixed and varied only $r$. The upper panels show the evolution of three Fourier modes that enter the sound horizon at different times. In the upper left panel, we can see how the sub-horizon evolution corresponds to a slow mode, while in the upper right panel the sub-horizon evolution oscillates, thus signalling the invalidity of the quasi-static approximation. We corroborate this in the lower panels, where we have compared the exact numerical solution with the solution that matches the super-horizon solution and the quasi-static solution at horizon crossing time $t_s$ defined as $c_s k=a(t_s) H(t_s)$. In all cases, however, we can observe how the clustering is reduced with respect to $\Lambda$CDM.
}
\label{Fig:density}
\end{figure}

Let us proceed with our analysis of sub-horizon scales. The quasi-static approximation amounts to having $\left\vert c_{\text{s}}^2k^2/(a^2H^2) \right\vert\gg1$, i.e., modes inside the sound horizon. Since $c_{\text{s}}^2$ is nearly constant in the matter era and smaller than 1 for small $\vert\alpham1\vert$ and $r$, the sound horizon is parametrically smaller than the Hubble horizon in that regime. This is an explicit example of the situation explained above where the sound horizon can be well below the Hubble horizon. In addition to being sub-horizon, the quasi-static approximation also requires the slow mode of $\zeta$ to dominate over its fast mode, but we will see that this does not need to be the case. Under the quasi-static approximation, the friction term and the effective Newton's constant at leading order in the EFT functions are given by 
\ba
\mathcal{C} &\simeq& 
2H\left[1+\frac{\alpham1(\alpham1+4\alpham2)}{8\alpham2\Omegac}(\Omegac-1)\right],\\
\frac{G_{\text{eff}}}{G_N} &\simeq&
1+\alpham1\frac{2\alpham2(\Omegac-1)(3\Omegac-10)+\alpham1(4-3\Omegac^2)}{12\alpham2\Omegac^2}\,,
\ea
where we have performed the expansion with 
respect to small $\alpham1$ and $\alpham2$. 
Deep in the matter-dominated epoch with $\Omegac=1$, we find $\mathcal{C}\simeq2H$ as in the standard case, while the effective Newton's constant is
\be
\frac{G_{\text{eff}}}{G_N}=\frac{\alpham2(6-\alpham1)^2}{12\alpham2(3-\alpham1)-3\alpham1^2}\simeq 1+\frac{\alpham1^2}{12\alpham2}\,,
\ee
where we have expanded for small values of 
$\alpham1$ and $\alpham2$. As advertised above, the effective Newton's constant in this regime is reduced provided $\alpham2<0$ and, thus, there is the suppression of the growth governed by the parameter $r = -\alpham1^2/(12\alpham2)$. In this regime, we can solve the equation for the density contrast to obtain the following solution:
\be
\deltac\big\vert_{\text{quasi-static}}\simeq C_{+}t^{c_+}+C_{-}t^{c_-},\quad\text{with}\quad c_\pm=\frac{1}{6}\left(-1\pm\sqrt{1+24\frac{G_{\text{eff}}}{G_N}}\right)
\,,
\label{eq:deltacquasistatic}
\ee
where we see again that the growing mode $C_+$ 
increases slowly due to a reduced effective 
Newton's constant. For small $\alpham1$ 
and $\alpham2$, 
we have $c_\pm \simeq (-1\pm 
5\sqrt{1-24r/25})/6$, 
so the effect in this regime is driven by the 
parameter $r$. Expanding Eq.~(\ref{eq:deltacquasistatic}) 
for small $r$, we obtain 
\be
\deltac\big\vert_{\text{quasi-static}}\simeq C_{+}t^{2/3}\left(1-\frac{2r}{5}\log t\right)+\frac{C_{-}}{t}\left(1+\frac{2r}{5}\log t\right)\,,
\ee
where we corroborate that the growing mode $C_{+}$ is logarithmically suppressed, similarly to the 
super-Hubble evolution, although now the suppressed evolution is determined by $r$. Thus, the considered model gives rise to slower 
growth of structures on super-Hubble scales driven by $m_{\text{eff}}^2$, while, on small (sub-horizon)  scales, the suppression of the growth rate is determined by $G_{\rm eff}/G_{N}$ that involves $\alpham2$. 

\begin{figure}[t!]
\centering    
\includegraphics[width=0.49\textwidth]{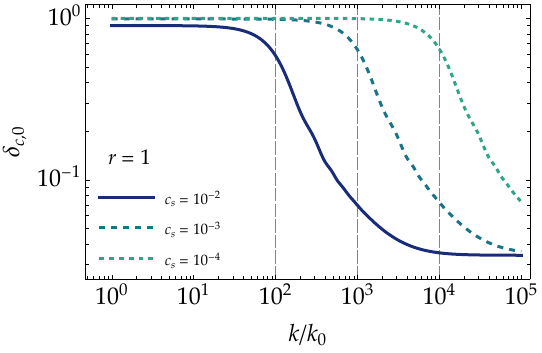}
\includegraphics[width=0.49\textwidth]{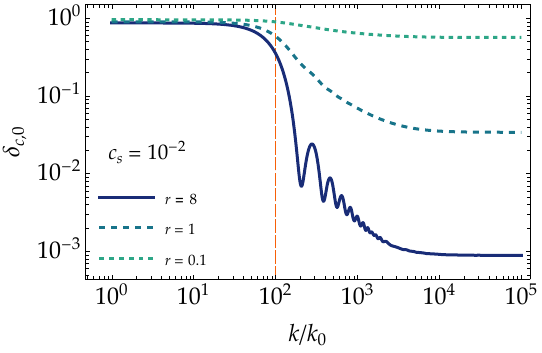}
\caption{In this Figure, we show the transfer function normalized to $\Lambda$CDM for different combinations of parameters. We normalize $k$ to the present Hubble horizon $k_0=a_0H_0$ and the vertical dashed lines correspond to the sound horizon scale. 
In the left panel, we keep $r$ fixed and plot 
today's CDM density contrast 
$\delta_{\rm c}$ for
several different values of $c_{\text{s}}^2$, while in the right panel we vary $r$ and keep the propagation speed fixed. We observe that $c_{\text{s}}$ determines the scales that undergo the 
suppression on small scales while $r$ drives the amount of suppression on those scales, in agreement with our analytical estimates in the main text. We also see the presence of acoustic oscillations that signal the failure of the quasi-static approximation. 
However, we can corroborate the suppressed growth on small scales in all cases. We also observe the small suppression on large scales due to the slower growth of super-Hubble modes.}
\label{Fig:TransferFunction}
\end{figure}

We can explicitly see the suppression of growth 
in the evolution of the density contrast depicted in Fig.~\ref{Fig:density}, where the full perturbation equations of motion have been numerically solved for a range of Fourier modes and different parameters. In that figure, we also illustrate the failure of the quasi-static approximation for some values of the parameters\footnote{In the upper and lower right panels, we keep the same wavelengths as the left panels but change the value of $r$ for illustrative purposes.}. This failure of the approximation originates from the fact that the fast modes dominate over the slow ones, thus leading to acoustic oscillations that can also be seen in the corresponding transfer function shown in Fig. \ref{Fig:TransferFunction}. We will not perform a detailed analysis of this explicit model here, but we simply want to illustrate how the suppression of the growth of structures can be achieved with a simple model as well as an explicit situation where the quasi-static approximation does not hold, but still the growth of structures is suppressed.

\section{Conclusions}

In this paper, we have constructed 
the EFT of coupled DE and DM, where DE interacts with DM through the energy and momentum transfers. For the DE sector, 
we have taken the unified description of 
vector-tensor and scalar-tensor theories 
by assuming the luminal 
propagation of gravitational waves.
This prescription allows us to accommodate 
not only GP theories 
but also shift-symmetric and 
non-shift-symmetric Horndeski theories as special cases.
The boundaries between different theories are characterized by the presence of a gauge coupling constant $g_M$ and the existence of consistency conditions. 
For the DM sector, we have used the effective description of fluids 
based on three scalar fields.

The EFT building blocks for DE are given 
by the three-dimensional intrinsic and extrinsic curvatures as well as 
$n_{\mu}$, $\tilde{g}^{00}
=g^{00}(1+g_M A_0)^2$, and 
$F_{\mu}=n^{\alpha}F_{\mu \alpha}$, where $F_{\mu \nu}=2\nabla_{[\mu}A_{\nu]}$ 
is the field strength of the vector 
field $A_{\nu}$.
The temporal part of the gauge 
transformation has been fixed 
to a unitary gauge in which 
the preferred timelike vector 
corresponds to $v_{\mu}=
\delta^{0}_{\mu}+g_M A_{\mu}$, 
with the configuration $A_{\mu}=
(A_0,{\bm 0})$.
For $g_M \neq 0$, this leads to 
a symmetry-breaking pattern 
different from that in scalar-tensor 
theories. In the DM sector, 
we have chosen the gauge condition 
$\phi^{i}= x^{i}$ for three scalar 
fields, under which the spatial part 
of the gauge transformation is fixed.
The DM Lagrangian is constructed 
from the energy density 
$\hat{m}_{\rm c}n$ 
with the bare mass $\hat{m}_{\rm c}$, 
by assuming that DM behaves as CDM 
with the vanishing pressure and 
sound speed. The building blocks 
in the DM sector are given by its
number density $n$ and 
four velocity $u^{\mu}$.

The interactions between DE and DM 
can be constructed from the EFT building 
blocks mentioned above. 
The coupling associated with 
the energy transfer is restricted 
to be linear in $n$ to avoid the 
non-vanishing propagation speed 
for DM. The leading operators describing 
the energy transfer between DE 
and DM consist of the products 
$\Delta m_{\rm c}(t)n$ and
$\delta \tilde{g}^{00} \delta n$, 
where $\Delta m_{\rm c}(t)$ is a 
time-dependent function. 
The other scalars associated 
with the momentum transfer are $n_{\mu}u^{\mu}$ and $F_{\mu}u^{\mu}$ or, equivalently $q_{\mu}q^{\mu}$  and $F_{\mu}q^{\mu}$, where 
$q^{\mu}=u^{\mu}+n^{\mu}
(n_{\nu}n^{\nu})$.
With these ingredients, we can write the total action for the EFT of coupled DE and DM in the form 
(\ref{Sfull}). The energy transfer 
is weighed by the two Lagrangians 
$-m_{\rm c}(t)n$ and 
$-m_1^4 (\delta n/\bar{n})
[\delta \tilde{g}^{00}
/(-\tilde{g}^{00}_{\rm BG})]$, 
whereas the momentum transfer 
consists of the two operators 
$-m_2^4 q^{\mu}q_{\mu}$ and 
$-\bar{m}_1^2 q^{\mu}F_{\mu}$.
In vector-tensor EFT, the  coefficients 
in the action (\ref{Sfull}) need to satisfy the consistency conditions
(\ref{consistency1Mr})-(\ref{consistency4MD}).
The shift-symmetric scalar-tensor 
EFT can be obtained by taking the limit $g_M \to 0$, while the 
non-shift-symmetric scalar-tensor 
theories correspond to the limit 
$g_M \to 0$ without imposing 
consistency conditions.

By using the EFT action (\ref{Sfull}) 
of coupled DE and DM with additional 
matter source (baryons, radiation), 
we derive the background equations of motion in the forms (\ref{back1}) and (\ref{back2}) with the matter continuity equation (\ref{coneq2}).
In our EFT setup of vector-tensor theories we have $\tilde{c}=0$, under which 
the background equations are 
simplified to Eqs.~(\ref{back1a}) 
and (\ref{back2a}).
The shift-symmetric scalar-tensor 
theories correspond to the limit 
$g_M \to 0$ with the consistency 
conditions (\ref{consistency1Mr})-(\ref{consistency4MD}).
Due to the consistency condition $\dot{f}=0$ in these two theories, the background dynamics is governed by the two EFT parameters $\Lambda$ and $d$. 
In non-shift-symmetric scalar-tensor theories, the function $f$ can vary in time due to the existence of the nonminimal coupling $G_4(\phi)R$. 
Using the dictionary between $\tilde{c}$, 
$d$, $\Lambda$ and the coupling functions 
appearing in a subclass of Horndeksi theories, 
we have confirmed that the background Eqs.~(\ref{nonback1}) and (\ref{nonback2}) coincide with those derived in the literature. 

In Sec.~\ref{sec:EFTaction_dimensionless}, we first expressed the Lagrangian 
${\cal L}_{\rm D}^{(2)}$ in Eq.~(\ref{EFTLD2}) 
by using the perturbations of $A_0$, $N$ and 
the shift vector $N^i$. 
The perturbed field $\delta A_{0}$ can be 
integrated out from the action, 
after which the coefficients 
$\mu_i$ ($i=1,2,\cdots,6$) in 
Eq.~(\ref{eq:EFT_action_no_dA}) 
depend not only on time but also on space. 
We also derived the second-order action arising 
from the non-linear Lagrangian 
(\ref{eq:non-linearpart}) 
in the form (\ref{Scon2}).
The total second-order action 
(\ref{eq:EQ_EFT_alpha_action}) 
is expressed in terms of the 
dimensionless $\alpha$-basis 
parameters. 
The time variation of the CDM mass 
$m_{\rm c}(t)$, which mediates the 
energy transfer, appears as the term $-6\Omega_{\rm c}H^2 (\delta N/\bar{N})
(\delta n/\bar{n})$ 
in Eq.~(\ref{eq:EQ_EFT_alpha_action}).
The EFT parameter $\alpha_{m_1}$ characterizes the other type of energy transfer associated 
with the interaction 
$\delta \tilde g^{00} \delta n$. 
The momentum transfer is weighed 
by the two parameters 
$\alpha_{m_2}$ and 
$\alpha_{\bar{m}_1}$, 
which are related to the 
interactions $q^{\mu}q_{\mu}$ 
and $q^{\mu}F_{\mu}$, respectively. 

The parameter $\alpha_g$ is an additional 
quantity indicating the contribution from 
vector-tensor theories. In the limit 
$\alpha_g \rightarrow 0$, 
all the EFT coefficients in Eq.~(\ref{eq:EQ_EFT_alpha_action}) reduce to functions of time, resulting in 
those of shift-symmetric 
scalar-tensor theories. 
If we do not impose consistency conditions, we can also accommodate the second-order 
action of non-shift-symmetric scalar-tensor theories. The mappings between these EFT parameters and the coupling functions 
of GP/Horndeski theories with DE-DM 
interactions are given in 
Appendix~\ref{sec:Dictionary}.
In the scalar-tensor limit, the 
operator $q^{\mu}F_{\mu}$ vanishes and 
hence $\alpha_{\bar{m}_1}=0$.

In Sec.~\ref{sec:perturbation_EOMs}, 
we derived the second-order actions 
of tensor and scalar perturbations 
as well as their perturbation 
equations of motion. 
In our EFT setup, the tensor mode propagates with the speed of light, with the modified 
friction term induced by the 
effective Plank mass squared 
$M^2=M_*^2 f$. 
By choosing the unitary gauge, 
the scalar perturbations 
are eaten by the metric.
Thus, the CDM density contrast 
is expressed in the form 
$\delta_{\rm c}=-(3\zeta+\nabla^2 E)$, 
where $\zeta$ and $E$ are metric 
perturbations appearing in the 
line element of three spatial dimensions.
For the standard matter part,
we consider the perfect fluid described 
by the Schutz-Sorkin action, 
which is equivalent to the 
three-scalar formulation used for CDM.
The total second-order action 
(\ref{Ssfinal}) of scalar perturbations contains six perturbed fields $\alpha$, $\chi$, $v_{\rm m}$, $\delta_{\rm c}$, 
$\zeta$, and $\delta_{\rm m}$, 
whose variations lead to the 
full sets of perturbation 
Eqs.~(\ref{pereq1})-(\ref{pereq3}) 
and (\ref{pereq4})-(\ref{pereq6}).

In Sec.~\ref{stasec}, we derived conditions for the absence of ghosts and Laplacian instabilities of dynamical scalar perturbations 
$\delta_{\rm c}$, $\zeta$, and 
$\delta_{\rm m}$ by taking the 
small-scale limit. 
The no-ghost conditions for 
DM ($\delta_{\rm c}$) and 
DE ($\zeta$) are 
given, respectively, by 
Eqs.~(\ref{snoghost1}) and 
(\ref{snoghost2}), where 
the former contains the effects 
of energy and momentum transfers 
through the dimensionless parameters 
$\Omega_{\rm c}$ and 
$\alpha_{m_2}$, $\alpha_{\bar{m}_1}$.
We showed that DE has the 
propagation speed squared $c_{\rm s}^2$
given by Eq.~(\ref{cst}), while 
the effective sound speed of DM 
is vanishing. In particular, the 
mixing matrix component 
$B_{12}~(=-B_{21})$, which 
contains the interacting 
EFT parameters $\alpha_{m_1}$, 
$\alpha_{m_2}$, and $\alpha_{\bar{m}_1}$, 
affects $c_{\rm s}^2$ through 
the contribution 
$\Delta c_{\rm s}^2$, which 
is positive under the no-ghost 
conditions. The sufficient condition 
for the absence of Laplacian 
instabilities is that 
$\hat{c}_{\rm s}^2>0$, where $\hat{c}_{\rm s}$ 
is defined by Eq.~(\ref{hatcs}). 
It is important to keep in mind that the derived stability conditions have been obtained in the strict large $k$ limit, but are not sufficient to establish full linear stability on all scales. Instability may appear on certain scales, for instance, in situations with a hierarchy of horizons (e.g., when the sound horizon is well inside the Hubble horizon). 

Using the quasi-static approximation 
for perturbations deep inside the 
DE sound horizon and neglecting 
the contribution of baryons, 
we derived the closed-form 
second-order differential equation 
for $\delta_{\rm c}$ in the form 
(\ref{delcqu}) with the effective gravitational 
coupling given by Eq.~(\ref{Geff}).
We also showed that the gravitational slip 
between the two gravitational potentials 
$\Psi$ and $\Phi$ is induced by 
the time variation of $M$ alone.
In Sec.~\ref{nointsec}, we considered 
theories with $\alpha_{m_1}=0$,  
$\alpha_{m_2}=0$, $\alpha_{\bar{m}_1}=0$ 
and showed that $G_{\rm eff}$ can be 
expressed in a simple form (\ref{Geff2}). 
So long as $V_{\rm s}>0$, which 
is ensured under the stability 
conditions $q_{\rm s}>0$ and 
$\hat{c}_{\rm s}^2>0$, the energy 
transfer induced by the time 
variation of the CDM mass, which 
is weighed by the EFT parameter 
$\alpha_{m_{\rm c}}$, works to 
enhance the gravitational attraction. 

In Secs.~\ref{model1sec}-\ref{model3sec}, 
we computed $G_{\rm eff}$ for the cases 
in which one of the 
EFT parameters $\alpha_{m_1}$,  
$\alpha_{m_2}$, and $\alpha_{\bar{m}_1}$ are 
non-vanishing, with $\alpha_M=0$. 
In these three cases, we showed that it is 
possible to have $G_{\rm eff}<G_N$ 
during the epoch of DE domination. 
This suppression of $G_{\rm eff}$ is 
attributed to the negative term (\ref{geff1}) 
in the numerator of Eq.~(\ref{Geff}) and the 
positive term $b_{12}^2 \nu_{\rm s}$ 
in the denominator of Eq.~(\ref{Geff}), 
which are both induced by the matrix 
component $B_{12}$. 
Physically, the presence of a non-vanishing 
pressure of DE can lead to the suppression of gravitational 
clusterings of CDM through the EFT parameters 
$\alpha_{m_1}$, $\alpha_{m_2}$, and 
$\alpha_{\bar{m}_1}$ because the pressure tends to oppose the gravitational collapse. Barring the obvious differences, the mechanism at work is analogous to the one that occurs in the primordial baryon-photon plasma in the pre-recombination epoch that prevents the clustering of baryons. 
The suppressed clustering of CDM by means of said EFT parameters naturally permits 
reducing the $\sigma_8$ tension. 
In Sec.~\ref{examplesec}, we also studied 
a specific model in which several $\alpha$-basis 
parameters are constant and showed that 
the suppressed growth of $\delta_{\rm c}$ 
can occur by the combination of the parameters 
$\alpham1$ and $\alpham2$. For this example, 
we also discussed the validity of the quasi-static 
approximation by considering several different wavenumbers. Upon using this example, we aimed at illustrating that care must be taken when working in the quasi-static approximation, and that, even beyond the quasi-static approximation, the suppression of structures can be achieved.

Thus, we have constructed a convenient and 
versatile EFT framework of coupled DE and DM 
mediated by the energy and momentum 
transfers. The next natural step is to incorporate 
our EFT scheme into a Boltzmann numerical 
code and perform a proper confrontation to data.
By using the current and upcoming 
observational data, we can place 
constraints on the EFT parameters 
and probe the signatures of 
DE-DM interactions. 
In particular, it will be of interest to 
study how the $\sigma_8$ tension 
may be alleviated by the EFT parameters 
$\alpha_{m_1}$, $\alpha_{m_2}$, 
and $\alpha_{\bar{m}_1}$. 
Let us however stress that, irrespective of the $\sigma_8$ tension, the developed framework will serve to explore interactions in the dark sector and improve the exploitation of observational data with e.g., the design of new observables. For instance, it was shown in Ref.~\cite{BeltranJimenez:2022irm} that momentum interactions can produce a distinctive signature in the dipole of the matter power spectrum. Such an effect could be more generally studied within our EFT framework.

\section*{Acknowledgements}

The work of K.A. was supported by JSPS KAKENHI Grant Nos.~20H05852 and 24K17046. J.B.J. was supported by the Project PID2021-122938NB-I00 funded by the Spanish “Ministerio de Ciencia e Innovaci\'on” and FEDER “A way of making Europe” and the Project SA097P24 funded by Junta de Castilla y Le\'on.  M.C.P has been partially supported by supported by Mahidol University (Fundamental Fund: fiscal year 2025 by National Science Research and Innovation Fund (NSRF)). M.C.P is
also supported by the National Research Foundation of Korea (NRF) through grants RS-2023-NR077094 and RS-2020-NR049598 (Center for Quantum Spacetime: CQUeST). S.T. was supported by the Grant-in-Aid for Scientific Research Fund of the JSPS No.~22K03642 and Waseda University Special Research Project No.~2024C-474.

\newpage

\appendix

\section{Dictionaries between EFT parameters and concrete theories}\label{sec:Dictionary}

In this Appendix, we present 
the dictionaries relating EFT 
parameters with the coupling functions 
appearing in the sub-classes 
of Horndeski theories and GP theories.

\subsection{Horndeski theories}
\label{Hocorres}

We first consider a scenario 
of coupled scalar DE and DM 
given by the following Lagrangian 
\begin{align}
\mathcal{L}=\mathcal{L}_{\rm H}+\mathcal{L}_{\rm DM}+\mathcal{L}_{\rm int}+\mathcal{L}_{\rm m}\,,
\end{align}
where $\mathcal{L}_{\rm H}$ is the Horndeski Lagrangian \eqref{Horndeski}, 
$\mathcal{L}_{\rm DM}
=-\hat{m}_{\rm c}n$, and 
$\mathcal{L}_{\rm m}$ is the matter contribution. 
We consider the 
interaction between DE and DM given by~\cite{Kase:2019veo,Kase:2020hst,Amendola:2020ldb}
\begin{align}
\mathcal{L}_{\rm int}=-f_1(\phi,X,Z)\rho_{\rm DM} +f_2(\phi,X,Z)\,,
\label{Hointer}
\end{align}
where $X=-\nabla_{\mu} \phi \nabla^{\mu}\phi/2$ 
and $Z=u^{\mu}\nabla_{\mu}\phi$. 
Recall that the DM density and 
the four-velocity 
are given, respectively, 
by $\rho_{\rm DM}=\hat{m}_{\rm c} n$ and 
\eqref{eq:four-velocity}, 
with $n$ being the number density \eqref{eq:number_density}.
While Refs.~\cite{Kase:2019veo,Kase:2020hst,Amendola:2020ldb} used the Schutz-Sorkin description 
of the (irrotational) perfect fluid, 
we describe the DM fluid by the 
three scalars $\phi^i$.
The dust Lagrangian $\mathcal{L}_{\rm DM}=-\hat{m}_{\rm c} n$ can be absorbed into the definition of $f_1$. Then, we write the DM Lagrangian and the interacting part, as
\begin{align}
\mathcal{L}_{\rm DM}+\mathcal{L}_{\rm int}=-f_1(\phi,X,Z)n +f_2(\phi,X,Z)\,,
\label{LDMint}
\end{align}
where the constant 
$\hat{m}_{\rm c}$ is also absorbed into $f_1$.

The EFT form of the Lagrangian can be found in the following manner~\cite{Gleyzes:2013ooa}. 
Upon choosing the unitary gauge 
$\phi=t$, we have 
$X=-g^{00}/2=1/(2N^2)$ and $Z=u^0$.
The Horndeski Lagrangian can 
be expressed in the ADM 
form, as
\begin{align}
\mathcal{L}_{\rm H} =
G_2+2Xg_{3,\phi}-2(2X)^{1/2}(Xg_{3,X}+G_{4,\phi}) K
+G_4\left[\Rs+K_{\mu\nu}K^{\mu\nu}-K^2 \right]
\,,
\end{align}
where $g_3(\phi,X)$ is defined by 
$G_3=g_3+2Xg_{3,X}$, 
and the notations like 
$g_{3,\phi}:=\partial g_3
/\partial \phi$ and 
$g_{3,X}:=\partial g_3
/\partial X$ are used. 
The Lagrangian (\ref{LDMint}) yields
\begin{align}
\mathcal{L}_{\rm DM}+\mathcal{L}_{\rm int}=-f_1(t,X,Z)n +f_2(t,X,Z)\,.
\end{align}
From Eq.~(\ref{eq:q2}), the quantity 
${\cal U}=-N u^0=-N Z$ satisfies 
the relation 
${\cal U}^2=1+q_{\mu} q^{\mu}$. 
In this case, we can choose the 
branch ${\cal U}=-\sqrt{1+q^{\mu}q_{\mu}}$, so that 
\begin{align}
Z=\frac{1}{N} \sqrt{1+q^{\mu}q_{\mu}}
=\frac{1}{N} \left(1-\frac{N_iN^i}
{N^2}\right)^{-1/2}\,,
\end{align}
where we used Eq.~(\ref{qmuex}).
Then, we obtain the action \eqref{Sfull} 
with the correspondence
\begin{tcolorbox}[colback=black!5!white,colframe=black!75!white,title=Dictionary to Horndeski theories]
\begin{align}
M_*^2 f &=2G_4\,,\label{fre} \\
\Lambda&=-G_2 +G_{2,X}\bX+2g_{3,\phi X} 
\bX^2 -3H \sqrt{2\bX}
\Bigl(3 g_{3,X} \bX + 2 g_{3,XX}\bX^2
+ G_{4,\phi} \Bigr)
\nonumber \\ 
& ~~~
-\bar{n}\left(f_{1,X}\bX
+\frac{1}{2}\sqrt{2\bX} f_{1,Z} \right) - f_2 + f_{2,X}\bX +\frac{1}{2}\sqrt{2\bX}
f_{2,Z} \,, \label{Lambda0} \\
\tilde{c} &=\frac{1}{2}G_{2,X}+g_{3,\phi}
+g_{3,\phi X}\bX 
-\frac{3}{2}H\sqrt{2\bX} 
(3 g_{3,X}+2g_{3,XX}\bX)  
-3H (2\bX)^{-1/2}G_{4,\phi}
\nn
&~~~-\frac{1}{2}\bar{n}
\left[ f_{1,X}+(2\bX)^{-1/2} f_{1,Z} \right] 
+ \frac{1}{2}\left[ f_{2,X}+ (2\bX)^{-1/2}f_{2,Z} \right]
\,, \label{c0} \\
d&= 2\sqrt{2\bX}(g_{3,X} \bX 
+ G_{4,\phi})\,, \label{d0} \\
M_2^4 &=G_{2,XX}\bX^2+2\bX^2
(2g_{3,\phi X} + g_{3,\phi XX} \bX) 
\nn
&~~-\frac{3}{2}H \bX \sqrt{2\bX} \left(3 g_{3,X}+12 g_{3,XX}\bX + 4 g_{3,XXX}\bX^2 \right) + \frac{3}{2}H \sqrt{2\bX} G_{4,\phi}
\nn 
&~~-\bar{n}\left[ f_{1,XX}\bX^2 +\frac{1}{2}f_{1,ZZ}\bX-\sqrt{2\bX}
\left(\frac{1}{4}f_{1,Z}-f_{1,XZ}\bX \right)
\right] 
\nn
&~~
+f_{2,XX}\bX^2+\frac{1}{2}f_{2,ZZ}\bX 
-\sqrt{2\bX} \left(\frac{1}{4}f_{2,Z}-f_{2,XZ}\bX \right)
\,, \label{M2def} \\
\bar{M}_1^3&=-2\sqrt{2\bX}
\left(3 g_{3,X}\bX +2 g_{3,XX}\bX^2+G_{4,\phi}
\right)
\,, \\
m_{\rm c} &= f_1
\,, \\
m_1^4 &= -\bar{n}\left(f_{1,X} \bX + \frac{1}{2}\sqrt{2\bX} f_{1,Z}\right)
\,, \\
m_2^4 &=\frac{1}{2}\sqrt{2\bX}
\left(f_{1,Z}\,\bar{n}-f_{2,Z}\right)
\,,\label{m2def2}
\end{align}
and
\begin{align}
g_M=\gamma_1=\bar{m}_1^2=0\,,
\label{eq:gMmap}
\end{align}
\end{tcolorbox}
\noindent
where $\bX=-g^{00}_{\rm BG}/2$ is 
the background value of $X$.
The quantity $\sqrt{2\bX}$ should be interpreted 
as the scalar field derivative 
$\dot{\phi}$, which is positive 
upon the choice of the unitary gauge $\phi=t$.

The coupled quintessence scenario studied in 
Refs.~\cite{Amendola:2020ldb,Liu:2023mwx} 
corresponds to the functions 
\ba
& &
G_2=X-V(\phi)\,,\qquad g_3=0\,,\qquad 
G_4=\frac{\Mpl^2}{2}\,,\nonumber \\
& &
f_1= e^{Q\phi/\Mpl}
\hat{m}_{\rm c}\,,
\qquad
f_2=\beta Z^2\,,
\label{Gfun}
\ea
where $V(\phi)$ is a scalar potential. 
The constants $Q$ and $\beta$ characterize 
the strengths of energy and momentum transfers, 
respectively. In this case, we have 
\be
m_{\rm c}=e^{Q\phi/\Mpl} \hat{m}_{\rm c}\,,\qquad 
m_1^4=0\,,\qquad m_2^4=-\beta Z_{\rm BG} \sqrt{2X_{\rm BG}}\,,
\ee
where $Z_{\rm BG}=1/\bar{N}
=\sqrt{-g^{00}_{\rm BG}}$ is the background value of $Z$. 
Undoing the unitary gauge, we have 
$Z_{\rm BG}=\dot{\phi}>0$ and 
$\sqrt{2X_{\rm BG}}=\dot{\phi}$, 
so that 
$m_2^4=-\beta \dot{\phi}^2$.
For $Q \neq 0$, the energy transfer 
occurs through the 
time-dependent effective mass $m_{\rm c}(t)$, whose effect 
appears as the term $\alpha_{\rm c}=\dot{m}_{\rm c}/(Hm_{\rm c})$ 
in $G_{\rm eff}$. Since $m_1^4=0$ in the present 
case, there is no energy exchange associated with 
the EFT function $\alpha_{m_1}$. 
For $\beta \neq 0$, the momentum transfer occurs 
through the non-vanishing EFT function $\alpha_{m_2}$.

\subsection{GP theories}
\label{sec:GP}

As for a vector DE scenario, we consider 
the subclasses 
of GP theories~\eqref{eq:GP} with the 
following coupling to DM~\cite{DeFelice:2020icf,Pookkillath:2024ycd}:
\begin{align}
\mathcal{L}_{\rm DM}+\mathcal{L}_{\rm int}
=-f_1(\tilde{X},\tilde{Z},
\tilde{E})n+f_2(\tilde{X},\tilde{Z},
\tilde{E})\,,
\label{GPint}
\end{align}
with $\tilde{X}=-A_{\mu}A^{\mu}/2$, $\tilde{Z}=A_{\mu}u^{\mu}$, and 
$\tilde{E}=-A^{\mu}F_{\mu\nu}u^{\nu}$.

The difference from scalar-tensor 
theories is the absence of preferred 
hypersurfaces because 
the vector field $A_{\mu}$, which is supposed to be 
non-vanishing and timelike, is not hypersurface orthogonal, in general. Nonetheless, one can introduce the projection tensor to decompose tensors into parallel and orthogonal components with 
respect to the preferred vector. 
To make a clear connection to scalar-tensor theories, we introduce the St\"uckelberg field $A_{\mu}\to A_{\mu}+g_M^{-1} \partial_{\mu}\phi$ 
and fix the gauge to $\phi=t$, where 
$g_M>0$ is the gauge coupling. 
This implies the following replacement rule
\begin{align}
A_{\mu} \to - \frac{\sqrt{-\tilde{g}^{00}}}
{g_M}\tilde{n}_{\mu}\,, 
\end{align}
where $\tilde{n}_{\mu}$ is 
the unit vector, and
\begin{align}
\tilde{g}^{00}:=g^{00}+2g_M A^0 
+ g_M^2 A_{\mu}A^{\mu}\,.
\end{align}
Note that $\tilde{n}^{\mu}$ is a future-directed vector, meaning that the vector field of the GP theories is chosen to be past-directed 
in our construction.
The kinematical quantities are, as usual, 
defined by
\begin{align}
\tilde{K}_{\mu\nu}:=\tilde{h}_{(\mu}{}^{\alpha}\nabla_{\alpha} 
\tilde{n}_{|\nu)}\,, \qquad  \tilde{\omega}_{\mu\nu}:=\tilde{h}_{[\mu}{}^{\alpha}\nabla_{\alpha}
\tilde{n}_{|\nu]}\,,\qquad
\tilde{a}_{\mu}:=\tilde{n}^{\alpha}\nabla_{\alpha}\tilde{n}_{\mu}\,,
\end{align}
with the projection $\tilde{h}_{\mu\nu}:=g_{\mu\nu}
+\tilde{n}_{\mu}\tilde{n}_{\nu}$. 
We can also define an object analogous to the curvature with respect to the projection tensor $\tilde{h}_{\mu\nu}$. While details can be found in Appendix~A of Ref.~\cite{Aoki:2021wew}, the equation relevant to us is the Raychaudhuri equation
\begin{align}
R={}^{(3)}\!\tilde{R}+\tilde{K}_{\mu\nu}
\tilde{K}^{\mu\nu}-\tilde{K}^2-\tilde{\omega}_{\mu\nu}\tilde{\omega}^{\mu\nu}-2\nabla_{\mu}(\tilde{a}^{\mu}-\tilde{K}\tilde{n}^{\mu})\,.
\end{align}
Finally, the field strength is decomposed into
\begin{align}
\tilde{F}_{\mu}:=\tilde{n}^{\alpha}
F_{\mu\alpha}\,, \qquad \tilde{F}_{\mu\nu}:=\tilde{h}^{\alpha}{}_{\mu}\tilde{h}^{\beta}{}_{\nu}F_{\alpha\beta}\,,
\end{align}
where the ``magnetic'' part $\tilde{F}_{\mu\nu}$ is related to the vorticity $\tilde{\omega}_{\mu\nu}$, 
via $\tilde{\omega}_{\mu\nu}=-g_M \tilde{F}_{\mu\nu}/
2\sqrt{-\tilde{g}^{00}}$.
These ingredients allow us to take the 
$3+1$ decomposition of 
GP theories~\eqref{eq:GP}:
\begin{align}
\mathcal{L}_{\rm GP}=
\frac{1}{2}\tilde{F}_{\mu}
\tilde{F}^{\mu}
-\frac{1}{4}\tilde{F}_{\mu\nu}\tilde{F}^{\mu\nu}
+G_2-(2\tilde{X})^{3/2}g_{3,X}
\tilde{K}+\frac{\Mpl^2}{2}\left[{}^{(3)}\!\tilde{R}
+\tilde{K}_{\mu\nu}\tilde{K}^{\mu\nu}-\tilde{K}^2-\tilde{\omega}_{\mu\nu}\tilde{\omega}^{\mu\nu}\right],
\end{align}
and
\begin{align}
\tilde{X}=\frac{-\tilde{g}^{00}}{2g_M^2}\,, \qquad \tilde{Z}=\sqrt{2\tilde{X}(1+\tilde{q}^{\mu}\tilde{q}_{\mu})}\,, 
\qquad \tilde{E}
=-\sqrt{2\tilde{X}}\tilde{q}^{\mu}
\tilde{F}_{\mu}\,,
\end{align}
where $\tilde{q}^{\mu}:=u^{\mu}
+\tilde{n}^{\mu}(\tilde{n}\cdot u)$, and $g_{3,\tilde{X}}$ is defined by
$G_3:=g_3+2 \tilde{X} g_{3,\tilde{X}}$.

Now, we assume the irrotational ansatz 
\eqref{Amua}, corresponding to
$\tilde{\omega}_{\mu\nu} 
\propto \tilde{F}_{\mu\nu}=0$.
Then, the preferred vector $\tilde{n}_{\mu}$ is 
hypersurface orthogonal and all tilded 
quantities are reduced to the untilded ADM 
variables except for $\tilde{g}^{00}$ 
(or equivalently $\tilde{X}$):
\begin{align}
\mathcal{L}_{\rm GP} \to 
\frac{1}{2}F_{\mu}F^{\mu}+G_2-(2\tilde{X})^{3/2}g_{3,\tilde{X}}K+\frac{\Mpl^2}{2}\left[{}^{(3)}\!R+K_{\mu\nu}K^{\mu\nu}-K^2\right]\,,
\end{align}
and 
\begin{align}
\tilde{Z} \to \sqrt{2\tilde{X}(1+q^{\mu}q_{\mu})}\,, \qquad \tilde{E}\to-\sqrt{2\tilde{X}}q^{\mu}F_{\mu}\,.
\end{align}
Hence, we obtain the unified form of 
the EFT action \eqref{Sfull} with 
the following coefficients:
\begin{tcolorbox}[colback=black!5!white,colframe=black!75!white,title=Dictionary to GP theories]
\begin{align}
M_*^2 f &=\Mpl^2 \,, \label{eq:DfP}\\
\Lambda&=-G_2 +G_{2,\tilde{X}}\btX 
-3H \sqrt{2\btX}\left(3 g_{3,\tilde{X}} \btX 
+ 2 g_{3,\tilde{X}\tilde{X}}\btX^2 \right)
\nn&~~~
-\bar{n}\left(f_{1,\tilde{X}}\btX+\frac{1}{2} \sqrt{2\btX} f_{1,\tilde{Z}} 
\right) - f_2 + f_{2,\tilde{X}}\btX +\frac{1}{2}\sqrt{2\btX}f_{2,\tilde{Z}} \,, \label{Lambda}\\
g_M^2 \tilde{c} &=\frac{1}{2}G_{2,\tilde{X}} 
-\frac{3}{2}H\sqrt{2\btX}\,(3 g_{3,\tilde{X}}
+2g_{3,\tilde{X} \tilde{X}}\btX)  \nn
&~~~-\frac{1}{2}\bar{n}
\left[ f_{1,\tilde{X}}+(2\btX)^{-1/2} 
f_{1,\tilde{Z}} \right] 
+ \frac{1}{2}\left[ f_{2,\tilde{X}} 
+ (2\btX)^{-1/2}f_{2,\tilde{Z}} \right]
\,, \label{tildec}\\
d&=(2\btX)^{3/2}g_{3,\tilde{X}}\,, 
\label{dcorre}\\
M_2^4 &=G_{2,\tilde{X}\tilde{X}}\btX^2
-\frac{3}{2}H \btX \sqrt{2\btX} 
\left(3 g_{3,\tilde{X}}+12 g_{3,\tilde{X}\tilde{X}}\btX 
+ 4 g_{3,\tilde{X}\tilde{X}\tilde{X}}\btX^2 \right) 
\nn 
&~~~-\bar{n}\left[ f_{1,\tilde{X}\tilde{X}}\btX^2 +\frac{1}{2}f_{1,\tilde{Z}\tilde{Z}}\btX- (2\btX)^{1/2}\left(\frac{1}{4}f_{1,\tilde{Z}}-f_{1,\tilde{X}\tilde{Z}}\btX \right)
\right] 
\nn
&~~~
+f_{2,\tilde{X}\tilde{X}}\btX^2+\frac{1}{2}f_{2,\tilde{Z}\tilde{Z}}\btX - (2\btX)^{1/2}\left(\frac{1}{4}f_{2,\tilde{Z}}-f_{2,\tilde{X}\tilde{Z}}\btX \right)
\,,\label{M2d} \\
\bar{M}_1^3&=-2\sqrt{2\btX}
\left(3 g_{3,\tilde{X}}\btX 
+2 g_{3,\tilde{X}\tilde{X}}\btX^2\right)
\,, \\
\gamma_1 &=1
\,, \\
m_{\rm c} &= f_1
\,, \\
m_1^4 &= -\bar{n}\left(f_{1,\tilde{X}} \btX 
+ \frac{1}{2}\sqrt{2\btX}\,f_{1,\tilde{Z}}\right)
\,, \\
m_2^4 &=\frac{1}{2}\sqrt{2\btX}\,\left(f_{1,\tilde{Z}}\,\bar{n}-f_{2,\tilde{Z}}\right)
\,,\\
\bar{m}_1^2 &= 
-\sqrt{2\btX}\,(f_{1,\tilde{E}}\,
\bar{n}-f_{2,\tilde{E}})\,.\label{bm1} 
\end{align}
\end{tcolorbox}
\noindent
The DE-DM coupling studied in Ref.~\cite{DeFelice:2020icf} 
corresponds to $f_1=\hat{m}_{\rm c}
={\rm constant}$, 
$f_2=f_2(\tilde{X},\tilde{Z})$, which translates to
\be
m_{\rm c}=\hat{m}_{\rm c}\,,\qquad 
m_1^4=0\,,\qquad 
m_2^4=-\frac{1}{2}\sqrt{2\btX}\,
f_{2,\tilde{Z}}\,,
\qquad \bar{m}_1^2=0\,.
\ee
In this case, there is no energy transfer, 
but the momentum transfer occurs through 
the EFT function $m_2^4$. 
For the model investigated in Ref.~\cite{Pookkillath:2024ycd}, 
$f_2$ has the $\tilde{E}$ dependence and hence 
there is also the momentum transfer 
induced by $\bar{m}_1^2$.

\section{Ghost-free conditions for vector perturbations in the decoupling limit}
\label{vecsec}

In this section, we study ghost-free conditions of vector-type perturbations in the EFT of 
vector-tensor theories. Note that the EFT action \eqref{Sfull} is valid only under the irrotational ansatz \eqref{Amua}. 
We should first upgrade the untilded variables 
into the tilded variables introduced in Appendix~\ref{sec:GP} to incorporate 
vector perturbations:
\begin{align}
\mathcal{L}_{\rm D}^{\rm NL} &\to
\frac{M_*^2}{2}f(t) \left( 
{}^{(3)}\!\tilde{R}+\tilde{K}_{\mu\nu}\tilde{K}^{\mu\nu}-\tilde{K}^2 -\tilde{\omega}_{\mu\nu}\tilde{\omega}^{\mu\nu}
\right)
-\Lambda(t) - c(t) \tilde{g}^{00}-d(t)K - m_{\rm c}(t) n\,,
\\
\mathcal{L}_{\rm D}^{(2)}& \to
\frac{1}{2}M_2^4{} (t) \left(\frac{\delta \tilde{g}^{00}}
{-\tilde{g}^{00}_{\rm BG}} \right)^2 
-\frac{1}{2}\bar{M}_1^3(t) 
\left(\frac{\delta \tilde{g}^{00}}
{-\tilde{g}^{00}_{\rm BG}} \right) 
\delta \tilde{K} +\frac{1}{2}\gamma_1(t) \tilde{F}_{\mu}\tilde{F}^{\mu} 
 \nn
&~~~-m_1^4(t) \frac{\delta n}{\bar{n}} \left(\frac{\delta \tilde{g}^{00}}{-\tilde{g}^{00}_{\rm BG}} \right) - m_2^4(t) \tilde{q}^{\mu} \tilde{q}_{\mu} 
-\bar{m}_1^2(t) \tilde{q}^{\mu} \tilde{F}_{\mu}+\cdots\,, 
\end{align}
where $\cdots$ are operators that vanish under the irrotational ansatz \eqref{Amua}. The missing operator at leading order in the derivative equation is $\tilde{F}_{\mu\nu}\tilde{F}^{\mu\nu}$ alone.
As we will see shortly, however, this operator 
is unimportant for our discussion.

For simplicity, we take the decoupling limit of gravity. To do so, let us introduce the St\"uckelberg fields according to
\begin{align}
t \to t+\pi_0\,,\qquad A_{\mu} \to A_{\mu}-g_M^{-1}\partial_{\mu} \pi_0 \,, \qquad
x^i \to x^i + \pi^i\,,
\end{align}
and ignore metric perturbations. 
Up to the necessary orders of 
perturbations, we find
\begin{align}
\tilde{F}_{\mu} &\simeq 
(0, \partial_i A_0 - \dot{A}_i)\,, \qquad
\tilde{F}_{\mu\nu} \simeq
\begin{pmatrix}
0 & 0 \\
0 & 2\partial_{[i} A_{j]}
\end{pmatrix}
\,, \nonumber \\
n &\simeq \bar{n}\left\{ 1+\partial_i \pi^i -\frac{1}{2}\left[\dot{\pi}_i^2 + (\partial_i \pi_j)^2 -(\partial_i \pi^i)^2 \right] \right\}\,,
\qquad q^{\mu} \simeq (0, -\dot{\pi}^i)\,.
\end{align}
The expressions for $\tilde{g}^{00}, \tilde{K}_{\mu\nu}$ and ${}^{(3)}\!\tilde{R}$, 
which are irrelevant to vector stability 
conditions, can be found in 
Ref.~\cite{Aoki:2021wew}.
Focusing on vector perturbations, 
the relevant operators in the decoupling 
limit are thus
\begin{align}
- m_{\rm c}(t) n + \frac{1}{2}\gamma_1(t) \tilde{F}_{\mu}\tilde{F}^{\mu} - m_2^4(t) \tilde{q}^{\mu} \tilde{q}_{\mu} 
-\bar{m}_1^2(t) \tilde{q}^{\mu} 
\tilde{F}_{\mu}\,,
\label{op_vector}
\end{align}
and the corresponding quadratic 
Lagrangian is
\begin{align}
\mathcal{L}^{(2)}_{\rm V} = \frac{1}{2}
(\bar{\rho}_{\rm c}-2 m_2^4) \dot{\pi}_i^2 
+ \frac{1}{2}\gamma_1 \dot{A}_i^2 -\bar{m}_1^2 \dot{A}_i \dot{\pi}_i +\cdots\,,
\end{align}
where $\bar{\rho}_{\rm c}=m_{\rm c}\bar{n}$.
As a result, the ghost-free conditions 
for vector perturbations are
\begin{align}
\bar{\rho}_{\rm c}-2 m_2^4>0 \,, \qquad \gamma_1>0\,, \qquad 
\gamma_1 (\bar{\rho}_{\rm c}-2 m_2^4)
>\bar{m}_1^4\,.
\label{GF_vector}
\end{align}

Let us briefly comment on the condition 
to avoid the Laplacian instability. 
The gradient terms of vector perturbations 
are generated by $\tilde{F}_{\mu\nu}\tilde{F}^{\mu\nu}$, 
whose sign is thus fixed by the stability. 
However, the operator $\tilde{F}_{\mu\nu}\tilde{F}^{\mu\nu}$ does not contribute to the scalar and tensor perturbations, 
so that the analysis in the main text is not 
affected.
This is different from the ghost-free conditions \eqref{GF_vector}, especially $\gamma_1>0$, which should be taken into account in studying other sectors of perturbations.

\section{EFT formulation of the matter sector}\label{sec:3scalarMatter}

Let us assume a matter sector in the form of 
a perfect fluid described in the EFT formalism. Analogously to the DM sector, we will introduce a set of three scalar fields $\psi^a$ 
(where $a=1,2,3$), describing the Lagrangian coordinates of the matter fluid. 
We then construct the fundamental matrix $B_{(\text{m})}^{ab}=\partial_\mu\psi^a
\partial^\mu\psi^b$, so the action of 
the matter sector is simply
\be
\mathcal{S}_{\text{m}}=
\int\rd^4x\sqrt{-g}\,F(B)\,,
\ee
with $B=\det B_{(\text{m})}^{ab}$ that is related to the number density as $n_{\text{m}}=\sqrt{B}$. At first order in scalar perturbations, 
we have
\be
\psi^a=\lambda_{\text{m}}(x^a+\partial^a\pim)\,,
\ee
where $\lambda_{\text{m}}$ is a constant and $\pim$ denotes the longitudinal phonon perturbation of the fluid. The energy density and pressure of 
the fluid are given, respectively, 
by $\rho_{\text{m}}=-F(B)$ and $p_{\text{m}}=-2BF_{,B}+F(B)$, so that  $\rho_\text{m}+p_{\text{m}}=-2BF_{,B}$. 
The density contrast can be computed as
\be
\delta_{\text{m}}:=\frac{\delta 
\rho_{\text{m}}}{\rho_{\text{m}}}
=\frac{\delta F}{F}
=\frac{2B F_{,B}}{F}
\left(\nabla^2\pim-\nabla^2E-3\zeta\right)\,.
\ee
This relation permits the density contrast to be related to the phonon perturbation in any gauge. 
In particular, in the unitary gauge with $\pim=0$, the above expression relates the density contrast 
of the fluid with e.g., the metric 
perturbation $E$, as it has been exploited in the main text. It is important to notice however that the unitary gauge can only be imposed on one of the fluids and we will maintain our choice of unitary gauge for the DM component. Let us also 
note that, by taking $F(B)\propto \sqrt{B}$ 
and in the unitary gauge, we naturally recover the relations used for the DM sector. 
If we go to Fourier space, we can obtain the (non-local) relation of the phonon field $\pim$ and the corresponding density contrast:
\be
\pim=E-\frac{1}{k^2}\left(\frac{F}{2BF_{,B}}\delta_{\text{m}}+3\zeta\right)\,.
\label{eq:piTodelta}
\ee
This relation, together with the corresponding one for the DM sector obtained in the unitary gauge,
\be
E=\frac{1}{k^2}\left(\delta_{\text{c}}
+3\zeta\right)\,,
\ee
allows to use $\delta_c$ and $\delta_{\text{m}}$ instead of $E$ and $\pim$. The quadratic action for the matter sector reads 
\begin{align}
\mathcal{S}_{\text{m}}=&-\int\rd^4xNa^3BF_{,B}\Bigg\{a^2\Big(\partial_i\dot{\pi}_{\text{m}}-\frac{1}{a^2} \partial_i\chi\Big)^2- c_{\text{m}}^2 \Big(3\zeta- \nabla^2\pim+\nabla^2E\Big)^2\nonumber\\
&-\frac{F}{2BF_{,B}} \left[3\zeta^2-\alpha^2+6\alpha\zeta+\frac{1}{a^2}(\partial_i\chi)^2+2\nabla^2E\Big(\alpha+\zeta-\frac12\nabla^2E\Big)\right]\nonumber\\
& +\left[2\nabla^2E\Big(\zeta+\alpha-\frac12\nabla^2E\Big)-2\alpha\nabla^2\pi+6\alpha\zeta+3\zeta^2\right]\Bigg\}\,.
\end{align}
From this expression, we immediately see that the condition for avoiding a ghostly matter phonon is given by $BF_{,B}<0$, which, in terms of the density and pressure, recovers the condition $\rho_\text{m}+p_{\text{m}}>0$, i.e., 
the null energy condition. 
We can alternatively express the quadratic 
action in terms of the density contrast 
instead of the phonon field by using the relation \eqref{eq:piTodelta}. The resulting quadratic action coincides with Eq.~\eqref{SMS2} upon integrating out the non-dynamical field $v_{\text{m}}$. We can then complement the matter action with the remaining part describing the interacting EFT sector and obtain the corresponding full quadratic action. It is interesting to notice that the high-frequency limit leads to a diagonal kinetic matrix when using the density contrast variables. However, if expressed in terms of the original phonon field, the kinetic matrix exhibits off-diagonal terms mixing $E$ and $\pi_{\text{m}}$.

\bibliography{ref}

\end{document}